\documentclass[
  journal=pasa,
  manuscript=research-paper, 
  year=2024,
  volume=37,
  useAMS
]{cup-journal}

\usepackage[columnwise]{lineno}
\usepackage{pdflscape}
\usepackage{microtype,siunitx,booktabs,amssymb,amsmath,bm}
\usepackage{upgreek}
\usepackage{CJKutf8}
\newcommand{\dm}{pc\,cm$^{-3}$}

\newcommand{\HTR}{D.~Scott et al., in prep.}
\newcommand{\pox}{P_{\rm host}}
\newcommand{\halflight}{\ensuremath{\phi}}
\newcommand{\PO}{\ensuremath{P(O)}}
\newcommand{\POx}{\ensuremath{P(O|x)}}
\newcommand{\NFRB}{43}

\newcommand{\NEWFRB}{22}
\newcommand{\ALLLOCALISED}{37}
\newcommand{\ALLHOST}{30}
\newcommand{\NEWLOCALISED}{16}
\newcommand{\NEWHOSTS}{11}
\newcommand{\UNLOCALISED}{6}
\newcommand{\pccm}{\dm}

\newcommand{\hostwidth}{2.25in}

\title{The Commensal Real-time ASKAP Fast Transient incoherent-sum survey}

\author{R.~M.~Shannon}
\affiliation{Centre for Astrophysics and Supercomputing, Swinburne University of Technology, John St., Hawthorn, VIC 3122, Australia}
\email[R. M. Shannon]{rshannon@swin.edu.au}

\author{K.~W.~Bannister}
\affiliation{Australia Telescope National Facility, CSIRO Space \& Astronomy, Box 76 Epping, NSW 1710, Australia}
\alsoaffiliation{Sydney Institute for Astronomy, School of Physics, The University of Sydney, Camperdown, NSW 2006, Australia}

\author{A.~Bera}
\affiliation{International Centre for Radio Astronomy Research (ICRAR), Curtin University, Bentley, WA 6102, Australia
}

\author{S.~Bhandari}
\affiliation{ASTRON, Netherlands Institute for Radio Astronomy, Oude Hoogeveensedijk 4, 7991 PD Dwingeloo, The Netherlands}
\alsoaffiliation{Joint institute for VLBI ERIC, Oude Hoogeveensedijk 4, 7991 PD Dwingeloo, The Netherlands}
\alsoaffiliation{Anton Pannekoek Institute for Astronomy, University of Amsterdam, Science Park 904, 1098 XH, Amsterdam, The Netherlands}
\alsoaffiliation{Australia Telescope National Facility, CSIRO Space \& Astronomy, Box 76 Epping, NSW 1710, Australia}

\author{C.~K.~Day}
\affiliation{Department of Physics, McGill University, Montreal, Quebec H3A 2T8, Canada}

\author{A.~T.~Deller}
\affiliation{Centre for Astrophysics and Supercomputing, Swinburne University of Technology, John St., Hawthorn, VIC 3122, Australia}

\author{T.~Dial}
\affiliation{Centre for Astrophysics and Supercomputing, Swinburne University of Technology, John St., Hawthorn, VIC 3122, Australia}

\author{D.~Dobie}
\affiliation{Centre for Astrophysics and Supercomputing, Swinburne University of Technology, John St., Hawthorn, VIC 3122, Australia}
\alsoaffiliation{The ARC Centre of Excellence for Gravitational-Wave Discovery (OzGrav)}

\author{R.~D.~Ekers}
\affiliation{Australia Telescope National Facility, CSIRO Space \& Astronomy, Box 76 Epping, NSW 1710, Australia}
\alsoaffiliation{International Centre for Radio Astronomy Research (ICRAR), Curtin University, Bentley, WA 6102, Australia
}
\author{W.-f. Fong}
\affiliation{Center for Interdisciplinary Exploration and Research in Astrophysics (CIERA) and Department of Physics and Astronomy, Northwestern University, Evanston, IL 60208, USA}

\author{M.~Glowacki}
\affiliation{International Centre for Radio Astronomy Research (ICRAR), Curtin University, Bentley, WA 6102, Australia
}

\author{A.~C.~Gordon}
\affiliation{Center for Interdisciplinary Exploration and Research in Astrophysics (CIERA) and Department of Physics and Astronomy, Northwestern University, Evanston, IL 60208, USA}

\author{K.~Gourdji}
\affiliation{Centre for Astrophysics and Supercomputing, Swinburne University of Technology, John St., Hawthorn, VIC 3122, Australia}

\author{A.~Jaini}
\affiliation{Centre for Astrophysics and Supercomputing, Swinburne University of Technology, John St., Hawthorn, VIC 3122, Australia}

\author{C.~W.~James}
\affiliation{International Centre for Radio Astronomy Research (ICRAR), Curtin University, Bentley, WA 6102, Australia
}

\author{P.~Kumar}
\affiliation{Department of Particle Physics and Astrophysics, Weizmann Institute of Science, 76100 Rehovot, Israel}

\author{E.~K.~Mahony}
\affiliation{Australia Telescope National Facility, CSIRO Space \& Astronomy, Box 76 Epping, NSW 1710, Australia}

\author{L.~Marnoch}
\affiliation{School of Mathematical and Physical Sciences, Macquarie University, NSW 2109, Australia}
\alsoaffiliation{Astrophysics and Space Technologies Research Centre, Macquarie University, Sydney, NSW 2109, Australia}
\alsoaffiliation{Australia Telescope National Facility, CSIRO Space \& Astronomy, Box 76 Epping, NSW 1710, Australia}
\alsoaffiliation{The ARC Centre of Excellence for All-Sky Astrophysics in 3 Dimensions (ASTRO 3D)}

\author{A.~R.~Muller}
\affiliation{Maria Mitchell Observatory, Nantucket, MA 02554, USA}

\author{J.~X.~Prochaska}
\affiliation{Department of Astronomy and Astrophysics, University of California, Santa Cruz, CA 95064, USA}
\alsoaffiliation{ Kavli Institute for the Physics and Mathematics of the Universe (Kavli IPMU), 5-1-5 Kashiwanoha, Kashiwa, 277-8583, Japan}
\alsoaffiliation{Division of Science, National Astronomical Observatory of Japan, 2-21-1 Osawa, Mitaka, Tokyo 181-8588, Japan}

\author{H.~Qiu (\begin{CJK*}{UTF8}{gbsn}邱昊\end{CJK*})}
\affiliation{SKA Observatory, Jodrell Bank, Lower Withington, Macclesfield, SK11 9FT, UK}

\author{S.~D.~Ryder}
\affiliation{School of Mathematical and Physical Sciences, Macquarie University, NSW 2109, Australia}
\alsoaffiliation{Astrophysics and Space Technologies Research Centre, Macquarie University, Sydney, NSW 2109, Australia}

\author{E.~M.~Sadler}
\affiliation{Sydney Institute for Astronomy, School of Physics A28, University of Sydney, NSW 2006, Australia}
\alsoaffiliation{Australia Telescope National Facility, CSIRO Space \& Astronomy, Box 76 Epping, NSW 1710, Australia}

\author{D.~R.~Scott}
\affiliation{International Centre for Radio Astronomy Research (ICRAR), Curtin University, Bentley, WA 6102, Australia
}

\author{N.~Tejos}
\affiliation{Instituto de F\'isica, Pontificia Universidad Cat\'olica de Valpara\'iso, Casilla 4059, Valpara\'iso, Chile}

\author{P.~A.~Uttarkar}
\affiliation{Centre for Astrophysics and Supercomputing, Swinburne University of Technology, John St., Hawthorn, VIC 3122, Australia}

\author{Y.~Wang}
\affiliation{Centre for Astrophysics and Supercomputing, Swinburne University of Technology, John St., Hawthorn, VIC 3122, Australia}

\doi{XXXXX}

\received {dd Mmm YYYY}
\revised  {dd Mmm YYYY}
\accepted {dd Mmm YYYY}
\published{dd Mmm YYYY}

\keywords{galaxies: distances and redshifts, stars: general, radio transient source} 

\begin{document}
\begin{abstract}
With  wide-field phased array feed technology, the Australian Square Kilometre Array Pathfinder (ASKAP) is ideally suited to search for seemingly rare radio transient sources that are difficult to discover previous-generation narrow-field telescopes.  The Commensal Real-time ASKAP Fast Transient (CRAFT) Survey Science Project has developed instrumentation to continuously search for fast radio transients (duration $\lesssim$\,1 second) with ASKAP, with a particular focus on finding and localising Fast Radio Bursts (FRBs).
Since 2018, the CRAFT survey has been searching for FRBs and other fast transients by incoherently adding the intensities received by individual ASKAP antennas, and then correcting for the impact of frequency dispersion on these short-duration signals in the resultant incoherent sum (ICS) in real-time. This low-latency detection enables the triggering of voltage buffers, which facilitates the localisation of the transient source and the study of spectro-polarimetric properties at high time resolution.
Here we report the sample of $\NFRB$ FRBs discovered in this CRAFT/ICS survey to date. 
This includes $\NEWFRB$ FRBs that had not previously been reported:  $\NEWLOCALISED$ FRBs localised by ASKAP to $\lesssim 1$\,arcsec and $\UNLOCALISED$ FRBs localised to $\sim 10$\,arcmin.
Of the new arcsecond-localised FRBs, we have identified and characterised host galaxies (and measured redshifts) for $\NEWHOSTS$. 
The median of all $\ALLHOST$ measured host redshifts from the survey to date is $z=0.23$. 
We summarise results from the searches, in particular those contributing to our understanding of the burst progenitors and emission mechanisms, and on the use of bursts as probes of intervening media. 
We conclude by foreshadowing future FRB surveys with ASKAP using a coherent detection system that is currently being commissioned. This will increase the burst detection rate 
by a factor of approximately ten and also the distance to which ASKAP can localise FRBs.

\end{abstract}


\section{INTRODUCTION}
\label{sec:intro}

Fast radio bursts (FRBs) are millisecond duration bursts of radio emission  of extragalactic origin that have  emerged  as crucial tools for understanding fundamental physics.
The study of FRBs  earnestly commenced with the  discovery of the fast radio burst FRB\,20010724A \cite[the Lorimer Burst, ][]{2007Sci...318..777L} in archival {\em Murriyang} (the 64-m Parkes telescope)  observations recorded with the 20-cm multibeam system \cite[][]{1996PASA...13..243S}.   
The burst was exceptionally bright and had a dispersion measure (DM) well in excess of what could be accounted for by the Milky Way Galaxy  ionised gas content along a high Galactic latitude line of sight. 
Due to the tenuousness of the intergalactic medium, the excess dispersion implied that the burst could have arisen at distances far greater than their closest analogues, pulsars \cite[][]{2007Sci...318..777L}.
After the discovery of the Lorimer Burst,  progress in understanding the nature of the FRBs was slow for two reasons.   
Firstly, the systems capable of detecting bursts (pulsar-search systems) used single-dish telescopes, so localisations were poor. 
Secondly, the detection rate was low. 
It was only  when further examples were discovered in the {\em Murriyang} High Time Resolution Universe Survey \cite[][]{2013Sci...341...53T} that evidence tipped in favour of the bursts being astrophysical \footnote{Other signals with similar spectro-temporal structure were discovered in Murriyang data \cite[][]{2011ApJ...727...18B} that  were subsequently identified to be radio-frequency interference produced by an observatory microwave oven \cite[][]{2015MNRAS.451.3933P}.}. 
The poor localisations meant it was not possible to associate an FRB to any particular source (a star, galaxy, or other object) resulting in intense debate on the cause and distance scale to fast radio bursts. 
FRBs continued to be discovered by {\em Murriyang} and other telescopes.  
However, rare and bright FRBs were shown not to be as rare with the discovery of the FRB\,20150807A, which had a similar brightness to the Lorimer burst \cite[][]{2016Sci...354.1249R}. 
The discovery of a second bright FRB implied that 10\% of the Murriyang FRBs found to that date would have been detectable by a single 12-m antenna\footnote{The discovery also spurred on the development of the Deep Synoptic Array 10 element array \cite[][]{2019Natur.572..352R}}.

The wider field of view of smaller antennas gave hope of a reasonable detection rate. One way the field of view of a radio telescope can be widened further is through the use of multi-element receivers. The Australian Square Kilometre Array Pathfinder \cite[][]{2021PASA...38....9H},  a 36-antenna interferometer with 12-m antennas and phased array feed (PAF) receivers capable of observing a $\sim 30\,\deg^2$ field, is one such instrument.
With a field of view a factor of $50$ greater than the Parkes multibeam system, it was realised that ASKAP would be able to detect FRBs at a rate competitive with other existing and planned facilities.  
When used as an interferometer, it would be capable of localising FRBs and start answering many of the confounding questions that existed about them at the time, such as the distance scale to their progenitors and their utility as a probe of the intergalactic medium \cite[][]{2016arXiv160505890C}.

The first FRB detection systems were commissioned in 2016 by the Commensal Real-time ASKAP Fast Transient (CRAFT) Collaboration. 
Initial searches in 2017 and 2018 were conducted in a fly's eye mode, pointing a sub-array typically comprising $8-12$ ASKAP antennas, with each in individual directions. This enabled  shallow but wide field of view searches. This was also the first natural technical development towards localising FRBs. 
The searches were successful with the first burst found within $3.4$ days of observing with 8 antennas\footnote{The  majority of fly's eye searches were conducted on a separate commissioning subarray of antennas equipped with digital backend subsystems before the ASKAP hardware correlator was capable of ingesting  and processing data from all $36$ antennas.} \cite[][]{2017ApJ...841L..12B}.  Over the course of the next year we continued this strategy, focusing on a set of $45$ high Galactic latitude ($|b| \approx 50^\circ)$  fields discovering $20$\, FRBs \cite[][]{2018Natur.562..386S}.  This represented the first well-controlled sample of FRBs \citep{2019PASA...36....9J}, with the dense sampling of the ASKAP beams allowing for a good localisation of the bursts within the beam pattern, and a reliable estimate of burst fluence.
Further all-sky \cite[][]{2019MNRAS.486...70B}, Galactic plane \cite[][]{2019MNRAS.486..166Q}, and Galactic latitude $|b| \approx 20\deg$ \cite[][]{2019ApJ...872L..19M} searches also detected FRBs.

At the conclusion of these searches, we upgraded the ASKAP FRB search systems to enable the interferometric localisation of FRB detections.
The searches operate on the incoherent sum (ICS) of intensities from all beams of all antennas (with the antennas now pointing in the same direction).    Localisation is achieved by conducting low latency (sub-second) searches,  which enable the triggered download of $3.1$-s voltage buffers, which can be correlated, calibrated,  imaged, and beamformed to localise an FRB and study its spectropolarimetric properties.  

\begin{figure*}
 \centering
 \begin{tabular}{ccc}

\includegraphics[width=\hostwidth]{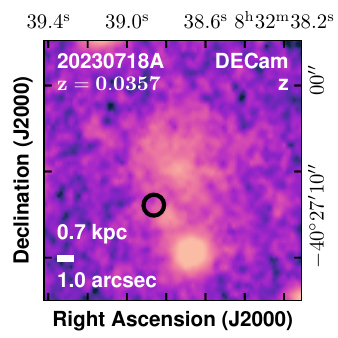}  &
\includegraphics[width=\hostwidth]{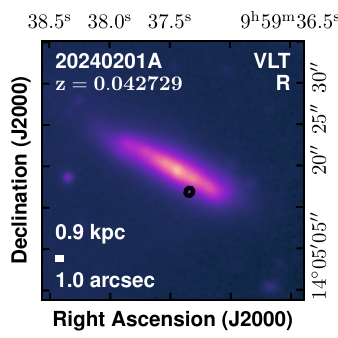} & 
\includegraphics[width=\hostwidth]{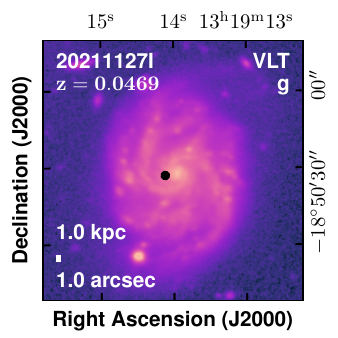} \\

\includegraphics[width=\hostwidth]{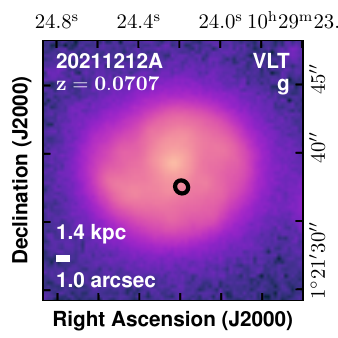} &
\includegraphics[width=\hostwidth]{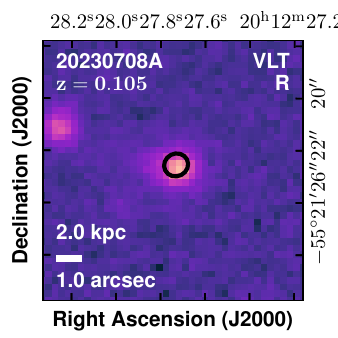}  &
\includegraphics[width=\hostwidth]{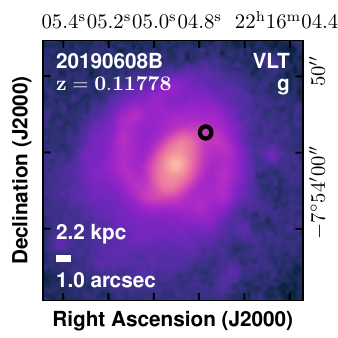} \\

\includegraphics[width=\hostwidth]{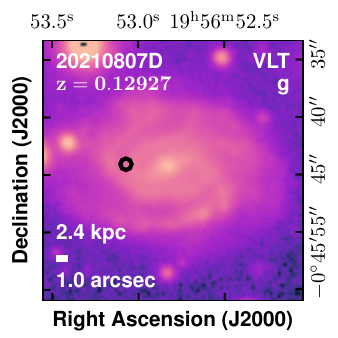} &
\includegraphics[width=\hostwidth]{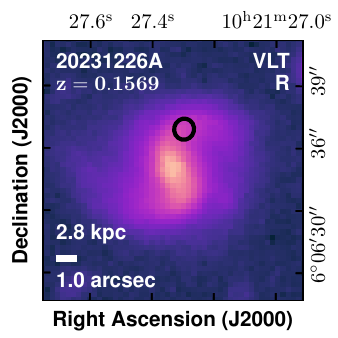}  &
\includegraphics[width=\hostwidth]{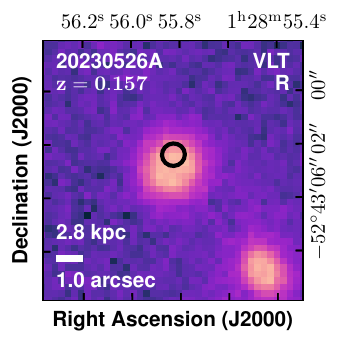}  \\

\includegraphics[width=\hostwidth]{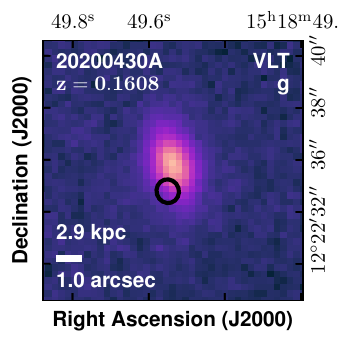} & 
\includegraphics[width=\hostwidth]{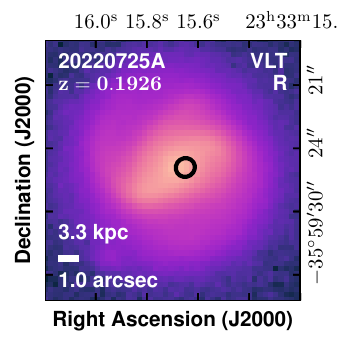} &
\includegraphics[width=\hostwidth]{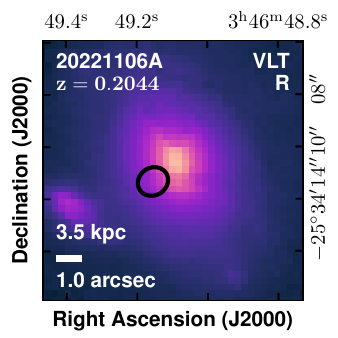} 

\end{tabular}
\caption{Montage of FRB host galaxies for ASKAP-localised FRBs with firm associations.  We present the FRBs in increasing redshift.  In the top left corner, we show the FRB name and redshift. In the top right corner, we list the telescope and observing band of the image.  The angular and physical scale at the host redshift are shown in the lower left corner. 
The 1-$\sigma$ localisation region of the
FRB is given by the black ellipse.
\label{fig:host1}
}
\end{figure*}

\begin{figure*}
 \centering
 \begin{tabular}{ccc}

\includegraphics[width=\hostwidth]{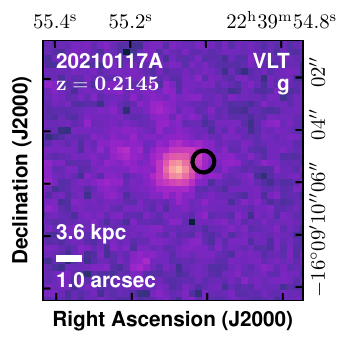} & 
\includegraphics[width=\hostwidth]{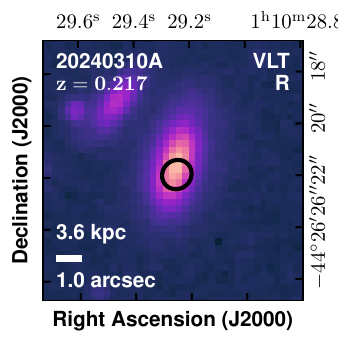} &
\includegraphics[width=\hostwidth]{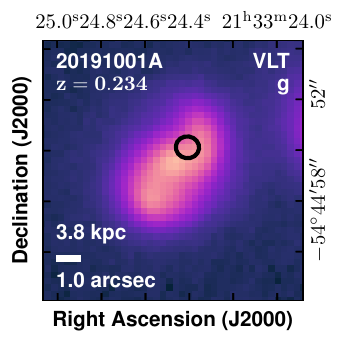} \\

\includegraphics[width=\hostwidth]{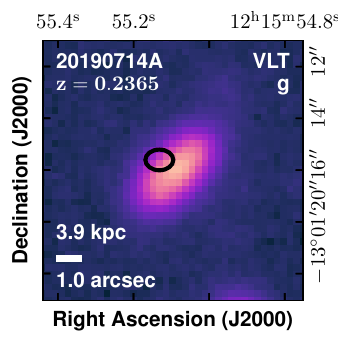} &
\includegraphics[width=\hostwidth]{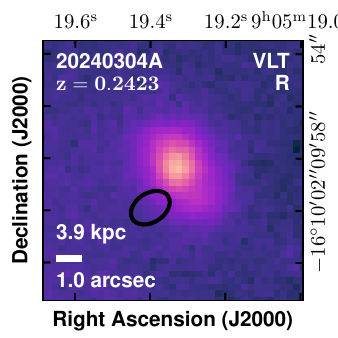} &
\includegraphics[width=\hostwidth]{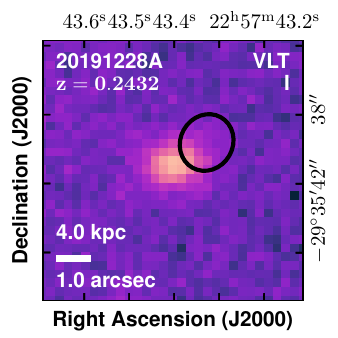} \\

\includegraphics[width=\hostwidth]{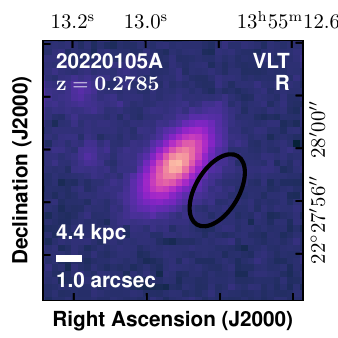} &
\includegraphics[width=\hostwidth]{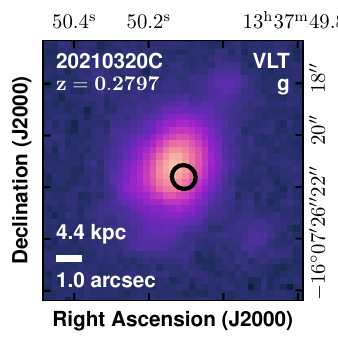} &
\includegraphics[width=\hostwidth]{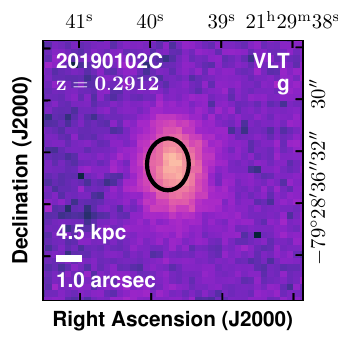} \\

\includegraphics[width=\hostwidth]{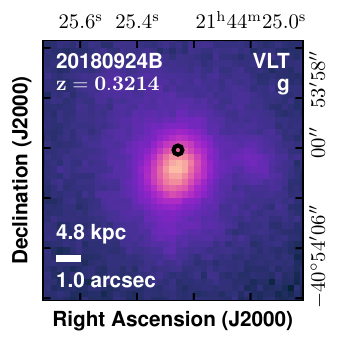} &
\includegraphics[width=\hostwidth]{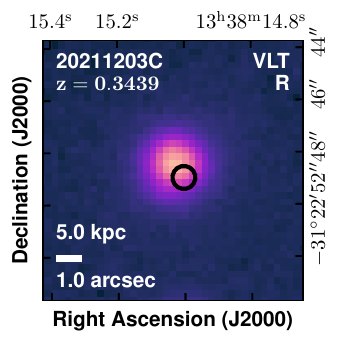} &
 \includegraphics[width=\hostwidth]{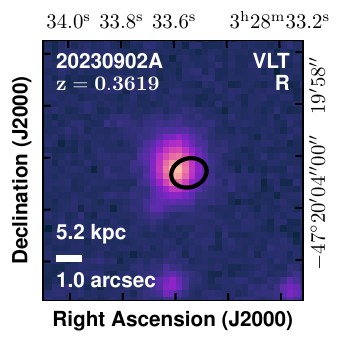} \\

\end{tabular}
\caption{FRB host galaxy montage (continued).\label{fig:host2}
}
\end{figure*}

\begin{figure*}
 \centering
 \begin{tabular}{ccc}

\includegraphics[width=\hostwidth]{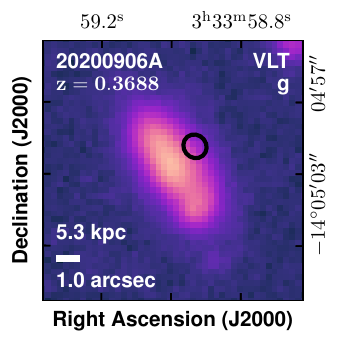} &
\includegraphics[width=\hostwidth]{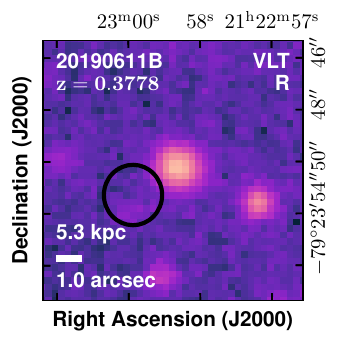} &
 \includegraphics[width=\hostwidth]{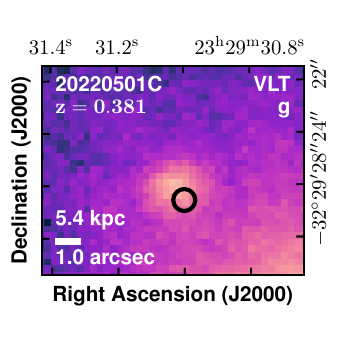} \\

\includegraphics[width=\hostwidth]{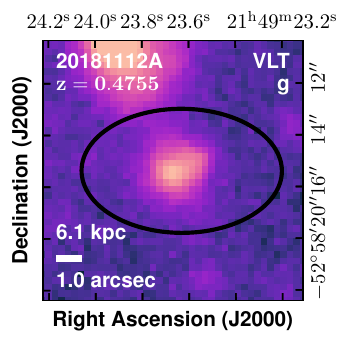} &
\includegraphics[width=\hostwidth]{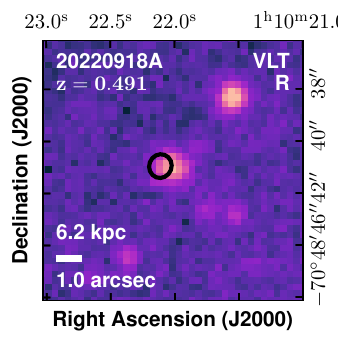} &
\includegraphics[width=\hostwidth]{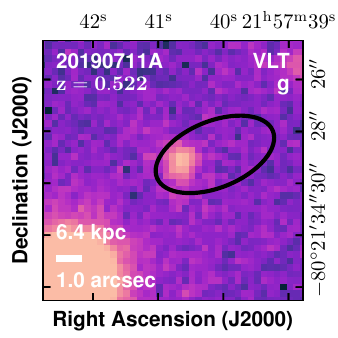} \\

& \includegraphics[width=\hostwidth]{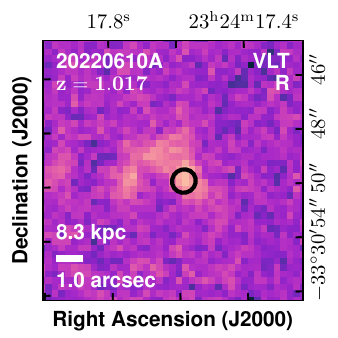} 

\end{tabular}
\caption{FRB host galaxy montage (continued). 
\label{fig:host3}
}
\end{figure*}

Here  we summarise the results of this CRAFT/ICS survey and describe the bursts detected. 
The host galaxies for many of the FRBs discovered are displayed in Figures \ref{fig:host1}, \ref{fig:host2}, and \ref{fig:host3}.
In Section \ref{sec:methods}, we describe the instrumentation and methods used to undertake the searches.
We motivate and describe the observing strategies we undertook to find the FRB sample in Section \ref{sec:strategies}
and our multi-wavelength follow up in Section~\ref{sec:followup}.
In Section \ref{sec:frbs}, we present the sample of \NFRB\ FRBs, of which \ALLLOCALISED\ have measured positions with precisions $\lesssim 1^{\prime\prime}$, of which  \ALLHOST\ host galaxies have been identified. 
We assess survey performance in Section~\ref{sec:survey_performance},
and in Section \ref{sec:science} we provide a review of the main scientific findings and outcomes of the searches. 
In Section \ref{sec:future}, we briefly describe the plan for new FRB search systems for ASKAP. 

\section{SEARCH AND LOCALISATION SYSTEMS AND METHODS}
\label{sec:methods}

\subsection{Hardware}

ASKAP is located at {\em Inyarrimanha Ilgari Bundara}, the CSIRO Murchison Radio-astronomy Observatory (MRO), a site selected for having low levels of radio-frequency interference (RFI).  The array comprises 36 12-m antennas.  The maximum baseline length of the antennas is 6\,km.  Each antenna is equipped with a phased array receiving system \cite[PAF;][]{iet:/content/conferences/10.1049/ic.2007.0899}, enabling multiple quasi-independent beams to be digitally formed on the sky.   This increases the field of view of the telescope by a factor of approximately 30 over an equivalent single-element system. 

\begin{figure*}
 \centering
\includegraphics[width=5in]{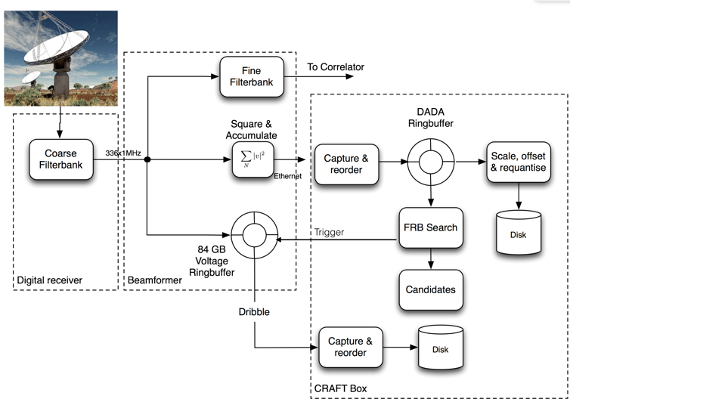}
\caption{CRAFT/ICS search system.  The dashed boxes indicate processesing stages operated within the telescope digital receiving system, antenna beamformers, and the CRAFT server (labelled `CRAFT box'), respectively}
\label{fig:craft_search} 
\end{figure*}

In this section, we briefly summarise aspects of the ASKAP hardware and signal path relevant to the searches described here.  \cite{2021PASA...38....9H} describe the general ASKAP hardware and standard ASKAP imaging system  in more detail.
A schematic diagram describing the FRB search and localisation systems is found in Figure \ref{fig:craft_search}.

Each PAF receiver comprises $188$ dipole elements arranged in a chequerboard pattern.
  The signal from each  dipole element is transmitted from the PAF to the MRO control building via  radio frequency over fibre systems \cite[][]{8168641}.
  Observing is undertaken in one of three bands (bands 1,2, and 3, with 1 being the lowest and 3 the highest). Within each band, it is possible to tune the observing frequencies best suited to the science goals. 
   The signal from each element is then digitised and channelised in a digital receiver \cite[][]{6903860}. The signals are channelised using a  polyphase filter into  coarse $1$~MHz channels oversampled by a factor of 32/27 \cite[][]{6328788}.  

  Digital beamformers then combine signals from the dipoles to form beams on the sky through appropriate weighting of each dipole element \cite[][]{1988IASSP...5....4V,6930062}. 
  In typical survey observations, $36$ beams are produced in either square or hexagonal close pack arrangements. 
  The separation between the beams is tailored to the goals of the observations, but varies between $0.75$ and $1.05$\,$\deg$, with beam separation largely guided by spacing of the frequency-dependent primary beam response, resulting in larger beam separation at lower frequencies.
  The beam weights are determined initially  through observations of the Sun.  
  While the weights are invalidated when certain changes to the observing system are made (e.g., changes between observing bands),
  they can be re-derived using an on-dish calibration solution \cite[][]{2021PASA...38....9H}.  Up until this point, the data products are common between FRB detection mode and standard ASKAP synthesis-imaging observations.

The beamformers produce two data products used by the CRAFT/ICS systems.  The first is total power summed over both polarisations for each coarse channel, integrated over $\approx 1$\,ms timescale, at $1$\,MHz frequency resolution.  These signals are broadcast over a network using unicast data protocol, and ingested by the CRAFT server where they are searched (as discussed in the next section).
  The second set of data products essential to the survey are voltages.  Each beamformer contains random access memory that is configured in a series of ring buffers.  For the searches presented here, a ring buffer is created for each beam for all of the antennas used in the searches. The voltages can currently be recorded with 1-bit, 4-bit, or 8-bit depth.  For all data presented here, we have used 4-bit depth, which enables 3.1\,s length voltage buffers.  The relatively short buffer length necessitates a low-latency search pipeline (discussed in the next section).  When the search pipeline detects a candidate FRB, the pipeline is stopped and the voltage buffers frozen.  Voltages for both polarisations from the candidate beam for each antenna are then downloaded to the CRAFT server.
  Both data products provide a total of $336$\,MHz of bandwidth, in contrast to the $288$\,MHz bandwidth available from the standard ASKAP hardware correlator. 
 Observations when the array operates in high-frequency resolution (zoom mode) observing are not compatible with the searches. 

\subsection{FRB Searches}

The primary motivation for implementing the ICS search method is the relative ease of computation compared to array-coherent approaches. The searches require only 36 data streams, fewer than those required for the previous fly's eye searches.  Array-coherent approaches require either imaging a large number of pixels for each phased array beam, or forming a similarly large number of tied array beams.
 The CRAFT server ingests the integrated intensities from each polarisation for each beam.  These are summed (after rescaling to ensure each stream had zero mean and the same variance) to create a pseudo total intensity, after synchronising data from each antenna beamformer to ensure temporal alignment. 
   
   Despite the MRO having low levels of nearby interference, RFI is present in the searches.
   The largest sources of interference are satellite interference from global navigation systems in band-2 observations and ionospheric ducting of mobile phone signals  (base towers and handsets) in band 1.  
   Data were also affected by interference associated with lightning and electrical storms (even when storms were more than $100$\,km from the observatory).
    Real-time  RFI flagging is conducted using a number of heuristics. Frequency channels with abnormally high kurtosis are set to zero, while remaining channels are normalised to a mean of zero and unit standard deviation. Impulsive RFI is removed by subtracting the frequency integrated signal at zero dispersion measure (DM). This is applied dynamically - additionally, a hard-coded list of known RFI channels is maintained, and these frequencies are zeroed for all observations.
    The methods were developed during commissioning of the ICS system (and the previous fly's eye survey), with the algorithms tuned to reduce the false positive rate.
The fraction of data flagged depended on the observing band, and was typically $10-20\%$.
  The level of false trigger varied between bands. In the band 1, after flagging there could be cases where there were no false triggers for many days.  In band 2, false triggers would occur at a rate of a few per day to one per hour.  In the presence of electrical storms, lightning produce false triggers at rate of many per hour.  The effectiveness of the mitigation was tested through observations of pulsars.  We confirmed that we were detecting the pulses at the expected rate and signal to noise ratio as had been conducted for the fly's eye survey \cite[][]{2019PASA...36....9J}.
   
   Searches are conducted using the FREDDA code \cite[a Fast Real-time Engine for De-Dispersing Amplitudes,][]{2019ascl.soft06003B,2023MNRAS.523.5109Q}, which is a graphical processing unit (GPU) implementation of a fast dispersion measure transform \cite[][]{2017ApJ...835...11Z}.
    Searches for all beams can be conducted on a single GPU.
    
     The searches produce a stream of raw candidates (time, signal-to-noise ratio, dispersion measure, width) which are broadcast over User Datagram Protocol.
     A burst (either RFI or a {\em bona-fide} FRB) produces candidates over a region of candidate parameter space. 
    A density-based spatial clustering algorithm \cite[][]{10.5555/3001460.3001507} is used to group raw candidates. This is important as it allows RFI (which typically has the greatest S/N at low DM) to be distinguished from  FRBs.
    Candidates were clustered using based on time (expressed as sample number within the scan), boxcar width, and dispersion measure (expressed as integer DM trial reported by FREDDA).  
    The density based spatial clustering algorithm used parameters $\epsilon=1$ and $n_{\rm min}= 1$ in the clustering.
    To ensure that the real-time clustering algorithm was working properly, raw candidates were also inspected off-line after being clustered using a friends-of-friends algorithm \cite[][]{1982ApJ...257..423H}.
    The friends of friends algorithm clustered based on DM (measured as integer DM units in FREDDA) and time (measured as samples within a scan). 
    The temporal clustering was normally set to a distance of $32$ samples.  DM clustering was set to a distance of $20$ integer DM units.
    
 The search latency ranges from a few hundred milliseconds to a few seconds.
  The time resolution of searches varied over the course of the survey, between $860\,\mu$s and $1.7$\,ms, as system performance was optimised. There is a trade-off between the number of antennas that can be ingested and the time resolution.  For the majority of the searching, a time resolution of $1.182$\,ms was chosen, which allowed the incoherent sum of intensities from $\sim$24 antennas to be formed.

\subsection{Localisations}
\label{sec:localisations}

 When a candidate FRB is detected, the voltage buffers are stopped and the beam-based buffers in which the burst is found are downloaded from the antenna beamformers to the CRAFT server. 
 They are then transferred offsite for further processing. 
     As the searches  stop on the first clustered candidate above a fixed signal-to-noise (S/N) ratio this is not necessarily the in which the burst has the most significant detection.
     In rare cases (e.g., very bright bursts, or bursts equally spaced between multiple beams), the beam downloaded was not the beam in which the FRB had the maximum S/N ratio.  
  Latency in the search pipeline and the short duration of the voltage buffers meant that for some FRBs the voltage buffer was of insufficient length to save the highest frequency emission. 
This is especially the case for the most dispersed FRBs, for which dispersed emission could extend over most of the voltage buffer. 
    If a candidate is confirmed (through visual inspection of the search data stream), additional observations are conducted for calibration. This included an observation of a bright compact radio galaxy (PKS~1934$-$638 or PKS 0407$-$658) to calibrate the bandpass, and a pulsar (PSR~J0834$-$4510, PSR~J1644$-$4559, or PSR~J2048$-$1616) as a polarisation reference. 
    A detailed description of the calibration, localisation, and astrometric performance is presented in \cite{2021PASA...38...50D}.
An end-to-end pipeline (CELEBI, The CRAFT Effortless Localisation and Enhanced Burst Inspection pipeline) processes voltage data, both to measure burst positions and produce high time resolution spectro-polarimetric data products (discussed in more detail in the next section). It is described in detail in \cite{2023A&C....4400724S}.

\subsection{High time resolution spectropolarimetry}

Access to the voltage data allows us to study in detail the spectro-temporal-polarimetric properties of the bursts. These studies provide insight into burst emission physics, as well as the effects of propagation through intervening ionised media, for example through the measurement of scatter broadening times and rotation measures. 
This was done through high time and frequency resolution imaging \cite[][]{2020MNRAS.497.3335D} and through high time resolution beamforming \cite[][]{2020ApJ...891L..38C,2023A&C....4400724S}. 
The technique of high time resolution imaging \cite[][]{2020MNRAS.497.3335D} did not provide the same time resolution possible through beamforming ($54\,\mu$s). It was a natural and (relatively) easily implemented extension of FRB burst localisation pipelines prior to the development of a tied array beamforming pipeline.
As the upstream ASKAP digital system implements channelisation through oversampled filterbanks, it is possible to invert the $1$\,MHz coarse channels of the voltage buffers to produce time series with temporal resolution as high as the Nyquist sampling rate of the digital receiving system ($1/(336\,{\rm MHz}) \approx 3$\,ns).
Using both methods, polarisation calibration was conducted by using a bright pulsar as a reference source to correct for polarisation leakage.   
Rotation measures can be inferred using a Bayesian methodology described in \cite{2019Sci...365..565B}. 

\section{SURVEY STRATEGIES}
\label{sec:strategies}

The CRAFT/ICS survey has been conducted throughout the final stages of ASKAP commissioning, the pilot surveys for the observatory-approved Survey Science Projects, and the early stages of full ASKAP survey science observations.  
The first searches were undertaken in  one month of observing time in Aug-Sep 2018 allocated in the pursuit of the first FRB localisation with ASKAP. 
Efforts were largely focused on observing the high Galactic latitude ($|b| =50$) fields that were the main targets of the fly's eye searches. These fields were chosen as they allowed constraints to be placed on burst repetition and demonstrate that  detections were originating from apparently non-repeating FRBs.
FRB searches benefited from not requiring data storage or processing at the Pawsey Centre, unlike other surveys. Thus FRB searches could be conducted when resources (storage or processing) at the Centre were limited.  

In early searches (prior to 2020), ASKAP scheduling was not fully automated. Occasionally we would choose to observe southern circumpolar fields that could be observed at any local sidereal time, so that searches could be scheduled for many days without requiring intervention. In particular, considerable time was spent observing a field centred at R.A. $=22^{\rm h}$ and Dec. $=-80\deg$. This field was chosen because it has a relatively low Milky Way DM contribution \cite[$\approx 50$ \pccm\ from the Milky Way disk;][]{2002astro.ph..7156C}. 

As ASKAP transitioned from scientific commissioning to a survey science instrument, FRB searches were more frequently run commensally with other projects. During the course of the ICS searches, these were largely pilot surveys for the main ASKAP Survey Science projects, and the observatory Rapid ASKAP Continuum Surveys \cite[RACS, ][]{2020PASA...37...48M}.

\begin{figure*}
    \centering
    \includegraphics[width=5in,trim={0 2.5cm 0 3cm},clip]{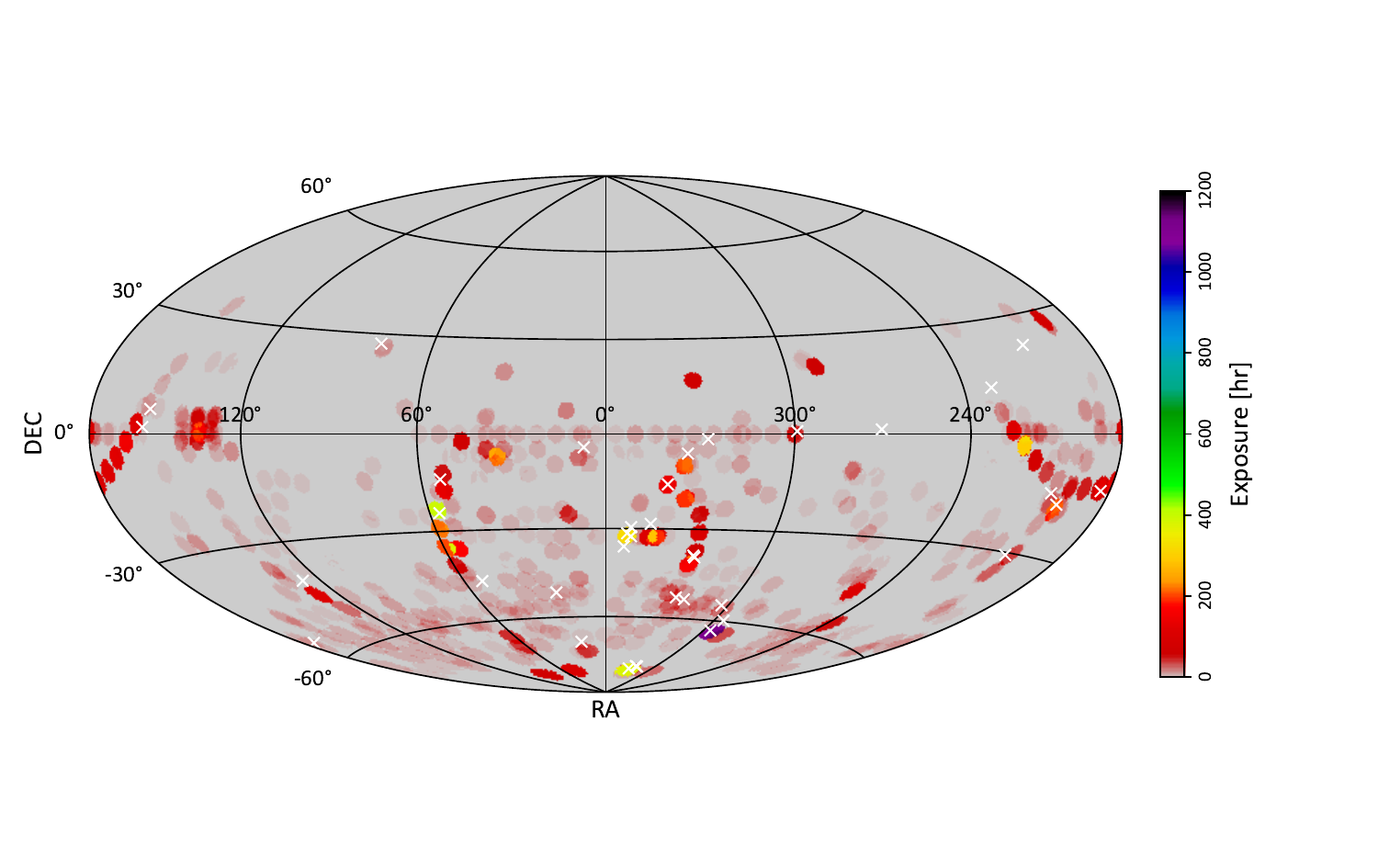}
    \caption{ICS exposure map.  The Hammer projection in J2000 coordinates shows the total time-per-field for ICS pointings for which we have records. Detected FRBs are shown as white crosses.}
    \label{fig:ics_healpix}
\end{figure*}

Figure \ref{fig:ics_healpix} shows the exposure map  for the survey,
 or the \num{12701.7}\,hr for which we have records through the end of 2023. The greatest exposure is near the ASKAP bandpass calibrator PKS~1934$-$638, with approximately \num{1150}\,hr of observations and one FRB detected. Other notable fields are the Deep Investigation of Neutral Gas Origins \cite[DINGO,][]{2009pra..confE..15M} fields near (R.A.,DEC.) $ = (339.0$ to $351.0, -35$ to $-30$) (199-347\,hr), and the aforementioned circumpolar (R.A.,DEC.) $=(22$\,h,$-80$) field (384.5\,hr).
 The sensitivity of the survey is discussed further in Section \ref{sec:survey_performance}.

\section{Multi-wavelength follow up}
\label{sec:followup}

The (sub-)arcsecond positions yielded by the ICS system enabled multi-wavelength follow up of the FRB host galaxies, which was the major focus of the survey. Less emphasis was placed on identifying prompt emission temporally coincident with the detected FRBs.
Below we identify the motivations and strategy employed to identify host galaxies and measure host/burst redshifts to advance FRB science.

\subsection{Optical and infrared photometry}

After obtaining high precision FRB  positions, we first searched for FRB host-galaxy candidates.
The FRB coordinates were first checked against available imaging archives, photometric and spectroscopic catalogues using tools such as the Data Aggregation Service (DAS) \cite[][]{2022SPIE12189E..2SM}.
While some of the lower-DM FRBs had an apparent host galaxy identified in wide-field surveys such as the  Sloan Digital Sky Survey (SDSS; \citealt{SDSS}), the Pan-STARRS 3$\pi$ survey (PS1; \citealt{PS1}), or the Dark Energy Camera Legacy survey (DECaLS; \citealt{DECALS}), the majority of ASKAP-detected FRBs had little or no reliable photometry or spectroscopy available for potential host-galaxy candidates.
We therefore obtained our own multi-band photometry to both identify the host galaxies and  model the host galaxy properties including total stellar mass and star formation history.
Photometric observations of the host galaxies were taken predominantly using the FOcal Reducer and low dispersion Spectrograph
 (FORS2) instrument on the VLT \cite[][]{1992Msngr..67...18A}; Gemini Multi Object Spectograph (GMOS-S) on Gemini-South \cite[][]{2004PASP..116..425H}; or the Low Resolution Imaging Spectrometer (LRIS) on Keck \cite[][]{1994SPIE.2198..178O}. 
Infrared imaging observations were also undertaken with the HAWK-I \cite[][]{2008A&A...491..941K} instrument on the VLT, usually in combination with a ground layer adaptive optics module (GRAAL).
Photometric data reduction strategies are described in detail elsewhere \cite[e.g.,][]{2020ApJ...903..152H,2023MNRAS.525..994M,2023ApJ...949...25G}.

\subsection{Host Associations}
\label{sec:path}

Central to establishing the redshift of an FRB is
to identify its host galaxy.
Our approach for the well-localised FRBs of the 
CRAFT/ICS survey has been to implement the
Probabilistic Association of Transients to
their Hosts (PATH) formalism introduced in \cite{2021ApJ...911...95A}.
The PATH methodology inputs the localisation region
of the FRB and the position and apparent magnitudes
of all candidate galaxies within or near that region.
The analysis requires prior assumptions on the
probability of the host being unobserved in the image,
the distribution of host galaxy magnitudes, and the
offset angular separations of FRBs from the centre of the 
galaxy. 
For galaxy magnitudes, we adopt an uninformative prior
that weights galaxies inversely proportional to their
number density on the sky.
For the separations,
\cite{2021ApJ...911...95A} adopted an exponential profile
with scale length equal to an angular size metric
of the galaxy, specifically the
{\tt semimajor\_sigma} parameter of the 
{\sc photutils} software package ($\phi$).

\begin{figure}
 \centering
\includegraphics[width=3. in]{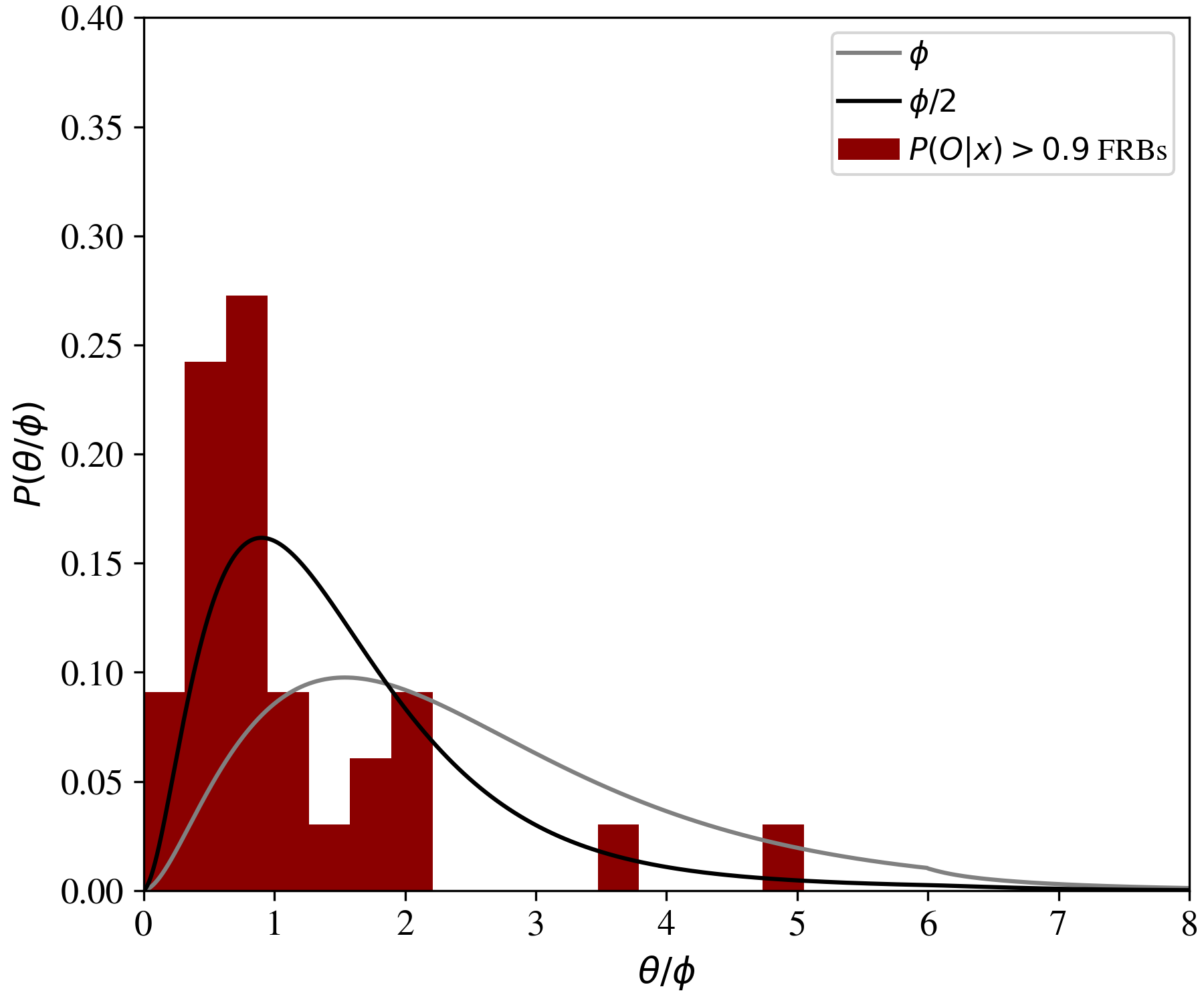}
\caption{Comparison of PATH prior distributions 
(curves) to FRB offsets (histogram)
obtained through CRAFT/ICS observations.  Here
we have restricted to those FRBs with PATH posteriors
$P(O|x)$
in excess of 0.9 using the original priors. This
corresponds to the gray curve which is an exponential
with  scale length equal to
the half-light radius of the galaxy $\phi$.  
We now advocate
the prior described by the black curve with
scale-length equal to $\phi/2$.
}
\label{fig:PATH} 
\end{figure}

We revisit this assumption and suggest an
updated formulation for this offset prior.
Figure~\ref{fig:PATH} shows the normalised
PDF of the angular offsets $\theta$ relative
to $\phi$ for 32~FRBs.
This FRB sample was restricted to the ICS sample
presented  in this manuscript with a secure association
(see section~\ref{sec:new_frbs}) and
with $\pox > 0.90$ using the original PATH analysis
and priors.
We also include the ASKAP-localised CHIME repeating source FRB\,20201124A \citep{2021ApJ...919L..23F}.
Approximately half of the sample shows
$\theta/\phi < 1$ with the remainder exhibiting
a tail to $\theta/\phi \approx 5$.

Overplotted on the observed distribution are
the exponential priors for host galaxy offset with scale
lengths of $\phi$ and $\phi/2$.
These have been convolved with the reported 
uncertainties in the FRB localisations.
They have also been weighted by a geometrical
factor ($2\pi\theta$)
which disfavours low offsets\footnote{We note that
Figure~11 of \cite{2021ApJ...911...95A} failed to include this factor.}.
Clearly, the exponential with scale-length of $\phi$
is disfavoured while the data are reasonably well-described by the smaller scale length ($\phi/2$).
We advocate adopting this new prior for future
work on FRBs with PATH, and we utilise it throughout
the manuscript.
It is also possible as the sample continues to increase
that one will adopt a different functional form
to better describe the distribution.

Reanalysing all of the FRBs presented in 
\cite{2020ApJ...895L..37B}, we find few changes in the
posterior probabilities.  Most of them were previously
$\pox > 0.95$ and the values increased towards 1.
The only notable changes were significant increases
in $\pox$ for FRB\,20181112A and FRB\,20191001A.  The former FRB is associated with a $z=0.4755$ which has a $z=0.3674$ galaxy in the foreground \cite[][]{2019Sci...366..231P}
The latter FRB is associated with a $z=0.2340$ galaxy which is separated by $\approx 5$''  from a galaxy at a similar redshift \cite[$z=0.2339$][]{2020ApJ...901L..20B}.
In Tables~\ref{tab:path} to \ref{tab:path4}  in the Appendix we present PATH probabilities for nearby host galaxies for all ASKAP-localised FRBs.

\subsection{Optical spectroscopy}

We obtained spectroscopic observations of the probable host galaxy candidates that lacked archival data, both to determine the redshift but also to study the global properties of the hosts. These were primarily obtained with FORS2 or X-shooter \cite[][]{X-shooter} on the VLT, or LRIS or the Deep Imaging Multi-Object Spectrograph (DEIMOS; \citealt{DEIMOS}) on Keck.  Brighter host galaxies were observed with the Goodman Spectrograph on the Southern Astrophysical Research Telescope \cite[SOAR,][]{2004SPIE.5492..331C}.
Spectroscopic data reduction is described in detail in \cite{2023ApJ...949...25G}, \cite{2023ApJ...954...80G}, \cite{2023Sci...382..294R},
and A.~Muller et al.\ (in prep).  Uncertainties in the redshifts derived from spectroscopy are approximately $0.0001$ for X-shooter and $0.005$ for other instruments.


Integral field unit (IFU) and multi-object spectroscopy  enabled more detailed spatially resolved studies of the host galaxies and the mapping of  structure foreground to the FRB host through spectroscopic identification and characterisation of galaxies close to the line of sight.
We have obtained IFU spectroscopy of some of the FRB hosts with the Multi Unit Spectroscopic Explorer on the VLT \cite[][]{2010SPIE.7735E..08B} and with the Keck Cosmic Web Imager \citep{2018ApJ...864...93M}. 
Wider-field spectroscopic study of host galaxy foregrounds was also taken with the Anglo Australian Telescope's AAOmega and two-degree field (2dF) fiber-fed multi object spectrograph \cite[][]{2004SPIE.5492..410S}, in collaboration with the FLIMFLAM project \cite[][]{2022ApJ...928....9L,2024arXiv240200505K}. 


\subsection{Radio and high energy follow-up}

We also searched for sources spatially coincident with the FRBs and their hosts across the electromagnetic spectrum.

Radio-wavelength follow-up entailed searches for persistent radio sources sometimes  associated with (repeating) FRBs, extended radio continuum emission (most likely associated with star formation) and spectral lines (especially targeting the 21-cm Hydrogen hyperfine transition at decimetre wavelengths and carbon monoxide transitions at millimetre wavelengths).  Observations at radio wavelengths of the FRB hosts were made with the Atacama Large Millimeter/submillimeter Array \cite[e.g,][]{ALMAobs}, Australia Telescope Compact Array \cite[ATCA, e.g.,][]{2022AJ....163...69B}, Jansky Very Large Array \cite[JVLA, e.g.,][]{2023ApJ...948...67B}, and MeerKAT \cite[e.g.,][]{Glowacki2024}. 
Some of these observations (particularly those made with ATCA) were also used for the first FRBs discovered in order to reference radio continuum sources in images produced from the 3-second voltage dumps with the International Celestial Reference Frame.  As our understanding of the astrometric  precision of the 3-second images improved this was no longer necessary \cite[][]{2021PASA...38...50D}. 

We also searched for X-ray emission from a sample of our hosts using the {\em Chandra} X-ray observatory \cite[][]{Chandra}. Such emission could be associated with coincident Active Galactic Nuclei, or Ultra-luminous X-ray sources.

\section{FAST RADIO BURST DISCOVERIES}
\label{sec:frbs}

To the end of April 2024, the CRAFT/ICS survey detected  \NFRB~unique sources.  This includes localisation of the  repeating FRB source FRB\,20201124A first identified by CHIME \cite[][]{2021ATel14497....1C}.
The FRBs range in dispersion measure from $206$~\pccm\ to 1780~\pccm.  The median dispersion measure was 440~\pccm, similar to that found in the fly's eye survey. However, notable higher DM FRBs (discussed below) were identified.
Compared to the fly's eye searches the FRBs detected were unsurprisingly fainter.   The median detected burst fluence was 30 Jy\,ms. 
Burst fluences ranged from 10 to 120 Jy\,ms. 
Bursts were detected in all three of the ASKAP observing bands, in observations centred at $832$\,MHz, $864$\,MHz, $920$\,MHz, $1272$\,MHz, and $1630$\,MHz. 

Table \ref{tab:frb_discoveries} in the Appendix summarises the key properties of the FRBs.
Table \ref{tab:frb_pos} in the Appendix lists the positions of the FRBs, derived interferometrically (where possible) and with the multi-beam localisation method \cite[][]{2017ApJ...841L..12B}. 
Dedispersed dynamic spectra of the bursts from the ICS search data stream, arranged from lowest to highest dispersion measure, can be found in Figures \ref{fig:dynspec1} to \ref{fig:dynspec5} in the appendix. Even with the relatively low time and frequency resolution of the search data stream the bursts show a variety of spectral and temporal morphologies. Their high-time-resolution properties are reported elsewhere (\HTR). We explore the spectral modulation of the bursts in Section \ref{sec:modulation}.

\subsection{Previously reported FRBs}
\label{sec:previous}
We briefly summarise the key properties of the previously published ICS FRBs, including motivation for the observations in which they were found.

{\em FRB\,20180924B --} The first FRB localised in the ICS survey was reported in \cite{2019Sci...365..565B}. It was discovered in one of the $|b|=50$ Galactic latitude fields used in the previous fly's eye survey \cite[][]{2018Natur.562..386S} during a dedicated time allocation from the ATNF to secure a FRB localisation.   The FRB host galaxy was initially speculated to be either a lenticular or early-type spiral galaxy; however subsequent observation and analysis indicates it to be the latter type from both detailed study of the host galaxy spectral energy distribution \cite[][]{2023ApJ...954...80G}, and high spatial resolution imaging with HST \citep{2021ApJ...917...75M} that showed the presence of spiral arms, with one of the arms coincident with the FRB localisation.

{\em FRB\,20181112A--} The discovery and localisation was first reported in \cite{2019Sci...366..231P}. The FRB was discovered in one of the $|b|=50$ Galactic latitude fields used in the previous fly's eye survey \cite[][]{2018Natur.562..386S}.   

{\em FRB\,20190102C --} The discovery and localisation was first reported in \cite{2020Natur.581..391M}. The FRB was discovered in the circumpolar R.A.$=22$h, Decl.$=-80$ field. 

{\em FRB\,20190608B--} The discovery and localisation was first reported in \cite{2020Natur.581..391M}.  The FRB was found during an attempt to localise FRB\,20171019A, an FRB discovered in the fly's eye survey \cite[][]{2018Natur.562..386S} from which repetitions had been detected with the GBT \cite[][]{2019ApJ...887L..30K}. A high-resolution imaging and kinematic study of the host galaxy by \citet{2021ApJ...922..173C} showed the FRB to be closely associated with star formation in a spiral arm.

{\em FRB\,20190611B--} The discovery and localisation was first reported in \cite{2020Natur.581..391M}. The FRB was discovered in the circumpolar R.A.$=22$h, Decl.$=-80$ field. 
The FRB has the largest DM deficit relative to the Macquart relation of any FRB discovered in the CRAFT/ICS survey.  The veracity of the host association has been questioned based on this \cite[][]{2022ApJ...931...88C}. 

{\em FRB\,20190711A--} The discovery and localisation was first reported in \cite{2020Natur.581..391M}. The FRB was discovered in the circumpolar R.A.$=22$h, Decl.$=-80$ field.  This is the only FRB discovered in the ICS survey that has been observed to repeat in our follow-up observations \cite[][]{2021MNRAS.500.2525K}.

{\em FRB\,20190714A--} The discovery and localisation was  reported in \cite{2020ApJ...903..152H}.  Only one of the voltage polarisation data streams was able to be downloaded for the FRB, making polarimetric analysis of the burst properties difficult to undertake\footnote{In principle, it would be possible to estimate a magnitude of rotation measure by searching for spectral modulation of the burst consistent with rotation measure.  However, FRBs also show spectral modulation that is either intrinsic to the burst emission or the result of diffractive scintillation.}. The FRB was discovered in one of the $b=50$ Galactic latitude fields used in the previous fly's eye survey \cite[][]{2018Natur.562..386S}. A wide-field spectroscopic survey around the FRB\,20190714A sight line \citep{Simha2023} shows a clear excess of foreground galaxy halos that contributes $\sim$2/3 of the observed extragalactic DM.

{\em FRB\,20191001A--}  The discovery and localisation was reported in \cite{2020ApJ...901L..20B}.   The FRB was detected commensally with pilot survey observations for the Evolutionary Map of the Universe (EMU) survey science project \cite[][]{2021PASA...38...46N}.  The burst was sufficiently bright to be detected as an image-plane transient in 10-s hardware correlator visibilities\footnote{The FRB was detected during the 2019 ATNF Radio Astronomy school which many CRAFT and ASKAP team members (Ekers, Hotan, Lenc, Moss, Shannon) were attending, during which time the hardware correlator position was measured.}. 

{\em FRB\,20191228A--} The discovery and localisation was reported in \cite{2022AJ....163...69B}. The FRB originated just 2~arcmin from Fomalhaut ($\alpha$ PsA), complicating  optical identification and redshift determination for the host galaxy.  The FRB was discovered commsensally in observations for the DINGO survey \cite[][]{2023MNRAS.518.4646R} targeting the GAMA-23 field. 

{\em FRB\,20200430A--}  The discovery and localisation was  reported in \cite{2020ApJ...903..152H}. The FRB was detected commensally in test observations for the first low-frequency epoch of the  Rapid ASKAP Continuum Survey \cite[RACS, ][]{2020PASA...37...48M}.

{\em FRB\,20200906A--}  The discovery and localisation was reported in \cite{2022AJ....163...69B}. The FRB was detected in observations searching for repetitions from FRB\,20171116A, which had been detected in the ASKAP Fly's Eye Survey \cite[][]{2018Natur.562..386S}. It had been selected for follow up because of its relatively large pulse width, which may be correlated with being a repeating FRB \cite[][]{2020MNRAS.497.3076C,2021ApJ...923....1P}.  Only seven antennas were recording CRAFT data during the observation.  The search system would select antennas for recording based on the number of antennas that were on-source at the beginning of an observation. Occasionally, antennas would arrive on source ascynchronously, resulting in smaller subsets of antennas being used in the searches and localisation. 

{\em FRB\,20210117A--} The discovery and localisation was reported in  \cite{2023ApJ...948...67B}. The FRB was detected commensally with observations for the mid-frequency RACS survey \cite[][]{2023PASA...40...34D}.
The FRB was found to be originating from a dwarf galaxy with a high host excess DM, similar to archetypal repeating sources FRBs~20121102A and 20190520B  \cite[][]{2017Natur.541...58C,2022Natur.606..873N}.

{\em FRB\,20210320C--}  The discovery and localisation were reported in \cite{2022MNRAS.516.4862J} and \cite{2023ApJ...954...80G}, respectively. The FRB was discovered in one of the $|b|=50$ Galactic latitude fields used in the previous fly's eye survey \cite[][]{2018Natur.562..386S}.

{\em FRB\,20210807D--} The discovery and localisation were reported in \cite{2022MNRAS.516.4862J} and \cite{2023ApJ...954...80G}, respectively. The FRB was discovered during time-lapse photography of ASKAP being undertaken for a documentary, demonstrating the  ability for the survey to be commensal with non-scientific observations. The FRB was wider than the width threshold for voltage download, but was sufficiently bright to be localised in hardware-correlator 10-s visibilities like FRB\,20191001A \cite[][]{2021ApJ...910L..18B}.

{\em FRB\,20211127I--} The discovery and localisation were reported in \cite{2022MNRAS.516.4862J} and \cite{2023ApJ...954...80G}, respectively.  The FRB was detected commensally in Widefield ASKAP L-band Legacy All-sky Blind Survey (WALLABY) \cite[][]{2020Ap&SS.365..118K} pilot survey observations towards the NGC 5044 group, but is not associated with the group.  \cite{2023ApJ...949...25G} present an analysis of the FRB and the commensal detection of HI from the host galaxy. 

{\em FRB\,20211203C--} The discovery and localisation were reported in \cite{2022MNRAS.516.4862J} and \cite{2023ApJ...954...80G}, respectively.  This FRB was detected commensally in POlarisation Sky Survey of the Universe's Magnetisation (POSSUM)  \cite[][]{2010AAS...21547013G} pilot-survey observations.

{\em FRB\,20211212A--}  The discovery and localisation were reported in \cite{2022MNRAS.516.4862J} and \cite{2023ApJ...954...80G}, respectively. The FRB was discovered in one of the $|b|=50$ Galactic latitude fields used in the previous fly's eye survey \cite[][]{2018Natur.562..386S}. This was also the first FRB detected in the ASKAP high band (at a central frequency of $1632.5$\,MHz)

{\em FRB\,20220105A--} The discovery and localisation was reported in \cite{2023ApJ...954...80G}.  The FRB was detected commensally with the first epoch of the RACS high band survey. 

{\em FRB\,20220610A--} The discovery and localisation was reported in \cite{2023Sci...382..294R}. Subsequent imaging with the \textit{Hubble Space Telescope} revealed the burst originated from a compact galaxy group at redshift $z=1.016$ \cite[][]{2024ApJ...963L..34G}.   The FRB was discovered in observations attempting to identify repetitions from ASKAP-discovered (but poorly localised) FRB\,20220501A (discussed further in  Section~\ref{sec:220501}).    

{\em FRB\,20230718A--} The discovery and localisation was reported in \cite{Glowacki2024}.  The host galaxy was identified through 21-cm HI emission using the MeerKAT radio telescope. The FRB was detected commensally with WALLABY survey observations, but the ASKAP hardware correlator data stream had a technical error and the WALLABY spectral line data was unusable.   A DECam image of the field is shown in Figure \ref{fig:host1}.

{\em FRB\,20201124A--}  In addition to FRBs discovered by CRAFT, we also localised one repeating FRB initially discovered by another facility. The detection and localisation of this repeating FRB source was reported in \cite{2021ApJ...919L..23F}.  The FRB was first detected by CHIME. In April 2021, \cite{2021ATel14497....1C} reported an episode of increased activity from the burst source, which motivated follow up with ASKAP. In total $11$ bursts were detected by ASKAP from the source \cite[][]{2022MNRAS.512.3400K}.  The brightest of these bursts was detected with high significance (S/N > 10) in 20 beams that spanned the phased array feed.
As the FRB detection system triggers off the first significant candidate, voltages of the first localised burst from this source were downloaded from a beam adjacent to the primary detection.  

\subsubsection{FRB\,20210912A}

The discovery and localisation was reported in \cite{2023MNRAS.525..994M}, but unusually no host galaxy has yet been associated with the FRB despite deep imaging follow up. The FRB was discovered commsensally in observations for the Deep Investigations of Neutral Gas Origins (DINGO) survey \cite[][]{2023MNRAS.518.4646R}. We also report new observations of this source with the Keck Cosmic Reionization Mapper (KCRM) and Keck Cosmic Web Imager (KCWI). 
 Observations were undertaken under poor weather conditions on 17 August 2023.
 We obtained $12\times300$\,s exposures with KCRM and $3\times1320$\,s exposures with KCWI that were usable.
  We carried out standard processing with the PypeIt software package \cite[][]{2020JOSS....5.2308P} to form a spectral cube covering approximately 340--570\,nm (KCRM) and 653--1030\,nm (KCWI). Neither cube shows any evidence for spectral lines, and the collapsed cubes do not show any evidence for continuum emission. 


\subsection{New localised FRBs}
\label{sec:new_frbs}

We summarise the properties and localisations of a few other FRBs discovered in the ICS searches, though detailed analyses of the bursts and their host galaxies are deferred to subsequent studies. 

\subsubsection{FRB\,20210407E}
\label{subsub:210407}

The FRB was discovered while monitoring the active repeating FRB source FRB\,20201124A \cite[][ Section~\ref{sec:previous}]{2021ApJ...919L..23F,2022MNRAS.512.3400K}.  The burst has the highest DM  ($1785.3 \pm 0.3$\,\dm) of any detected in a survey with ASKAP to date.
The FRB was discovered at relatively low Galactic latitude, $b=-6.7\deg$. 
The Milky Way's disk DM contribution is 154~\dm and 229~\dm assuming the NE2001 \cite[][]{2002astro.ph..7156C} and YWM16 \cite[][]{2017ApJ...835...29Y} Galactic electron density models, respectively,  indicating the burst is surely of extragalactic origin. 

Given the high DM (suggesting a high redshift source), we searched for a host galaxy for the burst, despite relatively high extinction along the line of sight with imaging in $Z$-band with DEIMOS at Keck (Prog ID O314; PI Blanchard) and  in $i$-band with BinoSpec at the MMT \cite[][]{2019PASP..131g5004F} (Prog ID UAO-G194-21A; PI Fong).
We identify no credible source 
at the position of the FRB localisation to a $5\sigma$ 
limiting magnitude $Z > 25.8$\,mag  and $i> 24.2$\,mag measured in 
a $1''$ radius aperture set by the seeing estimate.
Extinction is estimated to be approximately $1.5$ and $2.0$ magnitudes along this line of sight in $Z$ and $i$ band, respectively \cite[][]{2011ApJ...737..103S}.

We have also attempted to identify the host using integral field spectroscopy in the case that the host galaxy had strong emission lines but lower levels of continuum emission. 
We observed FRB\,20210407E with the Keck Cosmic Reionization Mapper (KCRM) and the Keck Cosmic Web Imager (KCWI) on 17 August 2023. Weather conditions were poor, but we obtained $12\times300$\,s exposures with KCRM and $3\times1320$\,s exposures with KCWI. 
Data reduction was identical to that for FRB\,20210912A, described above.
Like for the aforementioned FRB, the observations should no evidence for either spectral lines or continuum emission.

We are continuing to obtain follow-up observations due to the potentially high redshift nature of the FRB.
For this manuscript, however, we include the
FRB without a redshift despite a precise localisation.

\subsubsection{FRB\,20220501C}\label{sec:220501}

The FRB was detected in observations of the second epoch of the RACS-low survey \cite[][]{2020PASA...37...48M}. 
There was no catalogued galaxy coincident with the burst position. 
The  position is 15" from the $V=11.6$ magnitude star TYC 7514-122-1 \cite[][]{2014AJ....148...81M}.
VLT FORS2 $I$-band imaging identified a host galaxy coincident with the FRB position (Figure~\ref{fig:host2}).
VLT X-shooter spectroscopy of the host galaxy identified H$\alpha$ and [O\,{\sc iii}] $\lambda5007$ lines  at a redshift of $z=0.381$.
The FRB showed spectro-temporal morphology similar to repeating FRBs \cite[][]{2019ApJ...876L..23H,2021ApJ...923....1P}.  As a result, additional filler observations were scheduled to follow up the source to search for repetitions.  No repetitions were found, but two other unique FRB sources were discovered in these follow up observations. 

\subsubsection{FRB\,20220725A}

The FRB was detected while monitoring the FRB\,20220501C field. 
The FRB is coincident with the catalogued galaxy WISEA J233315.68-355925.0, which has an optical magnitude of  $b_J=19.0$ \cite[][]{1990MNRAS.243..692M}.
VLT/FORS2 imaging of the host  (Figure \ref{fig:host2}) shows spiral arm morphology common to many low redshift ASKAP/ICS FRBs. We obtained z-band imaging follow-up with SOAR (Prog ID SOAR2022B-007; PI Gordon). SOAR spectroscopy (Prog ID SOAR2022B-007; PI Gordon) identified a number of emission lines in the host spectrum, including from  H$\alpha$, [N\,{\sc ii}], and [S\,{\sc ii}] at a redshift of $z=0.1926$.

\subsubsection{FRB\,20220918A}

The FRB was detected during POSSUM \cite[][]{2010AAS...21547013G}  survey observations of the Large Magellanic Cloud (LMC).  The burst was detected $2.5^\circ$ from the nominal centre of the LMC, so unlikely to originate there. 
There was no catalogued host galaxy coincident with the FRB position.
VLT/FORS2 imaging identified a host galaxy coincident with the  position as shown in Fig.~\ref{fig:host2}. The host was also detected in HAWK-I $K_s$-band imaging. 
X-shooter spectroscopy of the host galaxy shows weak H$\alpha$ and [O\,{\sc iii}] $\lambda5007$ emission lines with a redshift of $z=0.491$.

\subsubsection{FRB\,20221106A}

The FRB was detected during filler observations of one of the fly's eye Galactic latitude $50$-degree fields.
The FRB was coincident with the galaxy  WISEA J034649.07-253411.7 which has a magnitude of $b_J=19.5$ \cite[][]{1990MNRAS.243..692M}.
A VLT/FORS2 image of the source is shown in  Fig. \ref{fig:host1}.  The galaxy was also identified in VLT/HAWK-I $K_s$-band imaging, while 
SOAR (Prog ID SOAR2022B-007; PI Gordon) and X-shooter spectra show H$\alpha$, [N\,{\sc ii}] and [S\,{\sc ii}] emission lines with a redshift of $z=0.2044$.

\subsubsection{FRB\,20230526A}

The FRB was detected during WALLABY survey observations \cite[][]{2020Ap&SS.365..118K}.  
The FRB is associated with a host galaxy identified in Dark Energy Survey (DES) data with a photometric redshift of $z=0.25 \pm 0.10$.
VLT FORS2 $R$-band imaging identified a host galaxy coincident with the FRB position.  The R-band observations was saturated at the nucleus of the host so only an upper limit can be place on the host magnitude.  The host was also detected in VLT/HAWK-I $K_s$-band imaging  (Table \ref{tab:photometry}).
X-shooter spectroscopy of the host galaxy showed a number of strong emission lines, including H$\alpha$, H$\beta$, [O\,{\sc ii}], [O\,{\sc iii}] $\lambda$5007, [N\,{\sc ii}] and [S\,{\sc ii}],  with a redshift of $z=0.1570$.  This is consistent with a previously reported photometric redshift from DES. 

\subsubsection{FRB\,20230708A}

The FRB was detected during EMU survey observations \cite[][]{2021PASA...38...46N}. 
VLT FORS2 $R$-band imaging identified a host galaxy coincident with the FRB position (Figure \ref{fig:host1}). The host was also detected in VLT/HAWK-I $K_s$ band imaging (Table \ref{tab:photometry}). 
X-shooter spectroscopic observations of the host galaxy shows emission lines associated with H$\alpha$, H$\beta$, [O\,{\sc ii}], and [O\,{\sc iii}]  with a redshift of $z=0.1050$.

\subsubsection{FRB\,20230731A}

The FRB was detected during WALLABY survey observations \cite[][]{2020Ap&SS.365..118K}.  The FRB was detected at relatively low Galactic latitude  $b=4.5^\circ$.  Given the dispersion measure of the FRB, it is unlikely that the host galaxy has detectable 21-cm emission in the commensal WALLABY observations. While deep optical and near-infrared imaging has been undertaken with the VLT, no host is obvious in R-band imaging. This field has a high density of stars, and all nearby objects appear point-like. Because this line-of-sight has a relatively low Galactic latitude ($\sim5^\circ$), Galactic extinction is expected to be high at $A_R\sim0.7$ \citep{2011ApJ...737..103S}.
No galaxy is found to be coincident with the position of the FRB in a VLT/FORS2 $R$-band image.
A faint extended source is close to the FRB position in VLT/HAWK-I 
$K_s$-band image.
Spectroscopic follow up sources in the field  has not been undertaken. 

\subsubsection{FRB\,20230902A}

The FRB was detected during  First Large Absorption Survey in HI \cite[][]{2022PASA...39...10A}  survey observations \cite[FLASH,][]{2022PASA...39...10A}. 
The host galaxy was identified in both VLT/FORS2 $R$-band and HAWK-I $K_s$ band images. 
X-shooter spectroscopy of the host galaxy shows emission lines  H$\alpha$, H$\beta$, [O\,{\sc ii}], [O\,{\sc iii}] $\lambda$5007, and [S\,{\sc ii}]  with a redshift of $z=0.3619$.

\subsubsection{FRB\,20231226A}

The FRB was detected during a VAST \cite[][]{2021PASA...38...54M} observation. 
The host galaxy was identified in Legacy Survey imaging, catalogued as WISEA J102127.29+060634.5.  A VLT image of the host galaxy can be found in Figure \ref{fig:host1}.
The FRB has been localised to a spiral arm in a host galaxy for which X-shooter spectroscopy shows emission from H$\alpha$, [O\,{\sc ii}], [O\,{\sc iii}] and [N\,{\sc ii}]  with a redshift of $z=0.1569$.

\subsubsection{FRB\,20240201A}

 The burst was detected in filler observations of a Galactic latitude-$50$ degree field \cite[][]{2018Natur.562..386S}.
The FRB is coincident with the galaxy WISEA J095937.44$+$140519.4. A VLT image of the host galaxy is presented in Figure \ref{fig:host1}.  The galaxy has been spectroscopically observed in  the Sloan Digital Sky Survey and has a catalogued redshift of $z=0.047279$ \cite[][]{2017ApJS..233...25A}.

\subsubsection{FRB\,20240208A}

 The FRB was detected in filler observations of a Galactic latitude-$50$ degree field \cite[][]{2018Natur.562..386S}.  Optical observations have not yet been undertaken.  The FRB appears to be coincident (separation $\lesssim 4''$) with the galaxy SDSS J103654.96$-$005712.2 which has a $g$-band magnitude of $g=22.8$ and a photometric redshift of $z=0.4\pm 0.1$ \cite[][]{SDSS}.	

\subsubsection{FRB\,20240210A}

 The FRB was detected in filler observations of a Galactic latitude-$50$ degree field \cite[][]{2018Natur.562..386S}, close to a null of the ASKAP primary beam. 
 As a result the fluence of the burst reported is likely underestimated, as our correction for primary beam assumes a Gaussian shape. 
 The burst has been localised to a spiral arm of the $r=14.9$ Seyfert~1 galaxy WISEA J003506.47$-$281619.1, with a redshift from the Southern Sky Redshift Survey of $z=0.023686$ \cite[][]{1998AJ....116....1D}.

\subsubsection{FRB\,20240304A}

The FRB was detected commensally with EMU. VLT/FORS2 imaging has been taken and a host galaxy identified (See Figure \ref{fig:host3}). Spectroscopic observations  undertaken with VLT/X-shooter show H$\alpha$,  H$\beta$, [O\,{\sc ii}], [O\,{\sc iii}], [N\,{\sc ii}], and  [S\,{\sc ii}], lines at a redshift of $z=0.2423$.

\subsubsection{FRB\,20240310A}

The FRB was detected in an EMU survey observation.
Optical imaging has been taken with VLT/FORS2, allowing a host to be identified (see Figure \ref{fig:host2}).
X-shooter observations of the host  show H$\alpha$, [O\,{\sc ii}], [O\,{\sc iii}], [N\,{\sc II}], and [S\,{\sc ii}] lines with a redshift  of $z=0.1270$.

\subsubsection{FRB\,20240318A}
This FRB was discovered in an observation of the third epoch of RACS-low \cite[][]{2020PASA...37...48M}.
Optical follow up of the burst has not yet been undertaken.  However the FRB appears to be  coincident  (separation $\lesssim 1.2''$)   with the galaxy WISEA J100134.45$+$373659.7 which has a $g$-band magnitude of $g=19.5$ and a  photometric redshift of $z=0.12\pm0.04$ \cite[][]{SDSS}.

\subsection{Poorly localised FRBs}

 During the ICS searches we detected a sub sample of FRBs for which we were unable to obtain interferometric positions.  This was usually due to the absence of a voltage download.  Fortunately, in one case (FRB\,20210807D), as noted above, the burst was sufficiently bright to be detected in commensal image-plane 10-s images.  This was not possible for other bursts because they were either too faint or the hardware correlator  was not recording data (as was the case in the filler observing mode).   Nonetheless, the bursts represent detections with the ICS search pipeline so should be considered when modelling the FRB population.
 
{\em FRB\,20200627A} was detected in a filler observation in a field targeting the fly's eye FRB\,20180131A \cite[][]{2018Natur.562..386S}.  The burst width exceeded that for the threshold for downloading voltages.

{\em FRB\,20210214A} was detected during a {\em pulsar check} observation of PSR\,B0031$-$07.  During {\em  pulsar check} observations  voltage downloading is disabled. While the majority of these were of two pulsars in the Galactic plane, PSR\,B0833$-$45 and PSR B1641$-$45 (being high DM so having more stable flux density in the ASKAP bands as they are less affected by diffractive scintillation), PSR~B0031$-$07 was observed when the others were not visible. 

{\em FRB\,20210809C} was detected commensally with time-lapse filming of the array for a documentary, as was the case for FRB\,20210807D.  As with FRB\,2020627A the burst duration was too long to trigger a voltage download.  However, the burst was insufficiently bright to be detected in 10-s visibilities. 

{\em FRB\,20220531A} was detected during a bandpass calibration observation for the hardware correlator visibilities.  
The array switched observing bands without having a voltage calibration observation conducted\footnote{The array switched frequencies immediately after the voltage download which destructively altered the delays and phases in the signal path; as a result, standard bandpass calibration was not possible.}.   
It may potentially be possible to calibrate the data using PKS~1934$-$638 as an off-axis calibrator, but this has not been attempted.

{\em FRB\,20230521A}  was detected during FLASH survey observations. This was the first FRB detected in the lowest frequency band used by ASKAP (the central frequency is set to $832.5$\,MHz). The burst width failed to meet the criteria for voltage download. The burst was confirmed in routine inspection of FRB candidates.  

{\em FRB\,20231006A} was detected during a bandpass observation. Only a subset of the voltages for the FRB were downloaded (for a subset of antennas and a single polarisation), as voltage downloads stopped at the end of the short scan.  No attempt has been made to further localise this FRB.

\section{SURVEY PERFORMANCE}
\label{sec:survey_performance}

Given the commensal nature of the survey, we assess detection rates as a function of changing observation parameters.
Through to the end of 2023 (MJD 60309), ASKAP  observed in ICS mode for a total of \num{15324.7}\,hr, for an average on-sky efficiency of $32\%$. As a commensal survey, this has resulted in a variety of pointing directions and frequency configurations. Full logging information was available only for \num{12701.7}\,hr of this, due to an early logging error.  This corresponds to the latter 30 FRBs detected until the end of 2023. Furthermore, the current version (`v3'; 28 FRBs) of the FREDDA FRB detection algorithm has been used only from April 2020 --- the initial version (`v1'; 7 FRBs) had slightly reduced sensitivity to high-DM FRBs, while `v2' (2 FRBs) had a bug that reported incorrect S/N values \citep{Hoffmann2024}. Therefore, some analyses below will be applied to a subset of the data. This will be noted in each case.

\subsection{Survey detection rates}
\label{sec:rates}

\begin{figure}
    \centering
    \includegraphics[width=\columnwidth]{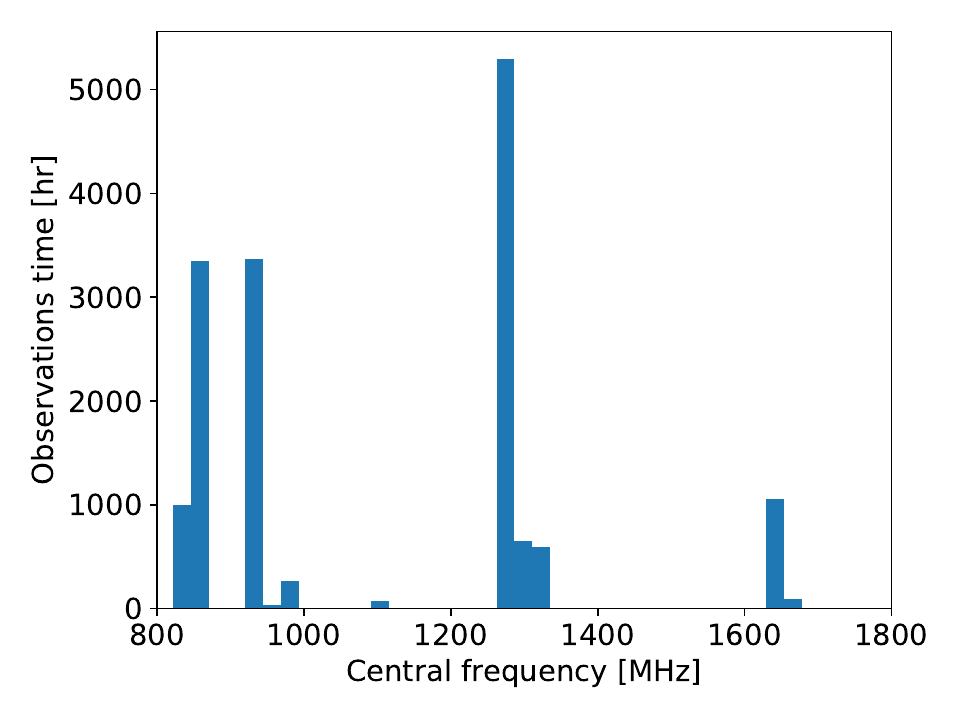}
    \caption{Histogram of observation time for ASKAP in ICS mode as a function of central observing frequency.}
    \label{fig:freq_hist}
\end{figure}

\begin{figure}
    \centering
    \includegraphics[width=\columnwidth]{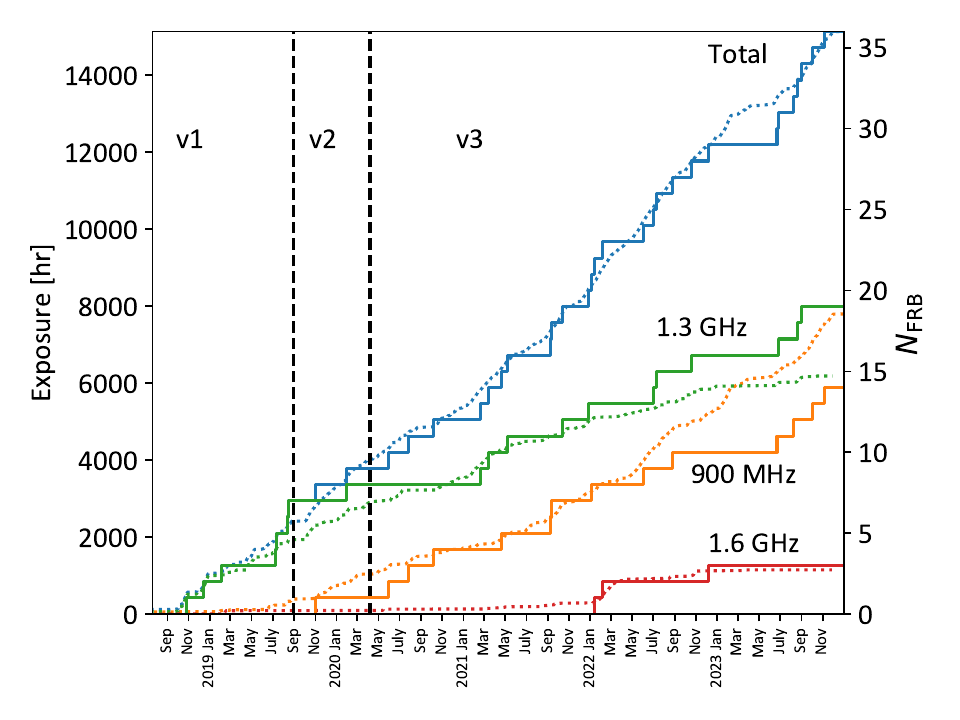}
    \caption{Cumulative total survey time for ICS observations to the end of 2023 as a function of date, and cumulative time divided according to approximate central frequency (dotted lines). This is compared to cumulative FRB discoveries (solid lines). The vertical dashed lines denote when different versions of FREDDA were used in the analysis.}
    \label{fig:freq_rates}
\end{figure}

Given the commensal nature of a large portion of the survey it is important to assess how detection rates depending on the nature of the underlying observing programs which observed over a wide range of Galactic latitudes across the entire available ASKAP band.

When assessing the detection rate, it is important to consider the range of central frequencies of the observations.
Figure~\ref{fig:freq_hist} shows the distribution of survey times per central frequency. Nominally, we classify observations into three categories: 900\,MHz ($\overline{\nu} < 1$\,GHz), 1.3\,GHz (1\,GHz$<\overline{\nu} < 1.5$\,GHz), and 1.6\,GHz ($\overline{\nu} > 1.5$\,GHz), reflecting the three bands used for ASKAP observing. While the Fly's Eye survey was conducted exclusively at 1.3\,GHz, the ICS survey was conducted at a range of frequencies --- many of the ASKAP survey projects have chosen to observe at lower frequencies to avoid interference from global navigation system satellites.

The cumulative detection rate is plotted against observation hours in Figure~\ref{fig:freq_rates}. ASKAP ICS observations have detected, on average, one FRB per 414\,hr, with the rates in each of the three frequency bands being once per 533\,hr, 326\,hr, and 383\,hr for 900\,MHz, 1.3\,GHz, and 1.6\,GHz respectively. These rates are lower than expected when compared to that found during observations in Fly's Eye mode \citep[20 FRBs in 1274.6 days, i.e.\ one FRB per \num{1530}\,hr;][]{2018Natur.562..386S,2019PASA...36....9J}. The sensitivity in ICS mode is expected to scale as $N_{\rm ant}^{0.5}$ with the number of antennas, and hence the rate should vary as $N_{\rm ant}^{0.75}$, assuming a Euclidian distribution of bursts. For $\sim25$ antennas being used simultaneously, this implies a rate of once per 137\,hr, i.e.\ 2.5 times higher than the 1.3\,GHz ICS rate. The cause of this deficit is so far unexplained.

\subsection{Modelling frequency dependence}

\begin{figure}
    \centering
    \includegraphics[width=\columnwidth]{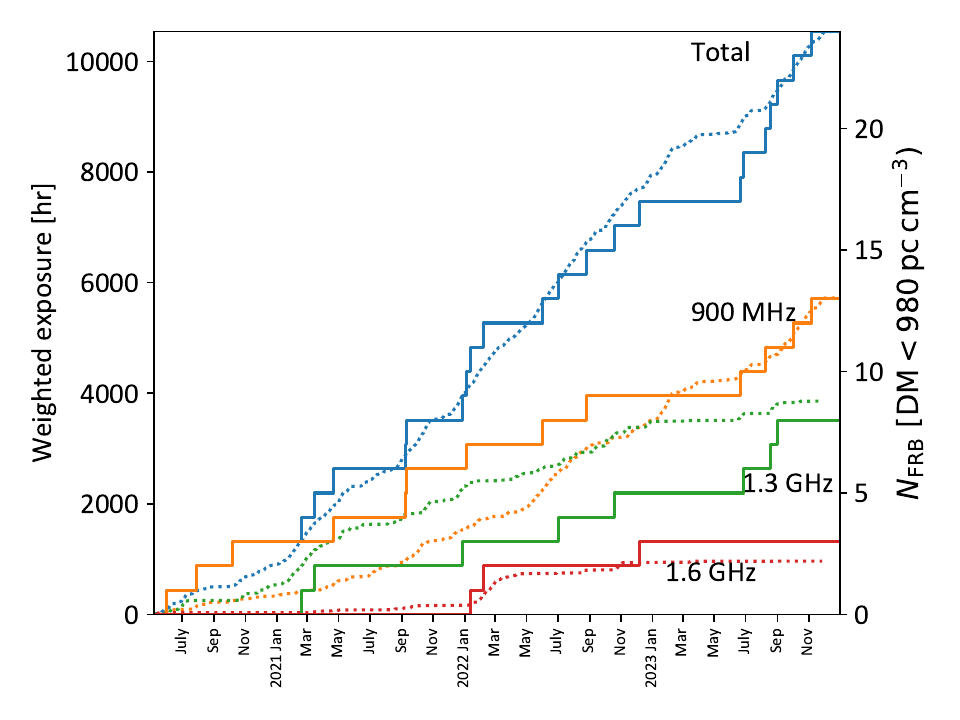}
    \caption{Weighted exposure (dotted lines; see text) against cumulative FRB detections (solid lines) for the period during which `v3' of our FRB detection algorithm was operating.}
    \label{fig:weighted_rates}
\end{figure}

We model the frequency dependence of the detection rate $R(\nu)$ to be
\begin{eqnarray}
R(\nu) & \propto & T(\nu) F(\nu) W(\nu) N_{\rm ant}^{0.75} \nu^{\alpha}\,, \label{eq:frb_rate}
\end{eqnarray}
where $T(\nu)$ describes the dependence on the observing system, $F(\nu)$ the dependence on the beam shape, $W(\nu)$ on the effective width, and the factor $\nu^\alpha$ represents the intrinsic dependence of the FRB rate on observation frequency with the rate index $\alpha$.
In all cases, we assume a Euclidean dependence of the rate on the detection threshold, i.e.\ $R \propto F_{\rm th}^{-1.5}$ (see Section~\ref{sec:source_counts}). The term $T(\nu) \propto T_{\rm sys}^{-1.5}$, reflecting the frequency-dependent system temperature from \citet{2021PASA...38....9H}. $F$ is the `footprint' term, calculated as
\begin{eqnarray}
F & = & \int B^{1.5} d\Omega,
\end{eqnarray}
for the beam values $B$ calculated as the envelope over 36 Gaussian beams (in either a hexagonal 'closepack 36' or `square 6x6' configuration, with pitch angles varying from 0.72$^\circ$--1.1$^\circ$), full width half maximum of $1.1 \lambda/D$, (where $\lambda$ is the wavelength of the emission and $D$ in the $12$-m antenna diameter) and peak amplitude depending on offset from boresight according to \citet{2021PASA...38....9H}, where $\lambda$ is the wavelength . The `width factor', $W$, is due to the effective width (and hence the threshold: $F_{\rm th} \propto W^{0.5}$) of FRBs changing with sampling time, FRB intrinsic width, degree of scatter broadening, and dispersion measure smearing; the latter two of which are frequency-dependent. This is calculated using the measured properties of FRBs in the high time resolution sample of (D. R. Scott, et al. in prep.), and the effective width according to \citet{Cordes_McLaughlin_2003}. $N_{\rm ant}$ is the number of antennas used in the observation, such that sensitivity scales as $N_{\rm ant}^{0.5}$, and thus rate as $N_{\rm ant}^{0.75}$.

We compare the cumulative integral of Eq.~\ref{eq:frb_rate} against FRB detections during the stable `v3' period in Figure~\ref{fig:weighted_rates}, assuming no intrinsic rate dependence ($\alpha=0$). To account for high-DM FRBs being undetectable in low-frequency observations due to only 4096 DM samples being searched, we remove all FRBs with a DM above 980\,pc\,cm$^{-3}$. The rates predicted by Eq.~\ref{eq:frb_rate} are re-scaled to have an average of unity, i.e.\ they convert real hours into weighted hours. The result is that the total number of detected FRBs closely matches the expected number in each frequency band, i.e.\ we see no evidence for an intrinsic rate dependence.

Our study of the spectral behaviour of 23 Fly's Eye FRBs \citep{2019ApJ...872L..19M} suggested that, on average, spectral fluence $F_{\nu} \propto \nu^{-1.6_{-0.2}^{+0.3}}$. However, as we note in \citet{2022MNRAS.509.4775J}, biases due to beam shape would revise this to $\alpha=-0.65 \pm 0.3$ should FRBs have extremely narrow bandwidths. Varying $\alpha$, and comparing the relative to predicted rates between the low and mid bands, produces $\alpha = -0.3_{-1.6}^{+1.4}$. Thus our observations cannot yet constrain the spectral dependence of FRBs.

The modelling of Eq.~\ref{eq:frb_rate} also predicts that the ASKAP configuration in Fly's Eye mode was --- when excluding the $N_{\rm ant}^{0.75}$ factor --- 33\% more efficient at detecting FRBs than the average ICS observation, and 22\% more efficient than ICS 1.3\,GHz observations. Allowing for this, the expected rate for ICS 1.3\,GHz observations becomes one per 167\,hr, i.e.\ the observed rate of once per 326\,hr is a factor of $\sim2$ below expectation.
 
\subsection{Galactic latitude dependence}

\begin{figure}
    \centering
    \includegraphics[width=\columnwidth]{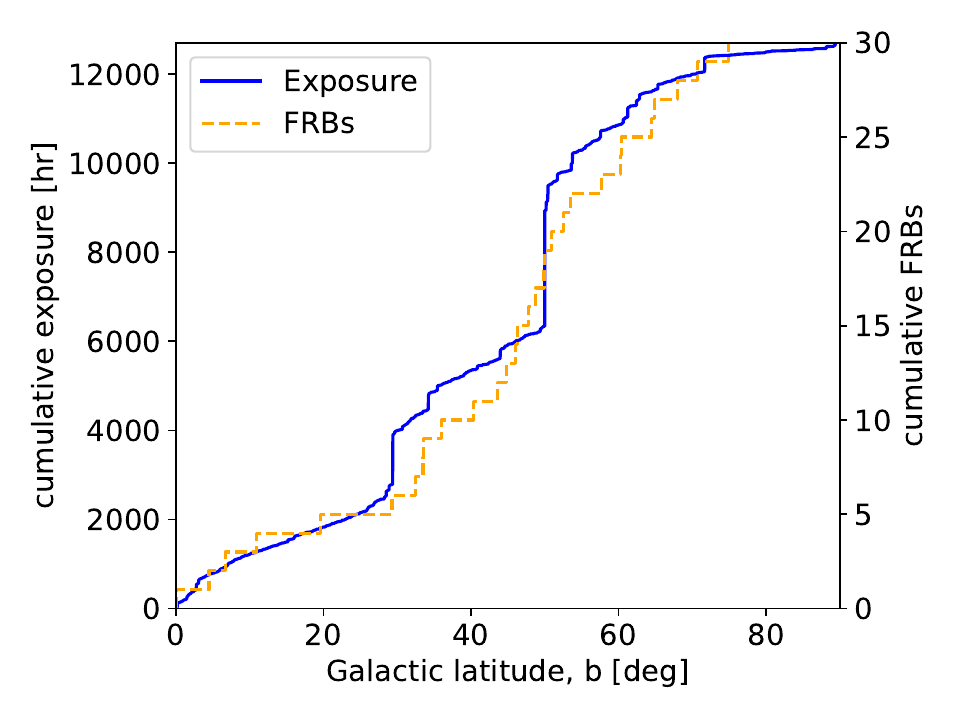}
    \caption{Cumulative exposure to Galactic latitude $b$ of \num{12701.7}\,hr of CRAFT data, against FRB discoveries in Galactic coordinates. The increase in exposure at $|b|$=50 is the result of  the latitude-50 CRAFT filler observations.}
    \label{fig:galactic_b}
\end{figure}

During the initial Murriyang surveys that established FRBs as a class of astrophysical transient, there were suggestions of a deficit of FRBs at mid-to-low Galactic latitudes \citep{2014ApJ...789L..26P}, which was speculated to be due to interstellar scintillation \citep{2015MNRAS.451.3278M}. Analysis after further detections did not find evidence for a strong effect \citep{2018MNRAS.475.1427B}. In Figure~\ref{fig:galactic_b} we compare Galactic latitude coverage against the 30 FRB detections for which we have logging data. We find no significant evidence for a latitude-dependent event rate, so discount this as an explanation for the rate deficit.

\subsection{Source counts analysis}
\label{sec:source_counts}

\begin{figure}
 \centering
\includegraphics[width=\columnwidth]{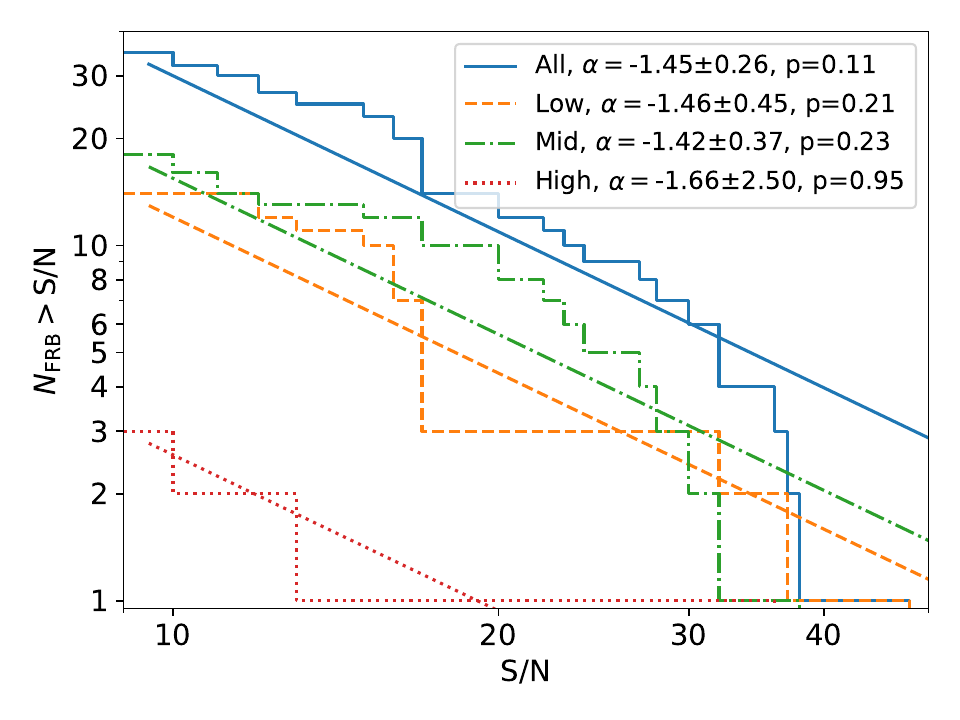}
\caption{Source counts for ASKAP/ICS FRBs. We show the counts using `all' frequencies, and splitting source counts into `low', `medium', and `high' ranges. The two FRBs detected during the period when the search algorithm returned incorrect S/N values have been excluded. Also shown is the best-fit value of the power-law slope $\alpha$ in each case, and the $p$-value of a KS-test against $\alpha = -1.5$.
}
\label{fig:source_counts} 
\end{figure} 

The FRBs detected in the fly's eye survey have been shown to have a S/N distribution consistent with the Euclidean expectation of $N_{\rm FRB}>{\rm S/N} \propto {\rm S/N}^{-1.5}$ \citep{James2019}. 
Figure~\ref{fig:source_counts} shows this distribution for all ASKAP ICS FRBs, excluding those two discovered during the period when FREDDA returned incorrect S/N values. Using the method of \citet{1970ApJ...162..405C} to estimate the power-law index $\alpha$ for all FRBs produces $\alpha = -1.50 \pm 0.27$, which becomes $-1.46 \pm 0.27$ when correcting for the expected bias. This suggests that if the source counts distribution does indeed flatten, as predicted by \citet{Macquart2018b} and potentially observed in Murriyang \citep{James2019} and MeerKAT\citep{Jankowski2023} data, this occurs below the detection threshold of CRAFT ICS observations.

The observed source counts distribution is also a good diagnostic tool for biases in the search pipeline: human vetting, RFI mitigation algorithms, or both, may reject candidates that are close to the nominal detection threshold, or well above it \citep{Macquart2018a}. We have performed Kolmogorov-Smirnov tests \citep{kolmogorov,smirnov} for consistency with a pure  $\alpha = -1.50$ power-law on ICS FRBs, with p-values quoted in Figure~\ref{fig:source_counts}, and find no strong statistical evidence for a deviation. The same conclusion is reached when dividing the sample into FRBs detected in low, medium, and high frequency ranges.

There is some evidence however for a reduced number of FRBs in the $S/N \le 14$ regime, and a deficit of FRBs with very high S/N, but this is not conclusive. Fitting to all FRBs with S/N>14 increases the expected total number of FRBs by a factor of 1.40, i.e.\ a lack of S/N$<14$ FRBs cannot explain the deficit found in Section~\ref{sec:rates}. Equivalently, scaling up the observed ICS rate at 1.3\,GHz by 1.4 to one FRB per 250 days, and comparing this rate to the fly's eye rate, implies a source-counts slope of $\alpha = -1.1$. Yet fitting to Figure~\ref{fig:source_counts} for S/N$>14$ produces $\alpha = -2.2 \pm 0.5$. In other words, a change in source-counts slope would also require our detection pipeline to miss high-S/N FRBs, or otherwise reduce the S/N of those detected.

\subsection{Non-linearities in detection}

Analysis of the source-counts slope using S/N assumes a linear relation between FRB fluence and S/N \citep{James2019}. There are several possible causes of non-linearities in our detection system, which could feasibly reduce the S/N of high-S/N events, which we consider below.

The FREDDA detection algorithm normalises channelised power according to the mean and standard deviation of each `block' of 256 samples, approximately $300$\,ms in duration. However, this normalisation is applied to data in a subsequent block, which means that a bright FRB cannot influence its own S/N estimates. Similarly, a check on the kurtosis of each channel to remove RFI is also applied to subsequent blocks. The only possible effect then would be for an extremely bright FRB to exceed the 8-bit dynamic range of the ICS data at detection. The RMS of each time--frequency scintle is typically set to 16 digital units (d.u.), meaning that the peak S/N of a narrow FRB, with 128 d.u.\ of power in all $336$ frequency channels, would be of $144$, above which the system response must be less-than-linear. FRBs will always be DM-smeared in time, which increases the S/N threshold beyond which the system becomes non-linear; while scintillation and narrow effective bandwidths will decrease the threshold, due to individual scintles exceeding 128 d.u. We estimate scintillated, low-DM FRBs to suffer non-linear effects for S/N$>102 \, (\Delta \nu /336\,{\rm MHz})$, where $\Delta \nu$ is the FRB bandwidth. However, such FRBs will only have their S/N reduced; they will not be missed, and could be readily identified in offline analysis. We have not detected any such FRBs yet in our sample.

We also observe that ASKAP beams typically overlap near the half-power points. Hence, any FRB with very high S/N, even if vetoed by some unforeseen part of the system, would be detected by adjacent beams, similarly to one burst from FRB 20201124A \cite[][]{2022MNRAS.512.3400K}.

The final possible cause of detection biases we consider is in the anti-RFI candidate vetting script that parses raw candidates, and determines whether or not to trigger the system. This has been developed by using ASKAP FRBs detected in Fly's Eye mode \citep{2018Natur.562..386S}, and in theory could introduce a S/N-dependent bias. However, no such bias was observed during tests. Therefore, if the observed deviations (at moderate significance) from a pure power-law in S/N are real, we cannot explain their origin with known systematic effects.

\subsection{Elevation dependence}

\begin{figure}
 \centering
\includegraphics[width=\columnwidth]{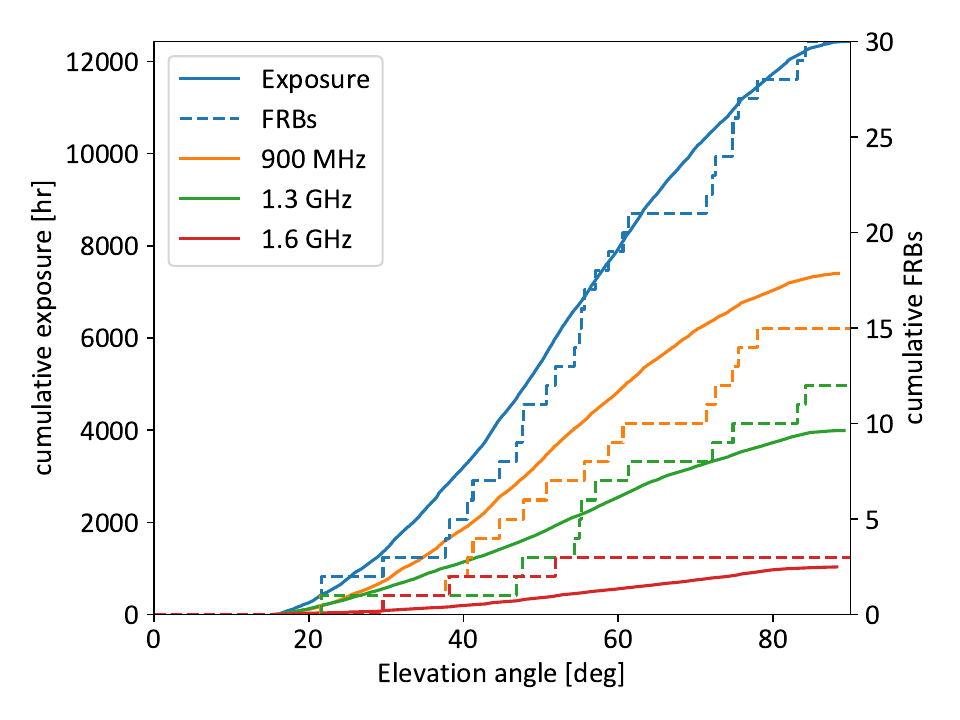}
\caption{Cumulative exposure as a function of local elevation angle (calculated at the beginning of each scan), compared to the elevation angles at which ICS FRBs have been detected, over the period Oct.\ 2019--Dec.\ 2023 for which we have records. The exposure and rates are also shown divided into the three frequency ranges described in Section~\ref{sec:rates}.}
\label{fig:cumulative_elev} 
\end{figure} 

The effect of RFI on FRB detection rates is expected to be elevation-dependent, as RFI sources --- particularly those on the horizon --- move in and out of ASKAP's sensitivity pattern. Figure~\ref{fig:cumulative_elev} plots the elevation dependence of FRB rates against the total exposure (approximated by the boresight elevation at scan start) for which we have records. We find no evidence for an elevation-dependence to the FRB detection rate.

\subsection{Assessing fly's eye localisation }

The capability to localise FRBs to arcsecond positions also enables us to examine the localisation and fluence measurements presented in the fly's eye surveys. 
For the fly's eye surveys, we leveraged the multiple-beam detections to improve on FRB localisation and to better determine FRB fluence.
The algorithm, described in \cite{2017ApJ...841L..12B}, used the relative signal-to-noise ratios of detections in multiple beams to determine the location of the burst position on the focal plane. The algorithm marginalised over uncertainties in beam shape, gain, and position when determining burst position and brightness, and uncertainties on the parameters. It utilised Bayesian inference, and the maximum a-posteriori parameters and their uncertainties were derived from posterior samples calculated from a nested sampling algorithm \cite[][]{2009MNRAS.398.1601F}.

\begin{figure}
 \centering
\includegraphics[width=\columnwidth]{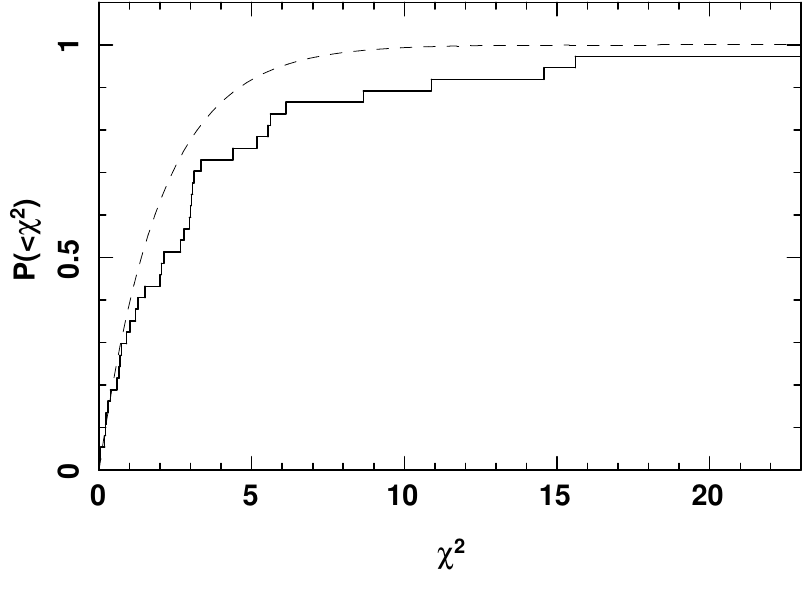}
\caption{Cumulative distribution function of multi-beam positions goodness of fit. The dashed line shows the expected $\chi^2$ distribution if the model was well specified.}
\label{fig:pos_chi2} 
\end{figure}

We assess the position performance using a $\chi^2$ test:  
\begin{equation}
\hat{\chi}^2 = \bm{\Delta} r \bm{C}^{-1} \bm{\Delta r}^T,
\end{equation}
where $\bm{\Delta r} = [ \Delta \alpha \, \Delta \delta ]$  is the vector difference between the interferometric and multi-beam positions and $\bm{C}$ is a covariance matrix parameterising the uncertainty in the multi-beam position.  
We assume the position uncertainty as determined using the multi-beam method to be a bivariate Gaussian, parameterised by the variance in right ascension and declination, and their covariance.  These were calculated directly from the posterior samples. 
We do not account for uncertainty in the interferometrically derived position as it is typically a factor of $\gtrsim 200$ smaller than that of the multi-beam-derived position. In Figure \ref{fig:pos_chi2}, we show the cumulative distribution function of the $\hat{\chi^2}$ values, and compare to the expected $\chi^2$ distribution with two degrees of freedom.  We find modest disagreement between the measurements and the expected distribution, with a Kolmogorov Smirnov test reporting a probability of $p=0.04$ that the distributions agree. This can be attributed to a few outlying burst localisations ($\hat{\chi}^2 > 10$).  FRB\,20240201A was found to originate in a null of the primary beam.  The beam model used in the multi-beam localisation method assumes a Gaussian beam so does not include nulls or side lobes.  FRB\,20210320C originated from an outer edge beam, which is also significantly non-Gaussian. The origin of the large disagreement between the interferometric and multi-beam positions observed for FRB\,20230902A is unclear; however it was the only FRB localised at the lowest standard ASKAP central observing frequency ($832$\,MHz). It is likely that at the lowest frequencies the ASKAP beams deviate greatest from Gaussian shape.   We find that if we increase the size of the uncertainties by $10 \%$ (which would reduce the $\chi^2$ by a factor of approximately $20\%$), the reported probability from the K-S test would increase to $0.3$.

\subsection{Burst positions within beams}

\begin{figure}
 \centering
\includegraphics[width=3. in]{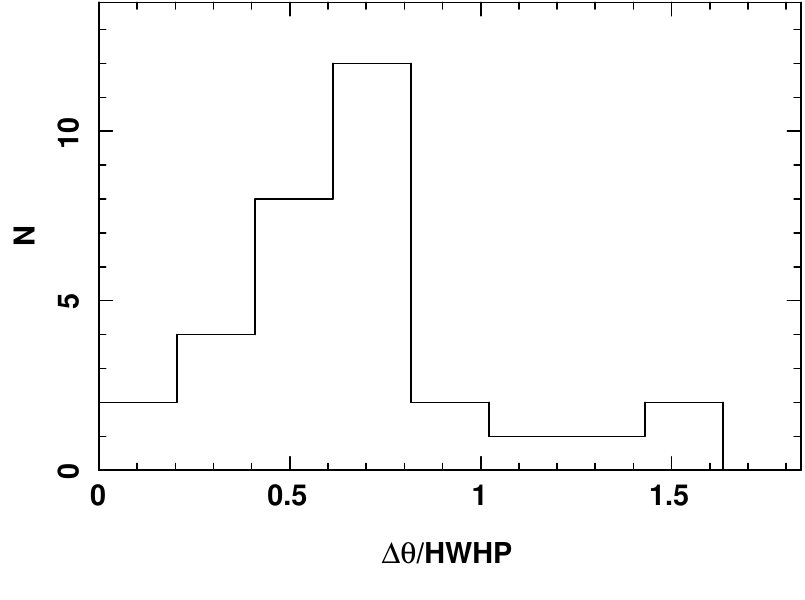}
\caption{Distribution of FRB localisations relative to beam centre.}
\label{fig:beam_offset} 
\end{figure}

Figure \ref{fig:beam_offset} shows the relative position of the FRB positions relative to the primary beam for the FRBs for which we obtained interferometric localisation. We have assumed a  beam of full width at half power (FWHP) following  $1.1 \lambda/D$ \cite[][]{2016PASA...33...42M}.
The number of bursts increases radially, matching the relative area which scales with the square of radial distance.
It then decreases at larger radius.  The inner beams are spaced at distances typically smaller than the FWHP. 
There is a population of bursts localised well outside the half power point of the primary beam.
These bursts were predominantly discovered in the outer beams of the PAF.
Future image-plane ASKAP detection systems (discussed below) have a more limited field of view, so are potentially not sensitive to these side-lobe detections.

\subsection{Burst modulation}
\label{sec:modulation}

Many of the ASKAP FRBs show spectral modulation: intensity variations  in frequency that could either be diffractive scintillation or intrinsic to the emission mechanism. This was identified in the fly's eye survey\cite[][]{2018Natur.562..386S}.
We investigated if there is any correlation between modulation and dispersion measure using the FRBs in the search data stream.  
We choose to investigate modulation and dispersion measure to consider both the fly's eye and ICS FRBs. 
We calculate the spectral modulation index to be 
\begin{equation}
m_I = \frac{\sigma_{\rm FRB}^2 - \sigma_n^2}{\sigma_n^2},
\end{equation}
where $\sigma_{\rm FRB}^2$ is the variance on pulse and $\sigma_n^2$ is the variance of (thermal) noise measured in a segment of data of equal temporal duration close to the FRB.
We estimate the uncertainty on $m_I$ using bootstrap \cite[][]{10.1214/aos/1176344552}.  For each FRB, we resample (with replacement) the spectrum and re-calculate $m_I$ on each of the realisations. We use the standard deviation of these resample modulation indices as the uncertainty.

While FRBs with DM  $\lesssim 500\,$\pccm show a variety of spectral modulation, those with DM $> 800\,$\pccm show little spectral modulation.
Spectra of two FRBs with high modulation and two FRBs with high DM are shown in Figure \ref{fig:frb_spectra}.
It is possible that spectral modulation is being quenched at higher DM.  
If scintillation in the Milky Way is causing the spectral modulation, quenching would occur if the bursts are spatially resolved by Milky Way scattering screens, i.e. had been scatter broadened by a second screen.
As there is a correlation between dispersion measure and distance \cite[][]{2020Natur.581..391M}, this would suggest the extra-Galactic scattering scr 
Studies of a larger sample of bursts in higher spectral and time resolution (such as using data products derived from the CELEBI pipeline) are required to assess this effect. 
Study of bursts produced by the CELEBI pipeline also benefit from the array-coherennt improvement in signal to noise ratio.
This could be complemented with bursts from CHIME/FRB, DSA-110 \cite[][]{2024Natur.635...61S}, or MeerKAT \cite[][]{Jankowski2023}.  The latter two would allow for comparison of large FRB samples detected at comparable central frequencies.

\begin{figure}
 \centering
\includegraphics[width=3. in]{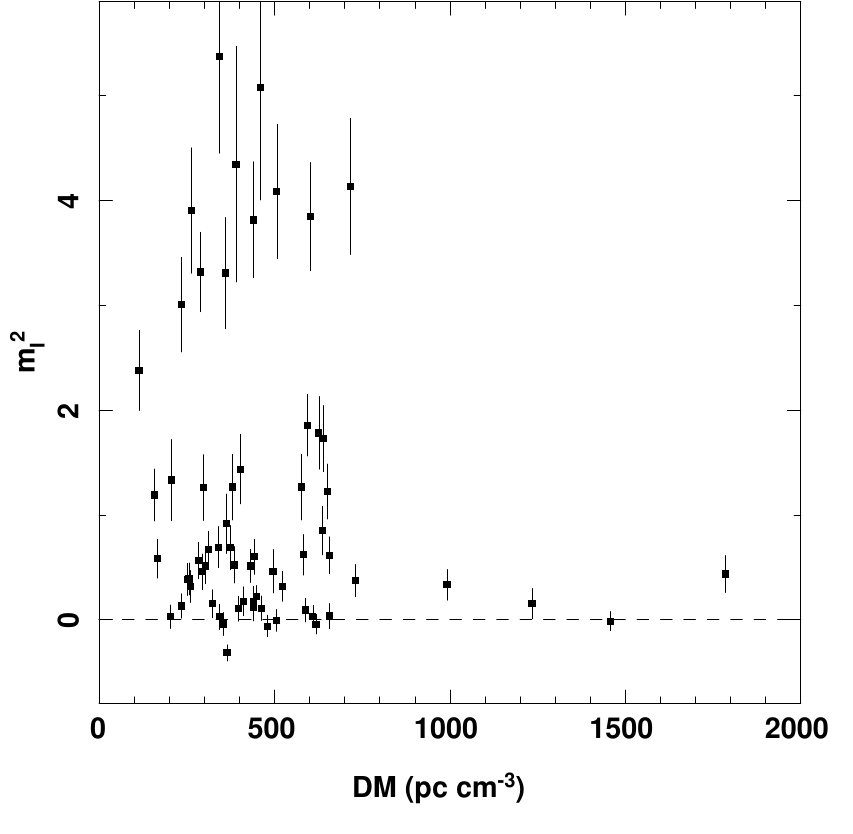}
\caption{FRB modulation indices $m_I$ for ICS and fly's eye FRBs.  The most dispersed FRBs show an absence of spectral modulation.}
\label{fig:mod_dm} 
\end{figure}

\begin{figure*}
 \centering
 \begin{tabular}{cc}
\includegraphics[width=2.5in]{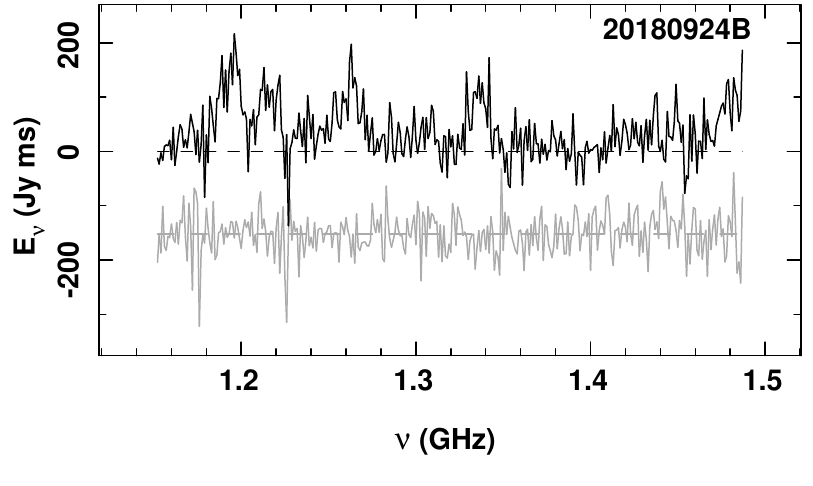} & \includegraphics[width=2.5in]{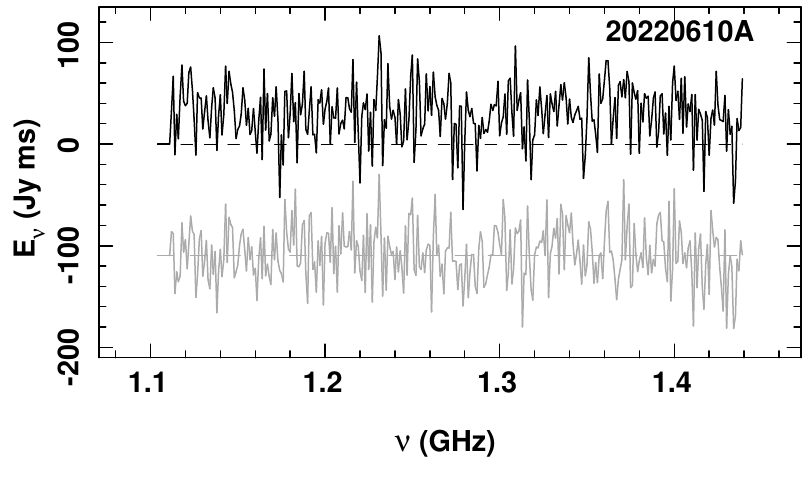} \\
\includegraphics[width=2.5in]{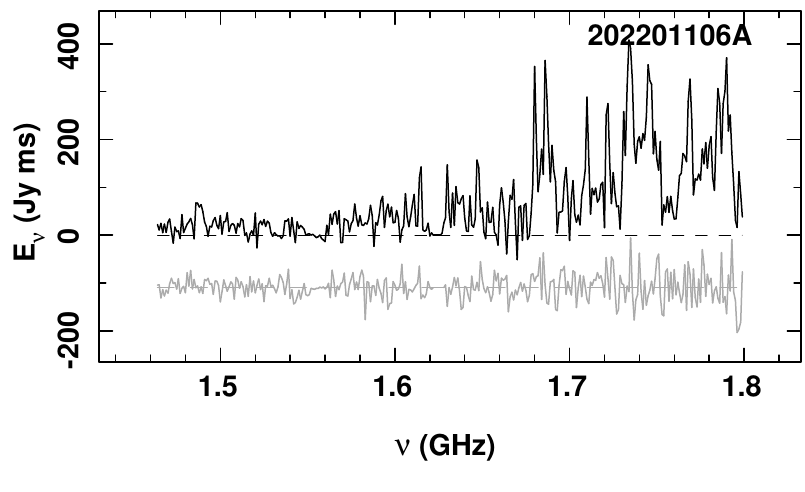} & \includegraphics[width=2.5in]{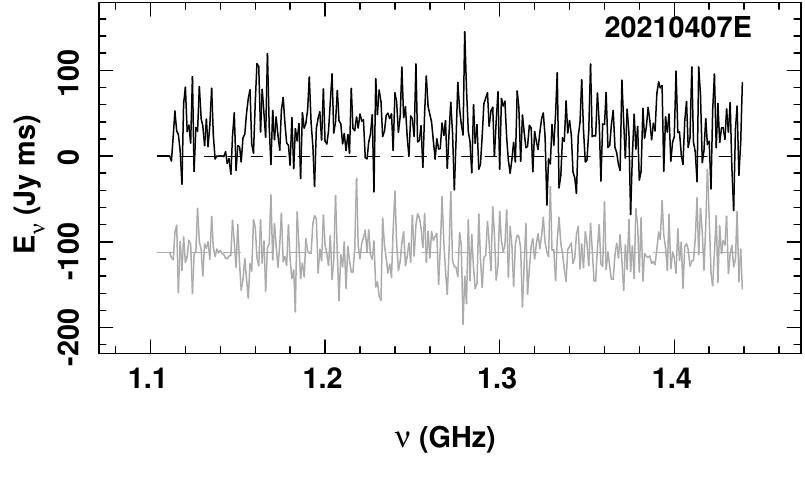} 
\end{tabular}
\caption{Pulse averaged spectrum of high and low DM FRBs.  The spectrum is shown as a black solid line.  An off pulse spectrum is shown offset in grey. The dashed horizontal lines show zero fluence.    FRB\,2020610A (DM $343.8$\,\pccm)  has the highest modulation index of the FRBs in our sample. FRB20180924B (DM $362.4$\,\pccm) also has high modulation.   FRBs 20220610 (DM $1458.1$\,\pccm ) and FRB20210407E (DM  $1785.3$\,\pccm) have the largest DM of the ICS FRB.   } 
\label{fig:frb_spectra} 
\end{figure*}

\section{SCIENTIFIC OUTCOMES OF THE ICS SURVEY}
\label{sec:science}

At the commencement of the ICS searches one of the key questions was the distance scale to FRBs.  The first results of the survey demonstrated that most FRBs do indeed originate from cosmological (gigaparsec-scale) distances.
Since then, the study of FRBs can be broadly bifurcated into answering two questions:
\begin{enumerate}
    \item What causes FRBs?
    \item How can FRBs be used as cosmological tools?
\end{enumerate}

\begin{figure}
 \centering
\includegraphics[width=\columnwidth]{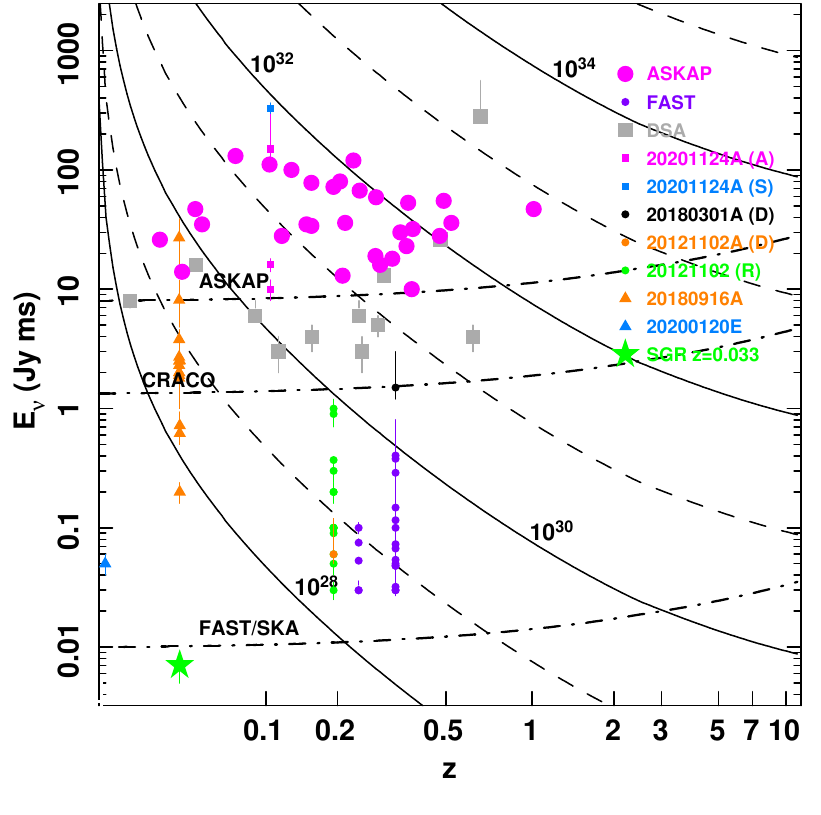}
\caption{ Fluence-redshift relation for localised FRBs with host galaxy associations. The legend lists the FRB surveys and specific FRBs  of interest.  The dash-dotted lines show the sensitivity of the ASKAP ICS Survey (ASKAP), the upgraded ASKAP coherent searches (CRACO), and surveys undertaken with the Square Kilometre Array or FAST (SKA/FAST) which have comparable sensitivity.  The solid and dashed lines are curves of constant energy, assuming concordance cosmology. In addition to the ASKAP-localised FRBs presented here, we show FRBs localised with the Deep Synoptic Array \cite[][]{2019Natur.572..352R, 2024ApJ...967...29L}, repeat bursts detected by the FAST telescope from FRB\,20180301A and FRB\,20190520B \cite[][]{2020Natur.586..693L,2022Natur.606..873N}. We  show  the initial detection of FRB\,20180301A with Murriyang \cite[][]{2019MNRAS.486.3636P}.  We also show bursts from the previously active repeater FRB\,20201124A from ASKAP \cite[][]{2022MNRAS.512.3400K} and the Stockert Radio Telescope \cite[][]{2021ATel14556....1H}; the initial detection of the first repeater, FRB\,20121102A \cite[][]{2014ApJ...790..101S} and a sample of its repetitions \cite[][]{2019ApJ...876L..23H}; and bursts from low-redshift repeating sources FRB\,20200120E \cite[][]{2023MNRAS.520.2281N} and  FRB\,20180916B \cite[][]{2020Natur.577..190M}.  Finally, we show the expected fluence of the bright FRB-like pulse emitted from the Galactic magnetar SGR~1935$+$21 \cite[][]{2020Natur.587...59B,2020Natur.587...54C} if it was emitted from the host galaxy of FRB\,20180916B, juxtaposing it with cosmological FRBs detected with surveys such as ours.} 
\label{fig:frb_pop} 
\end{figure}

Accumulating a population of localised FRBs is essential in answering both these questions. 
To determine what causes FRBs, it is necessary to identify the host galaxy and the FRB environments.
The ICS searches have delivered the first substantial sample of $\lesssim$1~arcsecond-localised FRBs, including the first large sample of localised one-off FRBs.
Figure \ref{fig:frb_pop} shows the redshift-fluence distribution of the FRBs discovered in the ASKAP surveys, and compares them to other FRBs localised to host galaxies.

\subsection{FRB host galaxies}

\begin{figure}
 \centering
\includegraphics[width=\columnwidth]{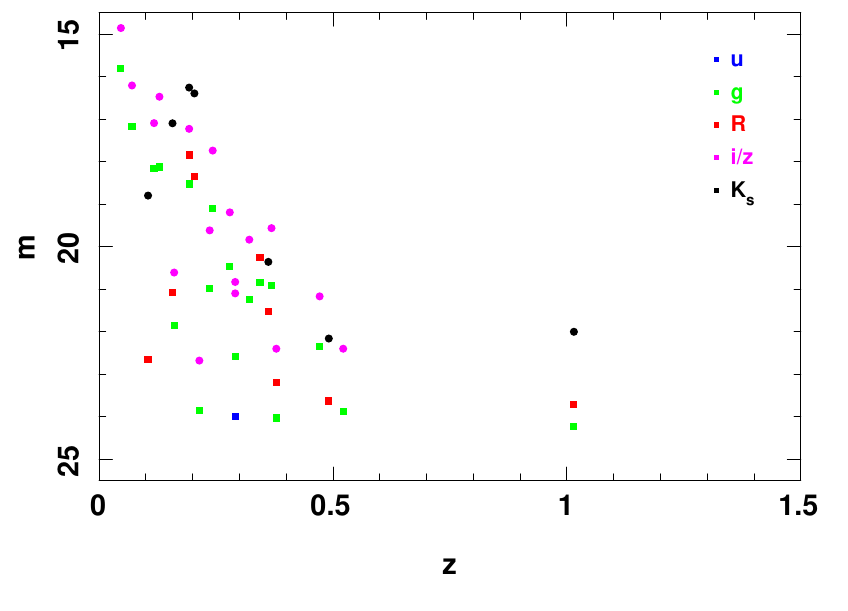}
\caption{VLT photometry of ASKAP/ICS fast radio burst host galaxies. 
}
\label{fig:photometry} 
\end{figure}

By virtue of delivering the largest sample of localised FRBs in its era, the CRAFT/ICS survey has provided a great opportunity to understand the demography of FRB host galaxies and the attribution of FRBs to potential progenitor populations.
Figures \ref{fig:host1} to \ref{fig:host3} show images of the sub sample of our FRB host galaxies observed primarily with the Very Large Telescope (VLT), arranged in increasing redshift.
Updated photometry for a sub sample of FRBs observed with the VLT is presented in Table \ref{tab:photometry}.
A plot showing the multi-band photometry as a function of redshift for this sub sample is shown in Figure \ref{fig:photometry}.

Over the course of the survey, we undertook a series of studies investigating the properties of FRB hosts.
\cite{2020ApJ...895L..37B} and \cite{2020ApJ...903..152H} conducted  the first demographic studies of FRB host galaxies, finding them to be largely late-type star forming galaxies.
\cite{2022AJ....163...69B} extended the analysis to include a larger sample of FRBs and compared the properties of the hosts of apparently one-off FRBs with those of repeating sources. The sample size was insufficient to distinguish any difference between the repeating and non-repeating sources.
Based on the Kolmogorov Smirnov tests reported in   \cite{2022AJ....163...69B} (which for many had $P_{\rm KS} \approx 0.25$) ,  which included a sample of $10$ apparently one-off FRBs and $6$ repeating FRBs, the expected sample size needed to be approximately a factor of  approximately two larger for each sub-population to reach $95\%$ confidence and a factor of approximately four larger in each to reach $99.9\%$ confidence assuming the distributions had identical distributions. 

\cite{2023ApJ...954...80G} undertook detailed modelling of host galaxy spectral energy distributions to assess the star formation histories of nearly all our FRB hosts identified prior to the beginning of 2022. 
From these studies it is now clear that the host galaxies of one-off bursts are markedly different from that of the first repeating FRB\,20121102A which was found to originate in a metal-poor dwarf galaxy \cite[][]{2017ApJ...834L...7T}.

The high angular resolution afforded by  ASKAP-detected FRBs can exceed that easily obtained on the ground with natural seeing,  warranting further follow up with the {\em Hubble Space Telescope} (HST).  In \cite{2021ApJ...917...75M} we presented the analysis of seven ASKAP localised FRBs (and one localised by the European VLBI Network, EVN), and showed that most of the bursts were located near spiral arm features in their hosts.
This implies that many FRB sources are associated with active star formation and young stellar populations.  
This was only possible because of the combination of the high precision ASKAP positions and the high resolution HST images. 
\cite{2023arXiv231201578W} used laser guide star adaptive optics with the Gemini South telescope to undertake a similar assessment of a further five ASKAP FRB host galaxies.
This provides support for the origin of FRB emission from young neutron stars, with the notable exception of the repeating CHIME FRB source 20200120E which has been localised to a globular cluster in the halo of M81 \cite[][]{2021arXiv210511445K}.

Very few of the host galaxies showed continuum radio emission.
No unresolved persistent radio emission has been identified coincident with any of the bursts discovered by ASKAP \cite[e.g.,][]{2019Sci...365..565B,2020ApJ...895L..37B} despite sensitive observations. 
However only one of the FRBs in the sample (FRB~20190711A) has been found to repeat, and originated at a redshift of $z=0.521$.
Diffuse emission attributed to star formation was observed for two host galaxies.
Hydrogen and carbon-monoxide line emission was also detected for the host galaxies of three FRBs \cite[FRBs 20230718A and 20180924B,][]{2023ApJ...949...25G,ALMAobs,Glowacki2024}.


\subsection{Burst-emission physics}

Access to the voltage buffers allowed us for the first time to study ASKAP-detected FRBs both at time resolutions much shorter than $1$\,ms, and with full polarisation.
This enabled improved insight into burst emission physics and the properties of intervening material along the line of sight.
\cite{2020ApJ...891L..38C} demonstrated the power of using the voltage buffers to form a tied-array beam time series at the position of the bursts. What appeared at low time resolution to be a single pulse in FRB\,20181112A was in fact four distinct bursts, with each showing different pulse morphologies and polarimetric properties.
\cite{2020MNRAS.497.3335D} used high time resolution imaging on a larger sample of ASKAP bursts, uncovering a diversity of polarimetric properties and morphologies. While some bursts were obviously scatter-broadened, many had multiple components with varying levels of linear and circular polarisation, including variations in both across  pulses.
This was in contrast to previous studies of repeating FRB sources which in general showed high degrees of linear polarisation, constant linear polarisation position angles, and less evidence for circular polarisation.

The spectropolarimetry enabled searches for conventional and generalised Faraday rotation \cite[][]{2024arXiv241114784B}, and spectral depolarisation.
The searches showed that most of the detected FRBs had low rotation measures (RMs) $\ll 10^2$ rad\,m$^{-2}$ \cite[][]{2023ApJ...954..179M}, in contrast to the first repeating FRB which has a high RM \cite[][]{2018Natur.553..182M}.
Similarly, constraints on spectral depolarisation suggest that the scattering media foreground to the ASKAP FRB sample are less magnetoionically active compared to that of repeating FRB sources \cite[][]{2024MNRAS.527.4285U,2023Sci...382..294R}.

\subsection{FRB repetitions} 
\label{sec:repetitions}

One of the fundamental (but potentially unproveable) questions about FRBs is whether they all (eventually) repeat. This uncertainty stems from the observational limitations and the stochastic nature observed in the repetition rates across the FRB population. In self-conducted follow up within the fly's eye survey, which spanned thousands of hours, none of the FRBs were found to repeat \citep{2018Natur.562..386S}. Initial studies of the first repeating FRB\,20121102A suggested a steeper luminosity function, highlighting the importance of using more sensitive radio telescopes for such investigations \citep{2017ApJ...850...76L, 2018ApJ...861L...1C}.

The presence of repetitions excludes cataclysmic progenitor models for producing some FRB emission.
The large volumetric rate of repeating FRBs is also inconsistent with cataclysmic progenitor models \cite[][]{2019NatAs...3..928R}.

Throughout the ICS survey, we have conducted an extensive monitoring program to search for repeat bursts from detected FRBs. This has been executed using some of the world largest single-dish radio telescopes, providing unprecedented time resolution ($\sim$20--80\,$\upmu$s) and frequency coverage ( as large as $\sim$3\,GHz bandwidth). Follow-up campaigns were conducted with the 110-m Robert C. Byrd Green Bank Telescope (GBT) using the L-band and the 800\,MHz receivers \citep{2015ursi.confE...4P}, and the {\em Murriyang} telescope using the 20-cm multi-beam \citep{1996PASA...13..243S} and the ultra wide-bandwidth low (UWL) receiving systems \citep{2020PASA...37...12H}. The GBT allowed us to monitor FRB sources in the northern sky (Decl. > $-46^\circ$) with higher sensitivity than Murriyang. The Murriyang telescope, equipped with the UWL receiver, offered unparalleled frequency coverage from 704 to 4032 MHz, unique among the existing facilities. Additionally, we searched for repetitions using the Five hundred meter Aperture Spherical radio Telescope (FAST) for the source FRB\,20171019A. It is pertinent to note that the majority of the ICS-detected FRBs are at southern declinations, not visible from FAST. Searches for repetitions were also undertaken with ASKAP through filler observations. These included repeated observations of the latitude-50 fields, targeted monitoring of FRB sources known to repeat, and specific targeting of ASKAP-detected FRBs exhibiting properties similar to known repeaters, notably the `sad-trombone' pulse morphology \citep{2019ApJ...876L..23H}.

In this monitoring program, we collected a total of 1070 hours of high-resolution data using the GBT and Murriyang. We monitored 42 FRB sources, with an average observation time of 25 hours per source. This included 26 FRBs detected in the Fly's Eye Survey and 15 sources detected in the ICS survey (up to March 2021). We also included the prolific repeater FRB 20201124A in our monitoring campaign to study its polarization properties. The bulk of our follow-up time, totaling 568 hours, was spent with the Parkes/UWL. We followed a standard search methodology, as described in \citet{2021MNRAS.500.2525K}.

Our efforts yielded several notable findings. We detected faint repetitions from one of the brightest FRBs found in the fly's eye survey, FRB\,20171019A, using the GBT \citep{2019ApJ...887L..30K}. Additionally, we discovered an extremely narrow-band repeat burst from FRB 20190711A, with a spectral occupancy of only $2\%$ \citep{2021MNRAS.500.2525K}. Furthermore, using the ASKAP/ICS and the Parkes/UWL, we observed multiple bursts from FRB\,20201124A during a period of heightened activity, which included significant circularly polarized emission in one of the bursts, a phenomenon not previously observed in repeating FRBs \citep{2022MNRAS.512.3400K}.

Among the 41 ASKAP-detected sources we monitored, only two showed definitive evidence of repetition. The remaining sources did not display any clear signs of repeat activity during our follow-up observations, suggesting either longer inactivity periods or that their repeat bursts are too faint to be detected with current generation of radio telescopes.

We have also used the lack of repetition observed in ASKAP FRB searches and follow up to show that the number density of very strong repeaters must be less than 27\,Gpc$^{-3}$ with 95\% confidence \citep{2019MNRAS.486.5934J}, and in follow-up observations to limit their repetition rates \citep{James2020a}. In particular, we have shown that if the likely nearest FRB detected in CRAFT observations \citep[FRB\,20171020, detected in Fly's Eye mode; ][]{Mahony2018} does repeat, its repetition rate must be less than 0.011 bursts per day above $10^{39}$\,erg \citep{LeeWaddell2023}. We have also used these measurements, under the assumption that all FRBs do repeat, to derive an FRB rate distribution $\mathrm{d}N_{\rm frb}/\mathrm{d}R \propto R^{-\gamma_r}$ with $\gamma_r < -1.94$ \citep{James2020b}. This has been shown by \citet{James2023} to be consistent with observations of both repeating and apparently one-off FRBs by the CHIME/FRB Collaboration \citep{CHIME_catalog1_2021}.

\subsection{Population modelling}

The distribution of FRB DMs, redshifts, and luminosities is a function of three factors: the properties of the FRB population (e.g., the intrinsic FRB luminosity function, source evolution, and DM contribution of host galaxies); cosmological parameters such as the Hubble Constant $H_0$ and the baryonic content $\Omega_b$; and properties of the detecting instrument, in particular its total sensitivity, beamshape, and DM-dependent biases \citep{Macquart2018a,Macquart2018b,Connor2019}. Due to ICS FRBs having accurate measurements of all three properties, they have been used in a number of ways to constrain both cosmological and FRB population parameters.

A statistical relationship between FRB DM and luminosity was first established by \citet{2018Natur.562..386S} using FRBs detected by \emph{Murriyang} and ASKAP in fly's eye mode, and further modelled by \citet{Arcus2020}. The {\sc zDM} code has subsequently been developed by the CRAFT and F4 Collaborations to model the redshift, DM, and fluence distribution of FRBs \citep{2022MNRAS.509.4775J}, as well as cosmological parameters (see Section~\ref{sec:IGM}). Using primarily ICS FRBs, it has found evidence for source evolution consistent with the star formation rate, a burst fluence ($\sim$luminosity) distribution consistent with a power-law with differential slope $\mathrm dN_{\rm FRB}/\mathrm dF \propto F^\gamma$ with $\gamma = -0.95_{-0.15}^{+0.18}$ \citep{2022MNRAS.516.4862J}, and a characteristic maximum FRB energy of $10^{41.7 \pm 0.2}$\,erg, assuming a 1\,GHz emission bandwidth \citep{2023Sci...382..294R}.

These studies have also prompted investigations into possible biases in {\sc FREDDA}, with \citet{2023MNRAS.523.5109Q} and \citet{Hoffmann2024} identifying deviations  from the DM-dependent sensitivity predicted by \citet{Cordes_McLaughlin_2003}. When excluding FRB\,20191128A from parameter estimation analysis (due to the aforementioned version 2 of FREDDA which detected it reporting incorrect values) however, these deviations result in a very small systematic error on parameter estimates, e.g.\ of 0.2\,km\,s$^{-1}$\,Mpc$^{-1}$ for $H_0$, which is small  compared to the statistical errors in the current FRB sample.

\subsection{Intergalactic and circumgalactic media} 
\label{sec:IGM}
 
 As a large sample of FRBs began to be amassed, it became clear that the burst dispersion measure is correlated with redshift. We first identified this trend by comparing high-fluence FRBs detected in fly's eye mode with lower-fluence FRBs detected by \emph{Murriyang} \cite[][]{2018Natur.562..386S}.
 This provided, for the first time, the opportunity to measure the baryon density $\Omega_b$ in the nearby low-redshift ($z<0.5$) Universe. 
 While $\Omega_b$ is well-measured at high redshift from Big Bang nucleosynthesis, at low redshift nearly 50\% of the baryons were undetected in optical and X-ray searches, but thought to reside in the diffuse IGM. The use of extragalactic radio bursts to detect this gas had been suggested as far back as 1965 \citep{1965PhRvL..14.1007H, 1973Natur.246..415G}, and quantitative predictions based on $\Lambda$CDM cosmology in the context of gamma-ray bursts had been made \citep{Ioka2003,Inoue2004}. However, it was the late Jean-Pierre Macquart who had most strongly espoused using ASKAP to detect FRBs and find these missing baryons as early as 2008, leading to CRAFT being formalised as an ASKAP survey science project in 2009 \citep{2010PASA...27..272M}.
 
 With a sample of just seven FRBs, we were able to measure the entirety of the baryon content of the Universe  \cite[][]{2020Natur.581..391M} and resolve the {\em missing baryon problem}. The relation between FRB dispersion measure and redshift is now referred to as the `Macquart Relation' in honour of Jean-Pierre Macquart's key contribution. The technique was extended, first to estimate the Hubble constant  \citep[$73^{+12}_{-8}$\,km\,s$^{-1}$\,Mpc$^{-1}$;][]{2022MNRAS.516.4862J}, and then to assess fluctuations in the Macquart relation about its mean, which is related to feedback processes \citep[][]{2023arXiv230507022B}.
   
 Our  discoveries have also enabled us to uniquely probe circumgalactic media, the similarly diffuse media in galaxy haloes that is difficult to study, but which plays an important role in galaxy formation and evolution.
   The second FRB we localised (FRB\,20181112A) was observed to pass through the halo of an intervening galaxy \citep[][]{2019Sci...366..231P}.  Using the resultant limits on pulse broadening and Faraday rotation, we constrained the turbulence and magnetisation in this galaxy halo. \citet[][]{2020ApJ...901..134S} investigated in detail the foreground galaxies towards FRB\,20190608B, and  \citet[][]{2021ApJ...921..134S} undertook a similar analysis towards FRB\,20180924B. We have also been working with the FLIMFLAM collaboration \citep{2022ApJ...928....9L} to detect galaxy halos along the sightlines to FRBs, with first results on ASKAP FRBs presented in \citet{2024arXiv240200505K} and \citet{2024arXiv240812864H}.

\begin{figure}
 \centering
\includegraphics[width=\columnwidth]{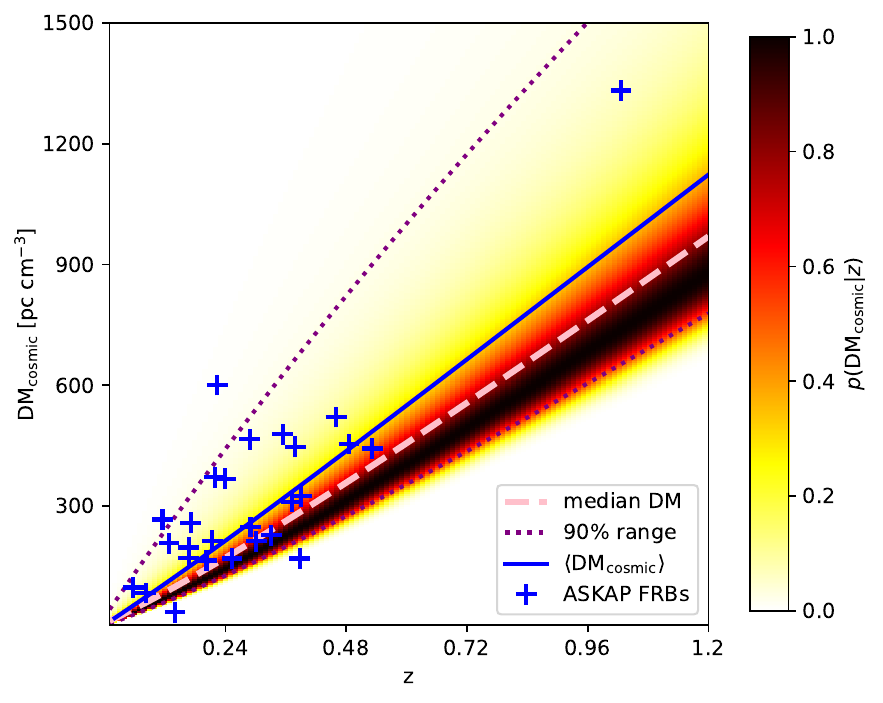}
\caption{The Macquart relation (solid blue line) compared to the $z$--DM distribution of CRAFT/ICS FRBs (blue crosses).  The shading shows the probability density of the DM for the cosmic web ${\rm DM}_{\rm cosmic}$, for which the median DM, and the range encompassing 90\% of the probability, is also given. 
}
\label{fig:frb_dmz} 
\end{figure}

Figure \ref{fig:frb_dmz} shows the Macquart relation for the FRBs detected in the ICS survey.
The dispersion measure of the FRBs has been corrected for the Milky Way contribution using a model for the Milky Way disk \cite[][]{2002astro.ph..7156C}, a Milky Way halo contribution assumed to be 50 \pccm, and a host galaxy contribution of 50\,\pccm. This is compared to expectations (median, and 90\% range, and mean) for $p({\rm DM}_{\rm cosmic}|z)$ using the model of \citet{2020Natur.581..391M} with $H_0 = 70$\,km\,s$^{-1}$\,Mpc$^{-1}$ and fluctuation parameter $F=0.32$. Our FRBs fluctuate significantly about the expectation value $\left< {\rm DM_{\rm cosmic}} \right>$ given by the relation. Two FRBs lie significantly below the minimum expectation, which we take as evidence of very small host contributions, under-fluctuations in Milky Way contributions, or a combination of the two. 
In the future it might be possible to use host galaxy or host environment properties to ascertain if DM excesses are local to the burst and tune FRB studies for studies of extragalactic matter.
Several FRBs lie significantly above the relation, notably FRB\,20210117A, the excess DM of which is attributable to the  local environment of the FRB \citep{2023ApJ...948...67B,Simha2023}. For a given observed DM, measured redshifts vary by at least a factor of two, which cautions against relying on the redshift inferred by the Macquart relation for modelling purposes. We note that all well-localised ICS FRBs with ${\rm DM}_{\rm cosmic}< 1000$\,\pccm\ have firm host galaxy identifications, while at larger DMs, host galaxy identification becomes redshift-dependent \citep[e.g.\ FRB\,20210912A;][]{2023MNRAS.525..994M}. This introduces potential biases when using high-DM data which must be accounted for in population models \citep{2022MNRAS.516.4862J,2023MNRAS.523.5006J}.


\section{FUTURE ASKAP FRB SURVEYS}
\label{sec:future}

 The incoherent searches were a computationally cheap extension to existing ASKAP data-recording systems intended for radio-interferometric imaging on $>10$\,s time scales. The searches could be conducted on a single GPU, but at the cost of reduced sensitivity relative to array coherent searches by factor $\approx \sqrt{N_{\rm ant}}$.
 This reduction in sensitivity provides the opportunity to undertake searches with higher yields with ASKAP by commissioning a new FRB detector.
    Assuming FRBs are a non-evolving population in a Euclidean Universe, we would expect the detection rate to increase by a factor of $N_{\rm ant}^{3/4} \approx 10$.  
   Coherent surveys are also likely to be able to access a population of FRBs at higher redshift. 
    A coherent FRB search system  \cite[the CRAFT Coherent backend, CRACO][]{2024arXiv240910316W} is in the final stages of being scientifically commissioned. The  system has started detecting FRBs in offline and real-time searches.
    Additionally,  coherent image plane searches are potentially more sensitive to long duration transients, which decorrelate interference and antenna-based gain variations. 
    New classes of long-duration transients of both Galactic \cite[][]{2022Natur.601..526H,2024arXiv240612352D} and extra-Galactic origin \cite[][]{2021arXiv210708463T} have been identified.
    However, the coherent search system will have a smaller field of view so will not be sensitive to FRBs in the outskirts of the primary beam or the side lobes.  This will affect FRB detection rates in the outer beams. 
    Even when it is operational we expect to continue with the incoherent sum searches to cross-validate both detectors.  
The CRACO detector will share the current voltage download system with the ICS detector, delivering high precision astrometry and high time resolution spectro-polarimetry.

   These new searches will complement those of other FRB localisation facilities currently operational or planned for the near future.
   FRB searches with MeerKAT  \cite[][]{Jankowski2023} and in the future with the SKA probe a narrower field of view at higher sensitivity while sharing access to the southern sky.
   In the northern hemisphere, outrigger stations on continental-length baselines are being commissioned for the CHIME \cite[][]{2024arXiv240207898L}.  This will provide the opportunity to localise FRBs to as good as tens of milliarcsecond precision.
   The DSA-110 interferometer \cite[][]{2019MNRAS.489..919K} located at Owens Valley Radio Observatory is the most comparable facility to CRACO in terms of sensitivity. ASKAP will deliver positions with precision a factor of two greater than DSA-110, increasing the reliability of host-galaxy association and enabling more detailed investigations of host galaxy environments.

\section{CONCLUSION}

The ASKAP incoherent sum survey has demonstrated the importance and significance of localising a population of fast radio bursts. 

The survey showed that there is a significant population of FRBs at $z \gg 0.1$, largely in star forming galaxies, with many originating coincident with spiral arms.
The discoveries will continue  to be used to study the structure of the intergalactic medium and the cosmology of the Universe.
The survey motivates further surveys for larger populations of FRBs, and FRBs at higher redshift.  We have recently started one such survey with ASKAP using new instrumentation, which will increase the burst detection rate and extend our reach in the Universe.

\begin{acknowledgement}
We are grateful to the ASKAP engineering and operations team for their assistance in developing fast radio burst instrumentation for the telescope and supporting the survey. 
This  work uses data obtained from Inyarrimanha Ilgari Bundara / the CSIRO Murchison Radio-astronomy Observatory. We acknowledge the Wajarri Yamaji People as the Traditional Owners and native title holders of the Observatory site. CSIRO’s ASKAP radio telescope is part of the Australia Telescope National Facility (https://ror.org/05qajvd42). Operation of ASKAP is funded by the Australian Government with support from the National Collaborative Research Infrastructure Strategy. ASKAP uses the resources of the Pawsey Supercomputing Research Centre. Establishment of ASKAP, Inyarrimanha Ilgari Bundara, the CSIRO Murchison Radio-astronomy Observatory and the Pawsey Supercomputing Research Centre are initiatives of the Australian Government, with support from the Government of Western Australia and the Science and Industry Endowment Fund.
Murriyang, the Parkes radio telescope, is part of the Australia Telescope National Facility (https://ror.org/05qajvd42) which is funded by the Australian Government for operation as a National Facility managed by CSIRO.
We acknowledge the Wiradjuri people as the Traditional Owners of the Parkes Observatory site.
The Australia Telescope Compact Array is part of the Australia Telescope National Facility (https://ror.org/05qajvd42) which is funded by the Australian Government for operation as a National Facility managed by CSIRO.
We acknowledge the Gomeroi people as the Traditional Owners of the Paul Wild Observatory site.
The Green Bank Observatory is a facility of the National Science Foundation operated under cooperative agreement by Associated Universities, Inc. 

Part of this work was performed on the OzSTAR national facility at Swinburne University of Technology. The OzSTAR program receives funding in part from the Astronomy National Collaborative Research Infrastructure Strategy (NCRIS) allocation provided by the Australian Government, and from the Victorian Higher Education State Investment Fund (VHESIF) provided by the Victorian Government.

Based on observations made with ESO Telescopes at the La Silla Paranal Observatory under programme IDs 0103.A-0101, 105.204W, and 108.21ZF.

Some of the data presented herein were obtained at Keck Observatory, which is operated as a scientific partnership among the California Institute of Technology, the University of California, and the National Aeronautics and Space Administration. The Observatory was made possible by the generous financial support of the W. M. Keck Foundation.
The authors wish to recognize and acknowledge the very significant cultural role and reverence that the summit of Maunakea has always had within the Native Hawaiian community. We are most fortunate to have the opportunity to conduct observations from this mountain. We thank Jeff Cooke for assistance with Keck observations.

Some of the observations reported here were obtained at the MMT Observatory, a joint facility of the University of Arizona and the Smithsonian Institution.

W. M. Keck Observatory and MMT Observatory access was supported by Northwestern University and the Center for Interdisciplinary Exploration and Research in Astrophysics (CIERA).

RMS, PAU, and YW acknowledge support through Australian Research Council Future Fellowship FT\,190100155.  RMS, ATD, and AJ acknowledge support through Australian Research Council Discovery Project DP\,220102305. MG and CWJ acknowledge support from the Australian Government through the Australian Research Council Discovery Project DP210102103. SB is supported by a Dutch Research Council (NWO) Veni Fellowship (VI.Veni.212.058). KG acknowledges support through Australian Research Council Discovery Project DP200102243.
JXP,  AM, ACG, WF, and NT\
as members of the Fast and Fortunate for FRB
Follow-up (F4) team, acknowledge support from 
NSF grants AST-1911140, AST-1910471
and AST-2206490. ACG, WF, and the Fong Group at Northwestern acknowledges support by the National Science Foundation under grant Nos. AST-1909358, AST-2308182 and CAREER grant No. AST-2047919.
WF gratefully acknowledges support by the David and Lucile Packard Foundation, the Alfred P. Sloan Foundation, and the Research Corporation for Science Advancement through Cottrell Scholar Award \#28284.  Parts of this work were undertaken with support from Australian Research Council Laureate Fellowship FL150100148 and Centre of Excellence CE170100004.
\end{acknowledgement}

\begin{appendix}
\section{FRB properties}

Table \ref{tab:frb_discoveries} lists the key properties of the FRBs discovered in the ASKAP-CRAFT incoherent sum survey. 
Table \ref{tab:frb_pos} presents FRB astrometry, derived interferometrically where possible, and using the multi-beam localisation method \cite[][]{2017ApJ...841L..12B}.  For most of the interferometrically localised FRBs we have reported updated positions derived using the CELEBI pipeline \cite[][]{2023A&C....4400724S}.  For the the earliest-discovered FRBs (2018094B, 20181112A, 20190102C, 20190608B, 20190611B, and 20190714A), we use positions reported in \cite{2021PASA...38...50D}. These FRBs had dedicated astrometric campaigns with the Australian Telescope Compact Array.

\begin{table*}
\caption{Key properties of ASKAP/ICS FRBs.  Redshifts ($z$) reported n/a can not be measured as only arcminute-precision localisations are available.  Redshifts reported n/h are not measured as no host galaxy has been identified. Redshifts reported p (pending) are are FRBs for which photometric and/or spectroscopic observations have not been executed.  Milky-Way dispersion measures (DM$_{\rm MW}$) are Galactic disk contributions estimated using the NE2001 model \cite[][]{2002astro.ph..7156C}.  Due to disagreements in metadata the arrival time of FRB~20240310A is only known to $\approx 15$\,s. 
 For the remaining FRBs the arrival time uncertainties are dominated by systematic error induced by burst morphology\label{tab:frb_discoveries}}
\begin{tabular}{ccrrrrlrrr}
\hline
FRB  & UTC & \multicolumn{1}{c}{$\nu_c$} & \multicolumn{1}{c}{$N_{\rm ant}$} & \multicolumn{1}{c}{DM} & \multicolumn{1}{c}{DM$_{\rm MW}$} & \multicolumn{1}{c}{$z$} & \multicolumn{1}{c}{S/N} & \multicolumn{1}{c}{$w$} & \multicolumn{1}{c}{$E_\nu$}  \\
 (TNS) &  & \multicolumn{1}{c}{(MHz)} &  & \multicolumn{1}{c}{(\pccm)} & \multicolumn{1}{c}{(\pccm)} &  & & \multicolumn{1}{c}{(ms)} & \multicolumn{1}{c}{(Jy\,ms)}  \\ 
\hline 

20180924B & 2018-09-24 16:23:12.561 & 1297.5 & 24 & 362.4(2) & 41 & 0.3214 & 21.1 & 2.6 & 18.4(9) \\ 
20181112A & 2018-11-12 17:31:16.099 & 1297.5 & 12 & 589.0(3) & 40 & 0.4755 & 19.3 & 3.5 & 28(2) \\ 
20190102C & 2019-01-02 05:38:44.002 & 1271.5 & 23 & 364.5(3) & 57 & 0.2912 & 14.0 & 2.6 & 16.0(9) \\ 
20190608B & 2019-06-08 22:48:13.367 & 1271.5 & 25 & 339.5(5) & 37 & 0.1178 & 16.1 & 8.6 & 28(2) \\ 
20190611B & 2019-06-11 05:45:43.417 & 1271.5 & 25 & 322.2(2) & 57 & 0.3778 & 9.3 & 3.5 & 10(1) \\ 
20190711A & 2019-07-11 01:53:41.689 & 1271.5 & 29 & 594.6(5) & 57 & 0.522 & 23.8 & 10.4 & 36(2) \\ 
20190714A & 2019-07-14 05:37:13.606 & 1271.5 & 28 & 504.7(3) & 39 & 0.2365 & 10.7 & 3.5 & 13(1) \\ 
20191001A & 2019-10-01 16:55:37.237 & 920.5 & 30 & 506.92(4) & 44 & 0.234 & 62.0 & 10.4 & 120(2) \\ 
20191228A & 2019-12-28 09:16:18.091 & 1271.5 & 28 & 297.5(5) & 33 & 0.2432 & 22.9 & 17.3 & 67(3) \\ 
20200430A & 2020-04-30 15:49:50.041 & 863.5 & 26 & 380.1(2) & 27 & 0.1608 & 16.0 & 13.8 & 35(2) \\ 
20200627A & 2020-06-27 19:23:42.754 & 920.5 & 31 & 294(1) & 40 & n/a & 10.8 & 31.1 & 28(3) \\ 
20200906A & 2020-09-06 21:40:53.600 & 863.5 & 7 & 577.8(2) & 36 & 0.3688 & 16.1 & 5.2 & 53(3) \\ 
20210117A & 2021-01-17 07:51:22.297 & 1271.5 & 25 & 730(1) & 34 & 0.2145 & 27.1 & 5.9 & 36(1) \\ 
20210214G & 2021-02-14 05:12:39.696 & 1271.5 & 26 & 398.3(7) & 32 & n/a & 11.6 & 4.7 & 13(3) \\ 
20210320C & 2021-03-20 18:38:08.508 & 863.5 & 24 & 384.8(3) & 42 & 0.2797 & 15.3 & 6.9 & 59(4) \\ 
20210407E & 2021-04-07 11:20:56.806 & 1271.5 & 24 & 1785.3(3) & 154 & n/h & 19.1 & 9.5 & 36(2) \\ 
20210807D & 2021-08-07 15:48:10.256 & 920.5 & 23 & 251.9(2) & 121 & 0.1293 & 47.1 & 17.7 & 100(3) \\ 
20210809C & 2021-08-09 10:03:02.954 & 920.5 & 23 & 651.5(3) & 190 & n/a & 16.8 & 23.6 & 45(3) \\ 
20210912A & 2021-09-12 13:30:05.680 & 1271.5 & 23 & 1234.5(2) & 31 & n/h & 31.7 & 7.1 & 70(2) \\ 
20211127I & 2021-11-27 00:03:51.573 & 1271.5 & 24 & 234.83(8) & 43 & 0.0469 & 37.9 & 3.5 & 35(1) \\ 
20211203C & 2021-12-03 02:21:35.468 & 920.5 & 24 & 636.2(4) & 63 & 0.3439 & 14.2 & 16.5 & 30(2) \\ 
20211212A & 2021-12-12 19:32:07.768 & 1631.5 & 24 & 206(5) & 27 & 0.0707 & 12.8 & 5.9 & 131(7) \\ 
20220105A & 2022-01-05 00:19:18.668 & 1631.5 & 22 & 583(2) & 22 & 0.2785 & 9.8 & 5.9 & 19(2) \\ 
20220501C & 2022-05-01 02:11:10.943 & 864.5 & 23 & 449.5(2) & 31 & 0.381 & 16.1 & 9.5 & 32(2) \\ 
20220531A & 2022-05-31 16:34:14.274 & 1271.5 & 23 & 727(2) & 70 & n/a & 9.7 & 10.6 & $30^{+800}_{-17}$ \\ 
20220610A & 2022-06-10 22:26:44.313 & 1271.5 & 22 & 1458.1(2) & 31 & 1.015 & 29.8 & 8.3 & 47(2) \\ 
20220725A & 2022-07-25 21:54:53.609 & 920.5 & 25 & 290.4(3) & 31 & 0.1926 & 12.7 & 8.3 & 72(6) \\ 
20220918A & 2022-09-18 17:33:33.933 & 1271.5 & 25 & 656.8(4) & 41 & 0.491 & 26.4 & 9.5 & 55(2) \\ 
20221106A & 2022-11-06 21:27:34.504 & 1631.5 & 21 & 343.8(8) & 35 & 0.2044 & 35.1 & 8.3 & 80(2) \\ 
20230521A & 2023-05-21 02:38:08.482 & 831.5 & 23 & 640.2(5) & 42 & n/a & 15.2 & 16.5 & 34(1) \\ 
20230526A & 2023-05-26 23:29:47.094 & 1271.5 & 22 & 361.4(2) & 50 & 0.1570 & 22.1 & 4.7 & 34(1) \\ 
20230708A & 2023-07-08 15:32:46.979 & 920.5 & 23 & 411.51(5) & 50 & 0.105 & 31.5 & 23.6 & 111(4) \\ 
20230718A & 2023-07-18 07:02:08.041 & 1271.5 & 22 & 477.0(5) & 396 & 0.035 & 10.9 & 3.5 & 14(1) \\ 
20230731A & 2023-07-31 05:28:41.587 & 1271.5 & 25 & 701.1(3) & 547 & p & 16.6 & 3.5 & 25(1) \\ 
20230902A & 2023-09-02 00:48:51.836 & 832.5 & 22 & 440.1(1) & 34 & 0.3619 & 11.8 & 5.9 & 23(2) \\ 
20231006A & 2023-10-06 08:14:45.849 & 863.5 & 24 & 509.7(2) & 68 & n/a & 15.2 & 8.3 & 25(1) \\ 
20231226A & 2023-12-26 18:46:19.997 & 863.5 & 22 & 329.9(1) & 145 & 0.1569 & 36.7 & 11.8 & 78(3) \\ 
20240201A & 2024-02-08 20:00:54.246 & 920.5 & 24 & 374.5(2) & 38 & 0.042729 & 13.9 & 9.5 & 47(3) \\ 
20240208A &  2024-02-08 20:00:54.246& 863.5 & 14 & 260.2(3) & 98 & p & 12.1 & 7.1 & 37(3) \\ 
20240210A & 2024-02-10 08:20:02.510 & 863.5 & 23 & 283.73(5) & 31 & 0.023686 & 11.6 & 9.5 & 26(2) \\ 
20240304A & 2024-03-04 17:44:55.155 & 863.5 & 24 & 652.6(5) & 30 & p & 12.3 & 11.8 & 34(2) \\ 
20240310A & 2024-03-10 07:38:50 & 920.5 & 25 & 601.8(2) & 36 & 0.1270 & 19.1 & 7.1 & 35(2) \\ 
20240318A & 2024-03-18 15:14:19.454 & 920.5 & 23 & 256.4(3) & 37 & p & 13.2 & 4.7 & 15(1) \\ 
\hline
\end{tabular}

\end{table*}

\begin{table*}
\caption{Astrometry of the CRAFT/ICS FRBs.  We list the FRB name, Dispersion Measure (DM), and redshift $z$ where one has been obtained. 
  We also list positions in Right Ascension ($alpha$) and Declination $\delta$ derived interferometrically (I) and using the multi-beam method (M). We list the uncertainties in both right ascension and declination for both methods: $\sigma_\alpha$, $\sigma_\delta$,  $sigma_{\alpha, M}$, and $\sigma_{\delta,M}$, respectively.  For interferometrically measured positions, we also include the major and minor axis of the error ellipse ($\sigma_{\rm maj}$ and $\sigma_{\rm min}$) as well as the position angle of the ellipse $\Psi$, measured East of North.  \label{tab:frb_pos}}
\begin{tabular}{crrrrrrrrrrrrr}
\multicolumn{1}{c}{FRB}  &  \multicolumn{1}{c}{DM}  & \multicolumn{1}{c}{$z$} &  \multicolumn{1}{c}{$\alpha$ (I)} &  \multicolumn{1}{c}{$\delta$ (I)} &  \multicolumn{1}{c}{$\sigma_{\alpha}$}  &   \multicolumn{1}{c}{$\sigma_{\delta}$} &  \multicolumn{1}{c}{$\sigma_{\rm maj}$} &  \multicolumn{1}{c}{$\sigma_{\rm min}$} & \multicolumn{1}{c}{ $\Psi$} &  \multicolumn{1}{c}{$\alpha$ (M)} &   \multicolumn{1}{c}{$\delta$ (M)} &  \multicolumn{1}{c}{$\sigma_{\alpha}^M$} &  \multicolumn{1}{c}{$\sigma_{\delta}^M$} \\ 
 \multicolumn{1}{c}{(TNS)} &  \multicolumn{1}{c}{(pc\,cm$^{-3}$)}  & & \multicolumn{2}{c}{(J2000)} &  \multicolumn{1}{c}{(")} &  \multicolumn{1}{c}{(")} &  \multicolumn{1}{c}{(")} &  \multicolumn{1}{c}{(")} &  \multicolumn{1}{c}{($^\circ$)} & \multicolumn{2}{c}{(J2000)} &  \multicolumn{1}{c}{(')} &  \multicolumn{1}{c}{(')} \\
\hline
20180924B & 362.4(2) & 0.3214 & 21:44:25.26 & $-$40:54:00.1 & 0.16 & 0.16 & $$-$$ & $$-$$ & $$-$$ & 21:44:25.5 & $-$40:54:23 & 3 & 3 \\ 
20181112A & 589.0(3) & 0.4755 & 21:49:23.63 & $-$52:58:15.4 & 3.8 & 2.4 & $$-$$ & $$-$$ & $$-$$ & 21:49:06.4 & $-$53:17:44 & 9 & 7 \\ 
20190102C & 364.5(3) & 0.2912 & 21:29:39.76 & $-$79:28:32.5 & 0.79 & 0.9 & $$-$$ & $$-$$ & $$-$$ & 21:30:43.3 & $-$79:29:47 & 3 & 3 \\ 
20190608B & 339.5(5) & 0.1178 & 22:16:04.77 & $-$07:53:53.7 & 0.33 & 0.3 & $$-$$ & $$-$$ & $$-$$ & 22:16:17.0 & $-$07:53:47 & 3 & 2 \\ 
20190611B & 322.2(2) & 0.3778 & 21:22:58.94 & $-$79:23:51.3 & 1.1 & 1.1 & $$-$$ & $$-$$ & $$-$$ & 21:23:46.5 & $-$79:21:28 & 3 & 3 \\ 
20190711A & 594.6(5) & 0.522 & 21:57:40.13 & $-$80:21:28.9 & 2.28 & 1.49 & 2.41 & 1.28 & $-$68.0 & 21:56:06.4 & $-$80:23:27 & 2 & 2 \\ 
20190714A & 504.7(3) & 0.2365 & 12:15:55.13 & $-$13:01:15.6 & 0.52 & 0.38 & $$-$$ & $$-$$ & $$-$$ & 12:15:36.5 & $-$13:00:44 & 3 & 2 \\ 
20191001A & 506.92(4) & 0.234 & 21:33:24.41 & $-$54:44:53.9 & 0.46 & 0.43 & 0.46 & 0.43 & 81.9 & 21:32:27.6 & $-$54:43:20 & 7 & 5 \\ 
20191228A & 297.5(5) & 0.2432 & 22:57:43.33 & $-$29:35:38.8 & 0.78 & 0.82 & 0.84 & 0.76 & $-$29.3 & 22:56:53.3 & $-$29:46:10 & 9 & 14 \\ 
20200430A & 380.1(2) & 0.1608 & 15:18:49.55 & +12:22:34.8 & 0.43 & 0.46 & 0.46 & 0.43 & 17.2 & 15:18:41.0 & 12:20:23 & 3 & 3 \\ 
20200627A & 294(1) & n/a & $$-$$ & $$-$$ & $$-$$ & $$-$$ & $$-$$ & $$-$$ & $$-$$ & 21:46:47.0 & $-$39:29:05.0 & 3 & 3 \\ 
20200906A & 577.8(2) & 0.3688 & 03:33:58.93 & $-$14:04:58.8 & 0.49 & 0.49 & 0.51 & $-$0.47 & 39.0 & 03:34:36.9 & $-$14:03:33 & 5 & 6 \\ 
20210117A & 730(1) & 0.2145 & 22:39:55.01 & $-$16:09:05.2 & 0.42 & 0.42 & 0.42 & 0.42 & 73.1 & 22:39:36.0 & $-$16:11:25 & 6 & 11 \\ 
20210214G & 398.3(7) & n/a & $$-$$ & $$-$$ & $$-$$ & $$-$$ & $$-$$ & $$-$$ & $$-$$ & 00:27:43.2 & $-$05:49:56 & 3 & 4 \\ 
20210320C & 384.8(3) & 0.2797 & 13:37:50.10 & $-$16:07:21.6 & 0.46 & 0.46 & 0.47 & 0.45 & 43.7 & 13:37:16.8 & $-$15:24:37 & 9 & 9 \\ 
20210407E & 1785.3(3) & n/h & 05:14:36.23 & +27:03:29.7 & 0.70 & 0.76 & 0.89 & 0.53 & $-$39.6 & 05:14:46.3 & 27:04:12 & 4 & 4 \\ 
20210807D & 251.9(2) & 0.1293 & 19:56:53.07 & $-$00:45:44.1 & 0.6 & 0.6 & $$-$$ & $$-$$ & $$-$$ & 19:56:49.0 & $-$00:48:51 & 2 & 2 \\ 
20210809C & 651.5(3) & n/a & $$-$$ & $$-$$ & $$-$$ & $$-$$ & $$-$$ & $$-$$ & $$-$$ & 18:04:37.7 & 01:19:44 & 3 & 3 \\ 
20210912A & 1234.5(2) & n/h & 23:23:10.44 & $-$30:24:20.1 & 0.44 & 0.43 & 0.44 & 0.43 & 66.2 & 23:24:40.3 & $-$30:29:33 & 9 & 5 \\ 
20211127I & 234.83(8) & 0.0469 & 13:19:14.12 & $-$18:50:16.5 & 0.51 & 0.51 & 0.51 & 0.51 & 65.1 & 13:19:09.5 & $-$18:49:28 & 3 & 2 \\ 
20211203C & 636.2(4) & 0.3439 & 13:38:15.00 & $-$31:22:49.0 & 0.44 & 0.44 & 0.44 & 0.44 & 80.7 & 13:37:52.8 & $-$31:22:04 & 3 & 3 \\ 
20211212A & 206(5) & 0.0707 & 10:29:24.19 & +01:21:37.6 & 0.48 & 0.48 & 0.50 & 0.46 & 45.6 & 10:30:40.7 & 01:40:37 & 5 & 4 \\ 
20220105A & 583(2) & 0.2785 & 13:55:12.81 & +22:27:58.4 & 1.05 & 1.37 & 1.51 & 0.82 & $-$30.5 & 13:54:51.4 & 22:29:20 & 11 & 7 \\ 
20220501C & 449.5(2) & 0.381 & 23:29:31.00 & $-$32:29:26.6 & 0.42 & 0.42 & 0.42 & 0.42 & 66.8 & 23:29:46.8 & $-$32:27:41 & 3 & 3 \\ 
20220531A & 727(2) & n/a & $$-$$ & $$-$$ & $$-$$ & $$-$$ & $$-$$ & $$-$$ & $$-$$ & 19:38:50.2 & $-$60:17:48 & 10 & 20 \\ 
20220610A & 1458.1(2) & 1.016 & 23:24:17.58 & $-$33:30:49.9 & 0.44 & 0.44 & 0.45 & 0.43 & $-$53.6 & 23:24:04.4 & $-$33:30:39 & 3 & 3 \\ 
20220725A & 290.4(3) & 0.1926 & 23:33:15.65 & $-$35:59:24.9 & 0.46 & 0.45 & 0.46 & 0.44 & $-$56.1 & 23:33:32.1 & $-$36:07:51 & 10 & 21 \\ 
20220918A & 656.8(4) & 0.491 & 01:10:22.11 & $-$70:48:41.0 & 0.43 & 0.45 & 0.46 & 0.43 & $-$20.4 & 01:10:57.9 & $-$70:47:06 & 2 & 2 \\ 
20221106A & 343.8(8) & 0.2044 & 03:46:49.15 & $-$25:34:11.3 & 0.58 & 0.55 & 0.60 & 0.53 & $-$58.1 & 03:46:38.1 & $-$25:39:45 & 3 & 3 \\ 
20230521A & 640.2(5) & n/a & $$-$$ & $$-$$ & $$-$$ & $$-$$ & $$-$$ & $$-$$ & $$-$$ & 21:51:00.3 & $-$02:23:10 & 3 & 4 \\ 
20230526A & 361.4(2) & 0.1570 & 01:28:55.83 & $-$52:43:02.4 & 0.43 & 0.42 & 0.43 & 0.42 & $-$65.3 & 01:29:27.5 & $-$52:46:08 & 2 & 2 \\ 
20230708A & 411.51(5) & 0.105 & 20:12:27.73 & $-$55:21:22.6 & 0.46 & 0.44 & 0.47 & 0.43 & $-$63.2 & 20:12:56.9 & $-$55:22:59 & 2 & 2 \\ 
20230718A & 477.0(5) & 0.035 & 08:32:38.86 & $-$40:27:07.0 & 0.60 & 0.61 & 0.61 & 0.60 & $-$20.4 & 08:30:27.1 & $-$41:00:13 & 18 & 18 \\ 
20230731A & 701.1(3) & p & 11:38:24.35 & $-$56:47:56.6 & 0.55 & 0.56 & 0.56 & 0.55 & $-$42.6 & 11:38:40.1 & $-$56:58:19 & 3 & 14 \\ 
20230902A & 440.1(1) & 0.3619 & 03:28:33.55 & $-$47:20:00.6 & 0.68 & 0.57 & 0.69 & 0.55 & $-$72.3 & 03:29:28.1 & $-$47:33:46 & 4 & 6 \\ 
20231006A & 509.7(2) & n/a & $$-$$ & $$-$$ & $$-$$ & $$-$$ & $$-$$ & $$-$$ & $$-$$ & 19:44:00.8 & $-$64:38:56 & 3 & 3 \\ 
20231226A & 329.9(1) & 0.1569 & 10:21:27.30 & +06:06:36.9 & 0.48 & 0.51 & 0.51 & 0.48 & $-$13.2 & 10:21:07.6 & 06:07:46 & 7 & 3 \\ 
20240201A & 374.5(2) & 0.042729 & 09:59:37.34 & +14:05:16.9 & 0.48 & 0.51 & 0.52 & 0.46 & $-$30.0 & 10:01:49.1 & 13:54:49 & 7 & 5 \\ 
20240208A & 260.2(3) & p & 10:36:55.02 & $-$00:57:11.4 & 0.81 & 1.36 & 1.37 & 0.80 & $-$6.5 & 10:36:46.5 & $-$00:33:50 & 4 & 10 \\ 
20240210A & 283.73(5) & 0.023686 & 00:35:07.10 & $-$28:16:14.7 & 0.53 & 0.51 & 0.55 & 0.49 & $-$54.1 & 00:39:55.0 & $-$27:39:35 & 14 & 7 \\ 
20240304A & 652.6(5) & p & 09:05:19.40 & $-$16:09:59.9 & 0.75 & 0.65 & 0.81 & 0.56 & $-$57.0 & 09:05:19.3 & $-$16:13:42 & 7 & 10 \\ 
20240310A & 601.8(2) &0.1270 & 01:10:29.25 & $-$44:26:21.9 & 0.57 & 0.57 & 0.59 & 0.55 & $-$48.6 & 01:10:57.7 & $-$44:24:05 & 3 & 5 \\ 
20240318A & 256.4(3) & p & 10:01:34.36 & +37:36:58.9 & 0.54 & 0.79 & 0.82 & 0.50 & $-$19.3 & 10:01:50.6 & 37:36:49 & 3 & 3 \\ \hline
\end{tabular}
\end{table*}

\section{FRB dynamic spectra}

\begin{figure*}
 \centering
 \begin{tabular}{ccc}
\includegraphics[width=2in]{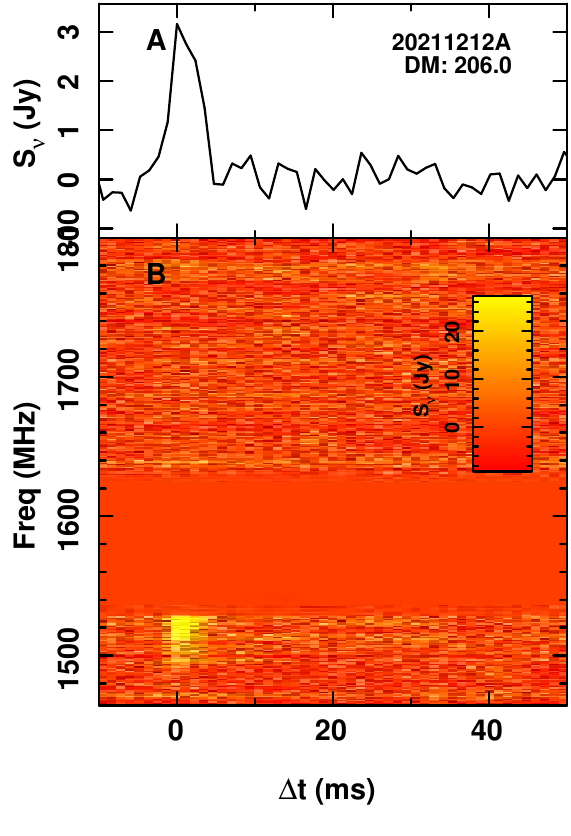} & 
\includegraphics[width=2in]{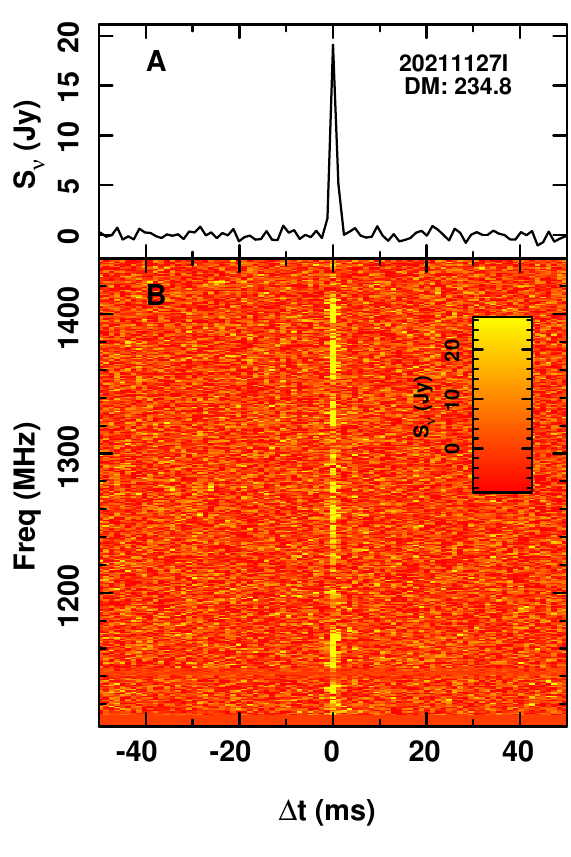} &
\includegraphics[width=2in]{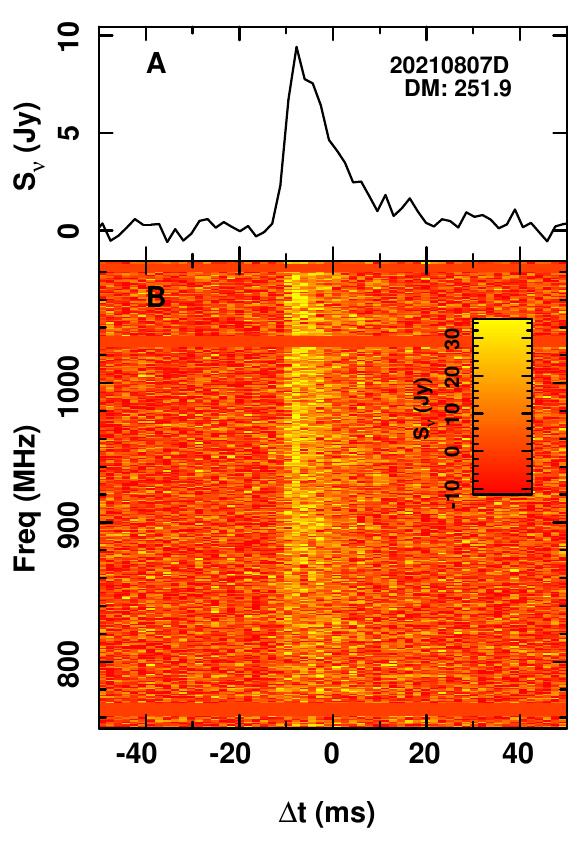} 
\\
\includegraphics[width=2in]{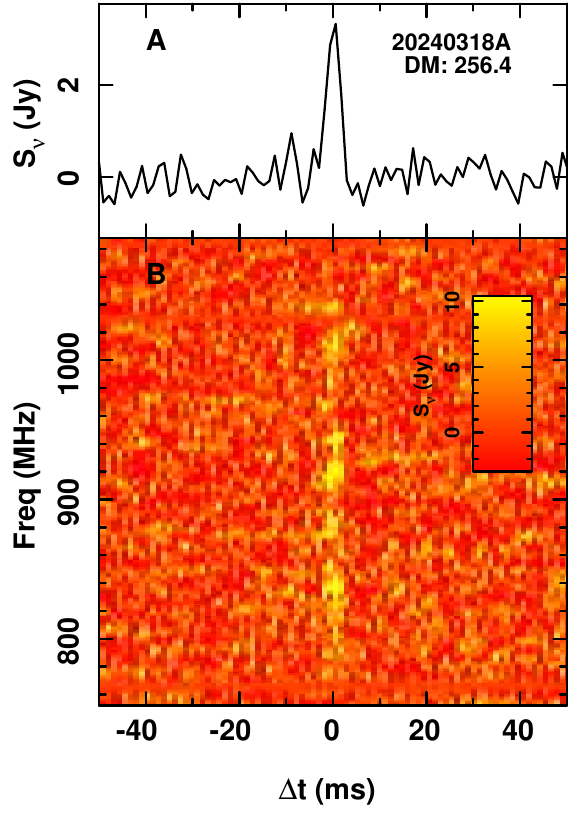} &
\includegraphics[width=2in]{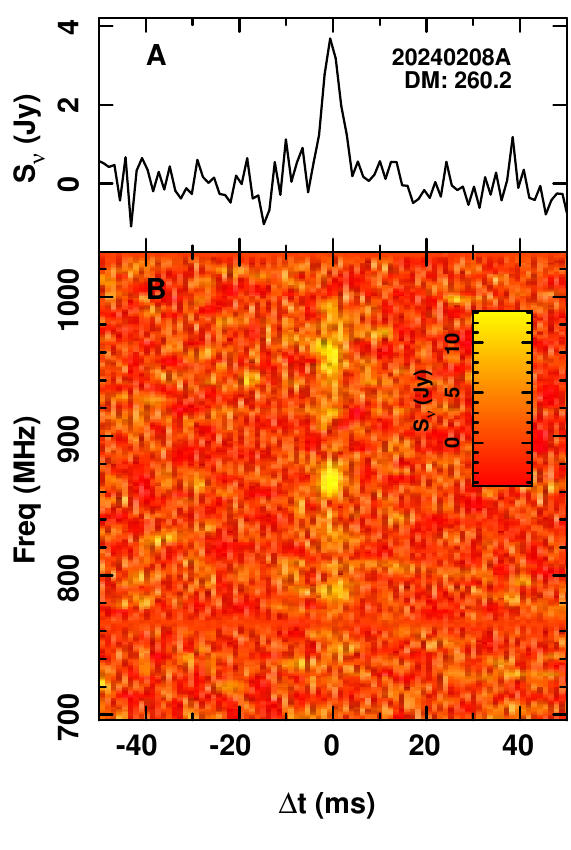} &
\includegraphics[width=2in]{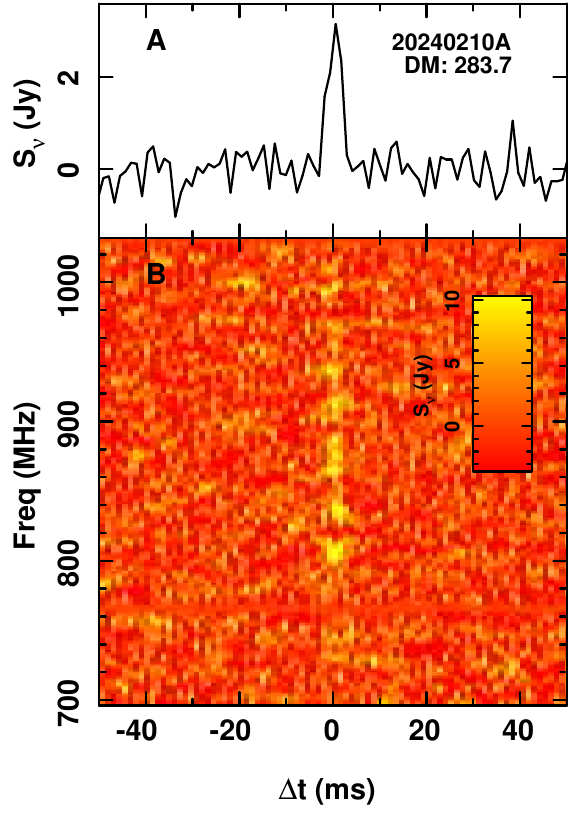} \\

\includegraphics[width=2in]{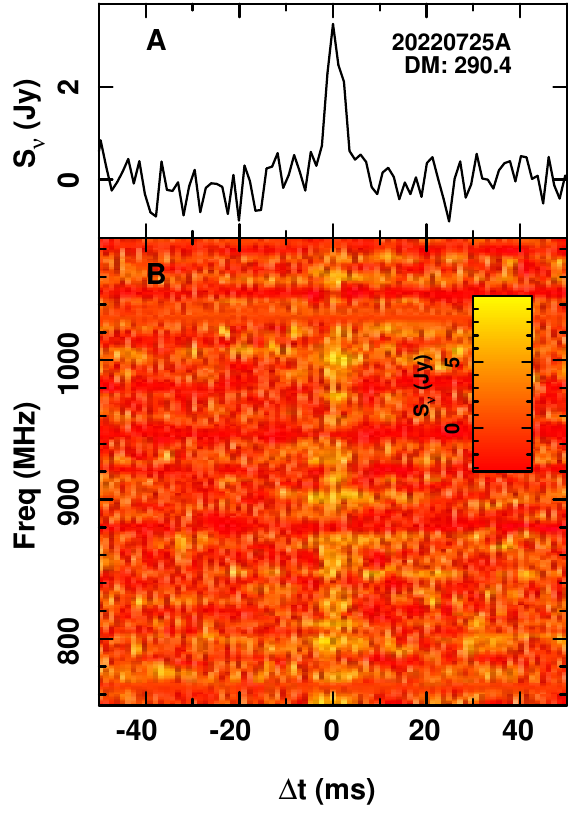} &
\includegraphics[width=2in]{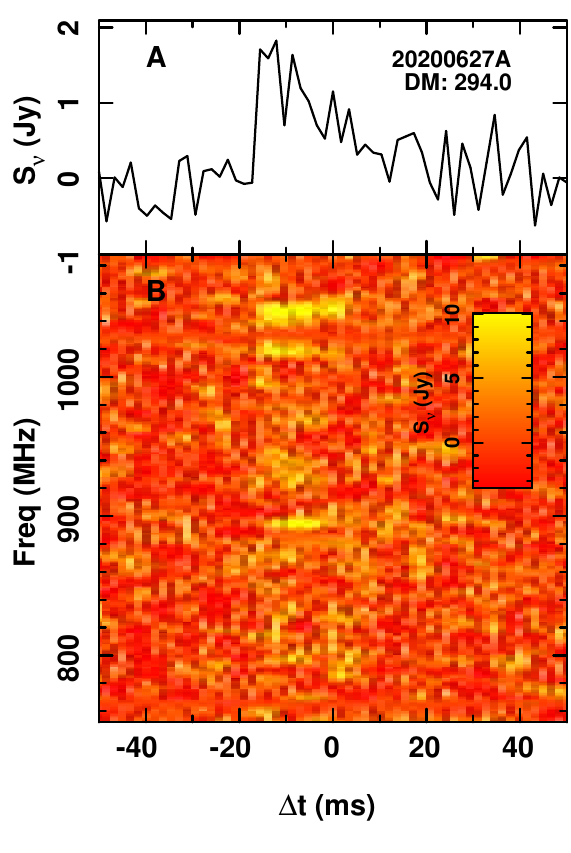} &
\includegraphics[width=2in]{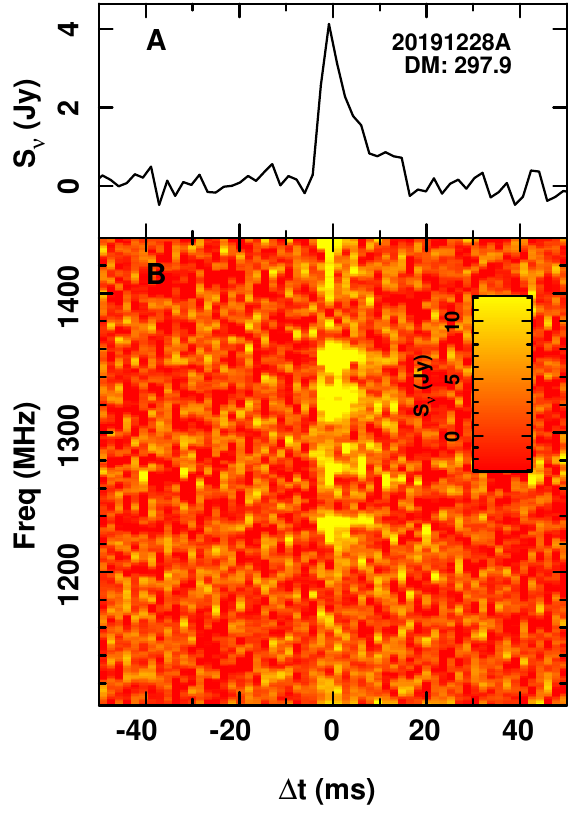} \\

\end{tabular}
\caption{FRB dynamic spectra.  The dedispersed dynamic spectra are produced from the search data stream and ordered by increasing DM. For each FRB the band averaged-pulse profile is displayed in panel A, and the dedispersed dynamic spectrum is shown in panel B.  Horizontal bands of constant intensity indicate channels flagged due to radio-frequency interference.
 \label{fig:dynspec1}}
\end{figure*}

\begin{figure*}
 \centering
 \begin{tabular}{ccc}

 \includegraphics[width=2in]{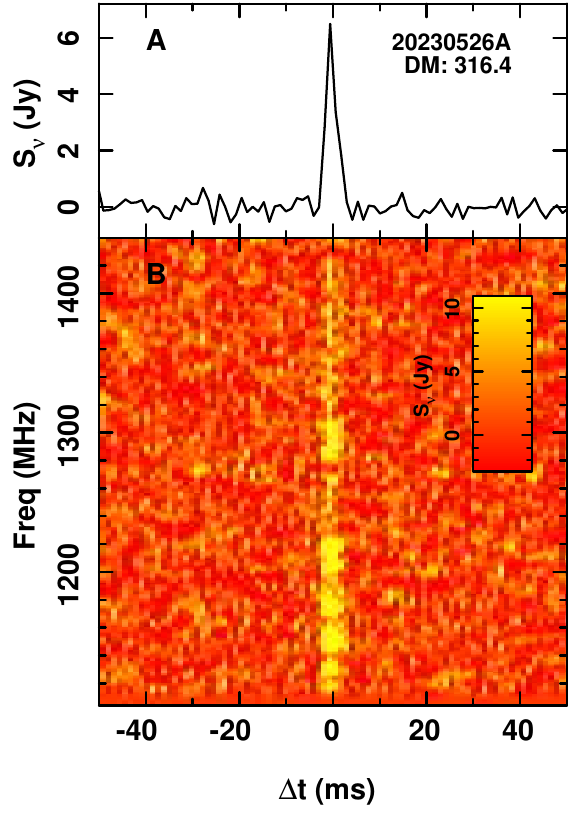}& 
 \includegraphics[width=2in]{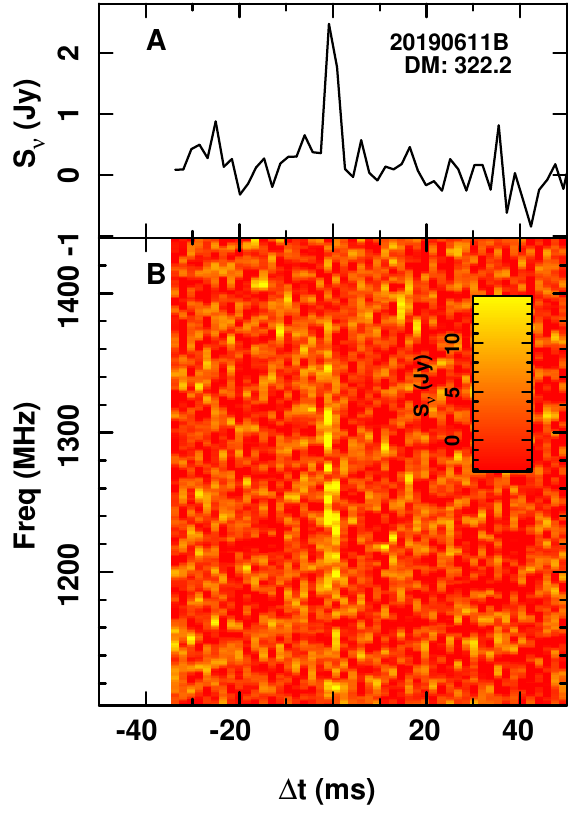} &
\includegraphics[width=2in]{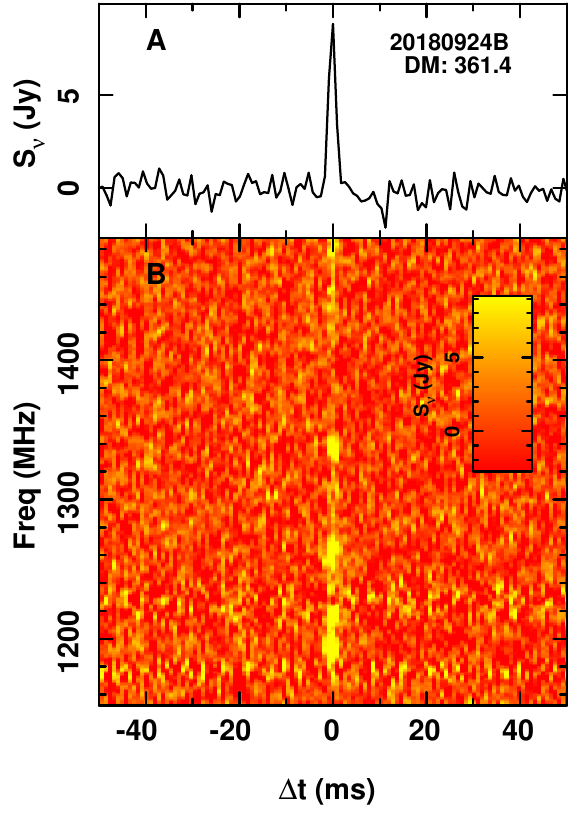} \\

\includegraphics[width=2in]{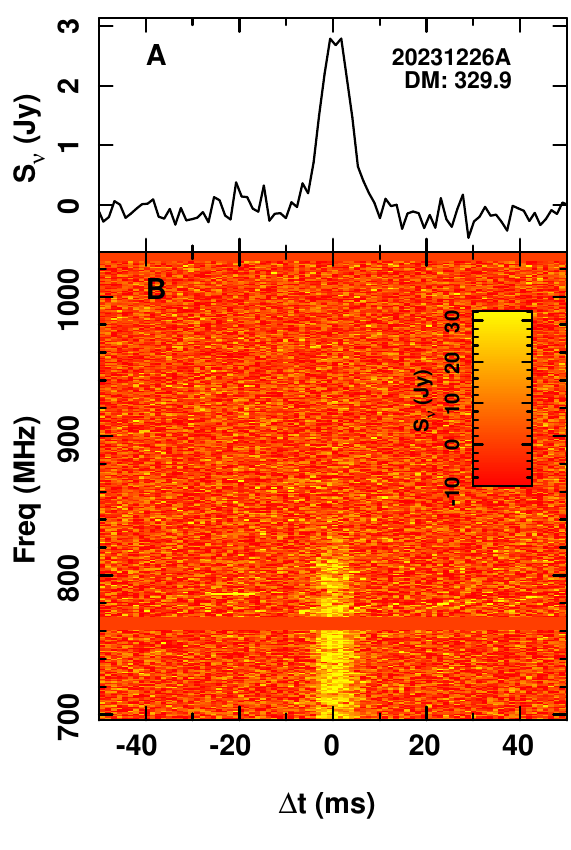} &
\includegraphics[width=2in]{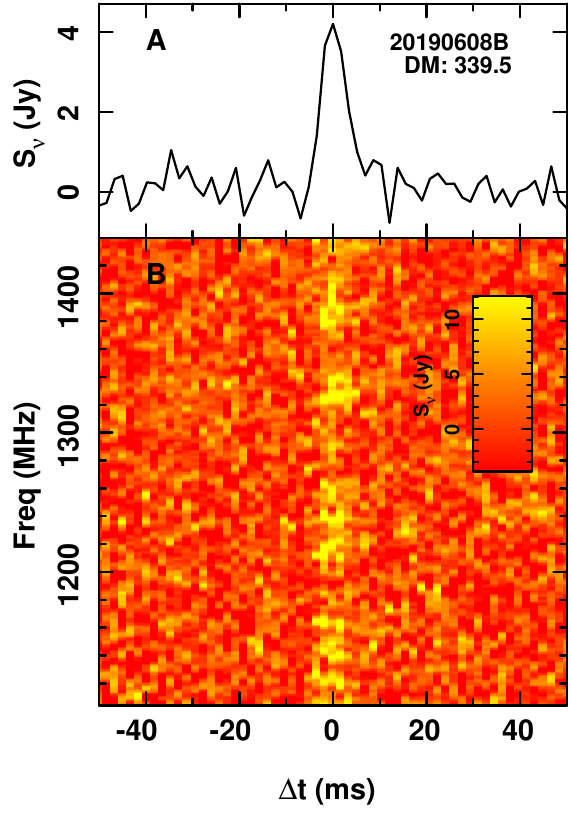} &
\includegraphics[width=2in]{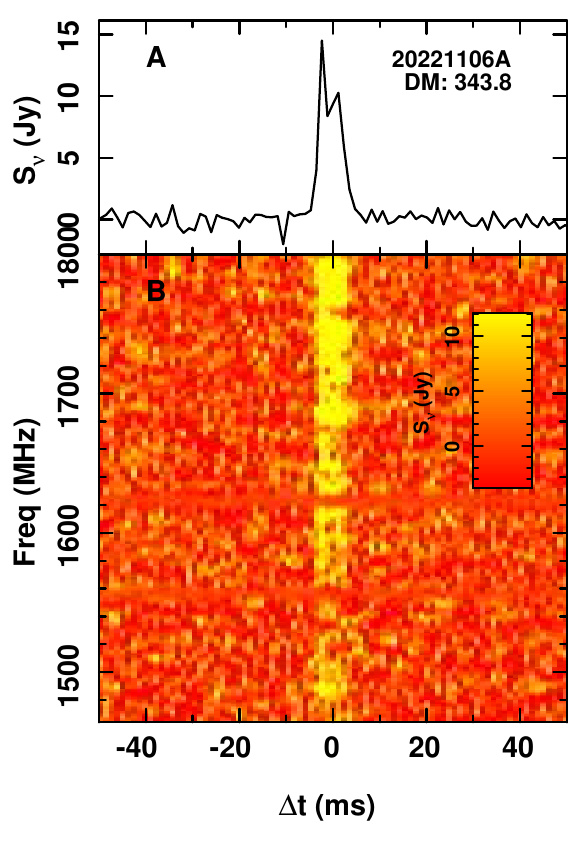}  \\

\includegraphics[width=2in]{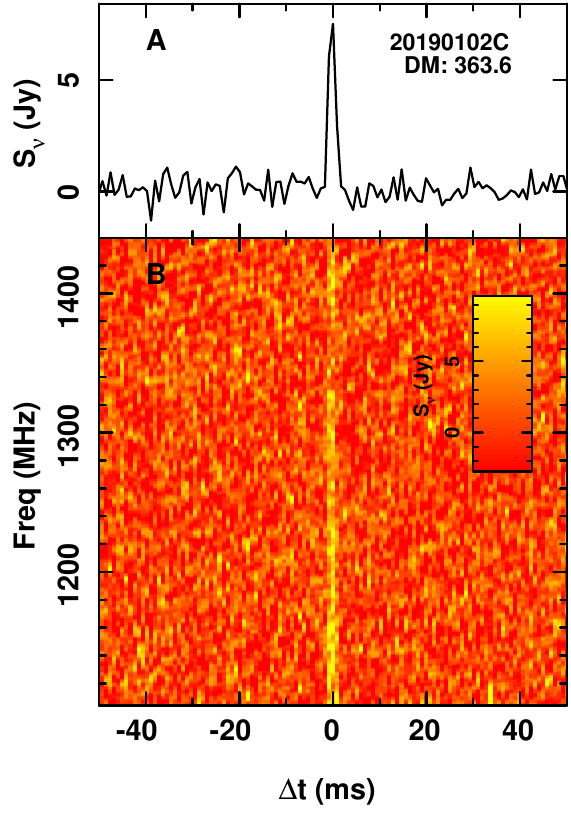} &
\includegraphics[width=2in]{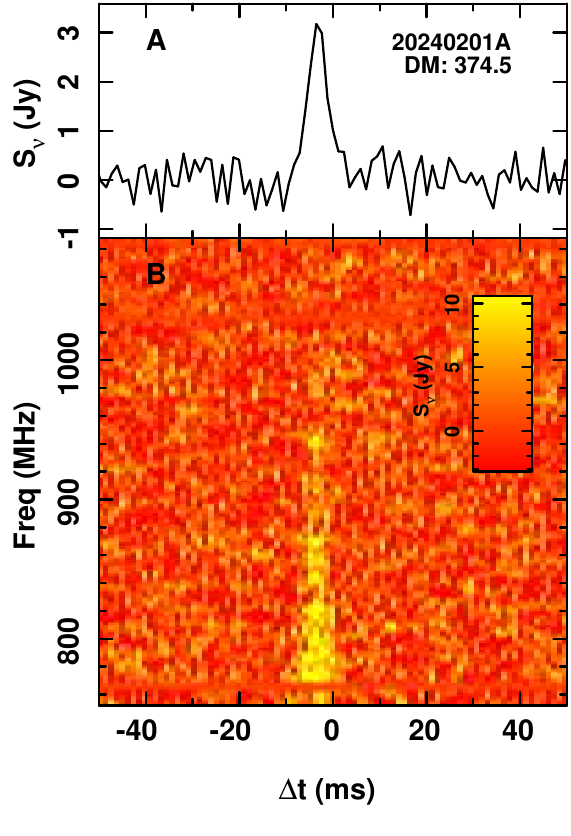}  &
\includegraphics[width=2in]{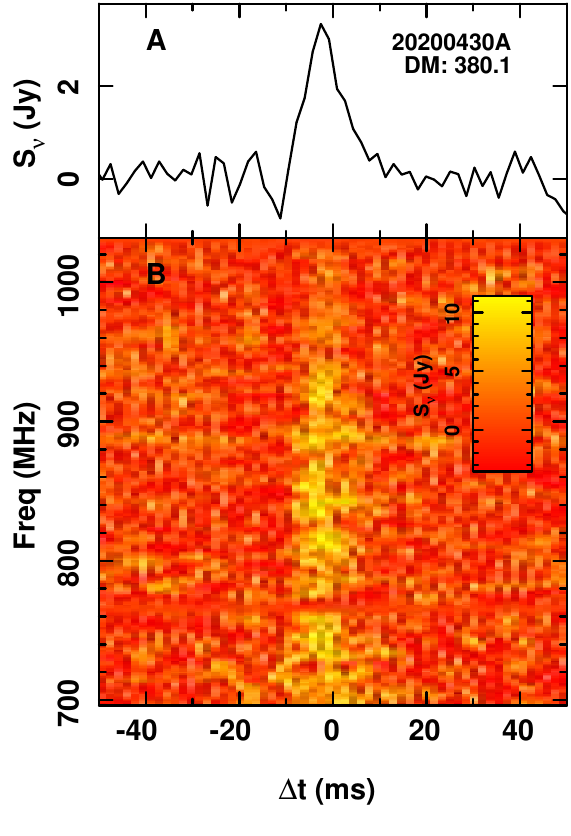}  \\

\end{tabular}
\caption{FRB dynamic spectra (continued). \label{fig:dynspec2}
}
\end{figure*}

\begin{figure*}
 \centering
 \begin{tabular}{ccc}

 \includegraphics[width=2in]{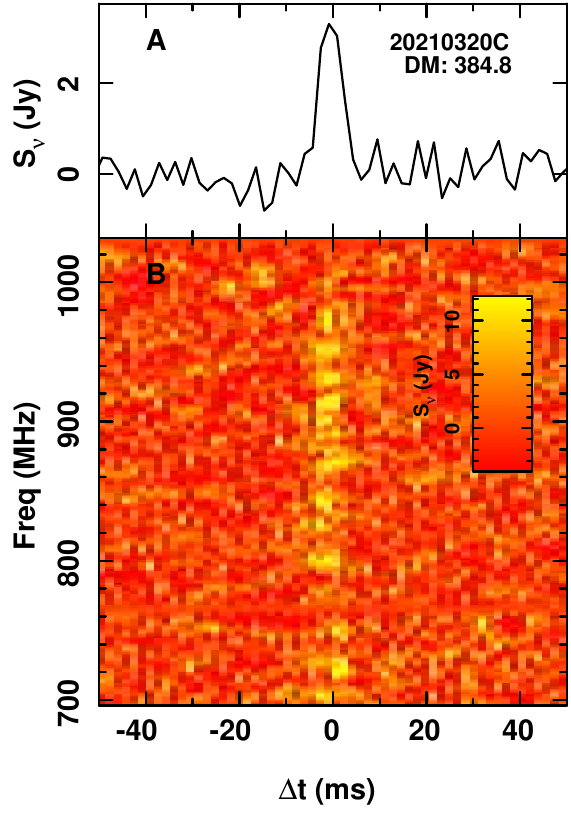} &
\includegraphics[width=2in]{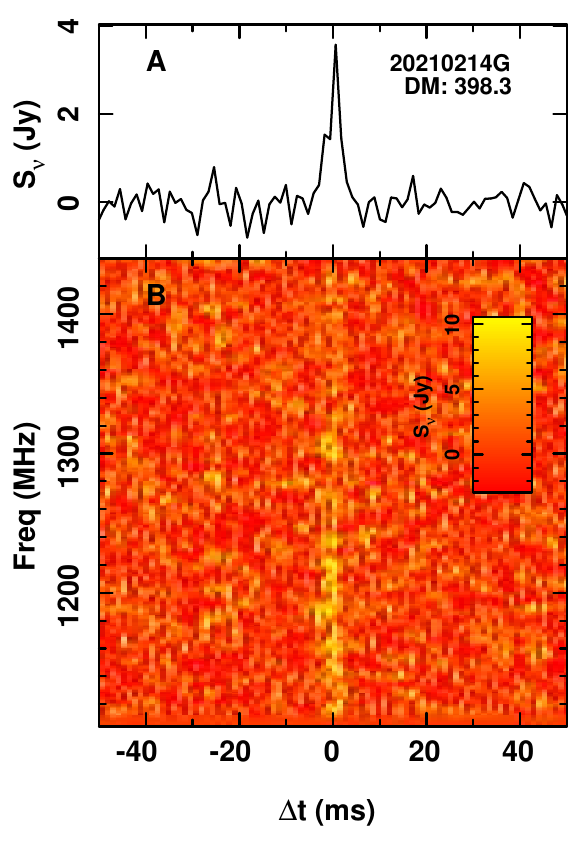} &
\includegraphics[width=2.1in]{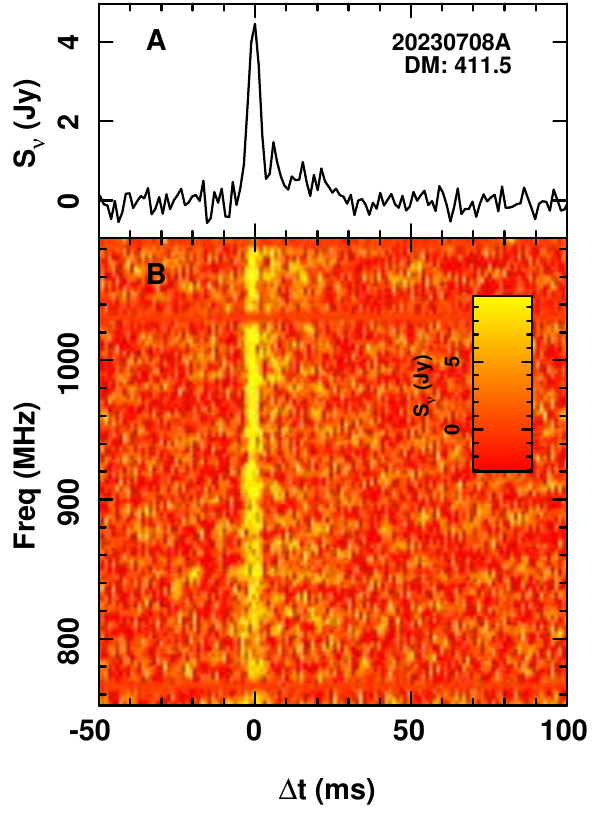} \\

\includegraphics[width=2in]{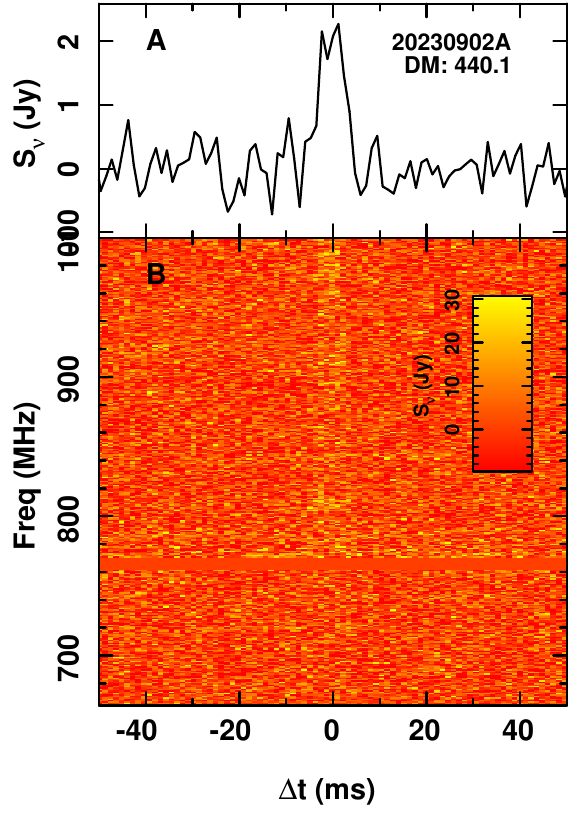} &
\includegraphics[width=2in]{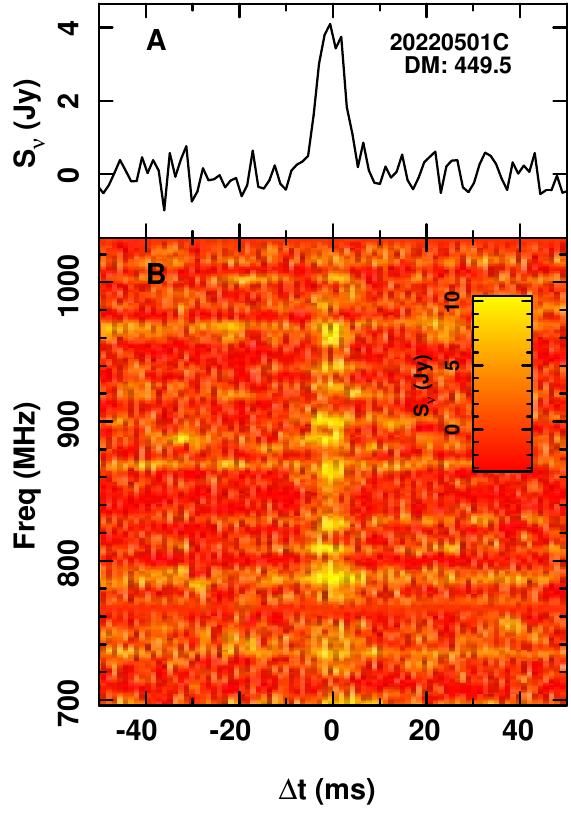} &
\includegraphics[width=2in]{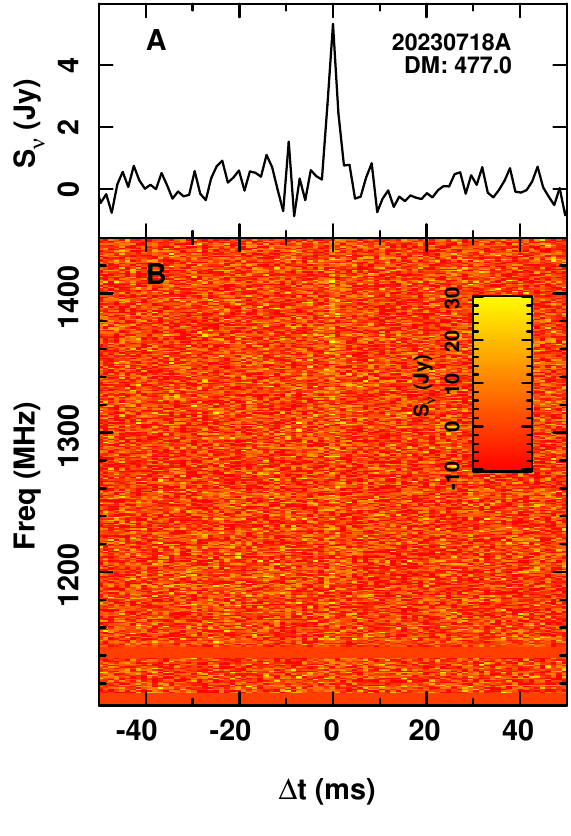} \\

\includegraphics[width=2in]{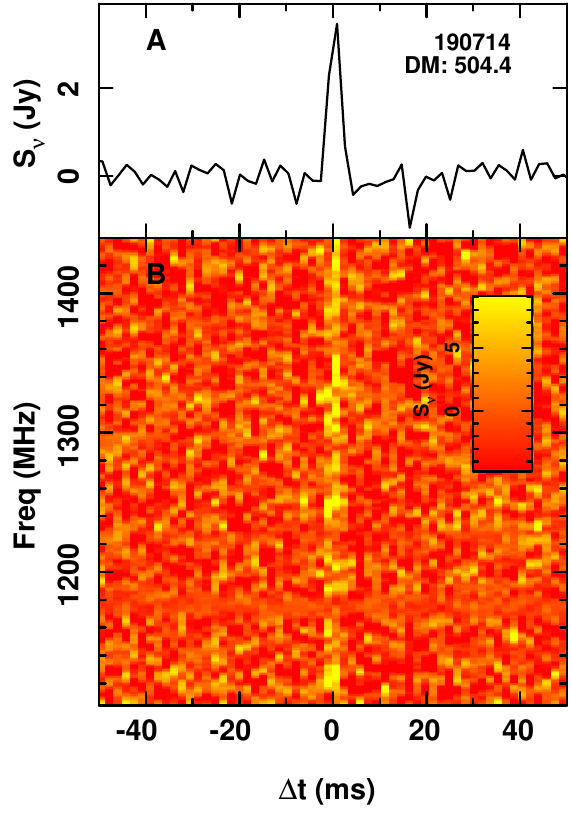} & 
\includegraphics[width=2in]{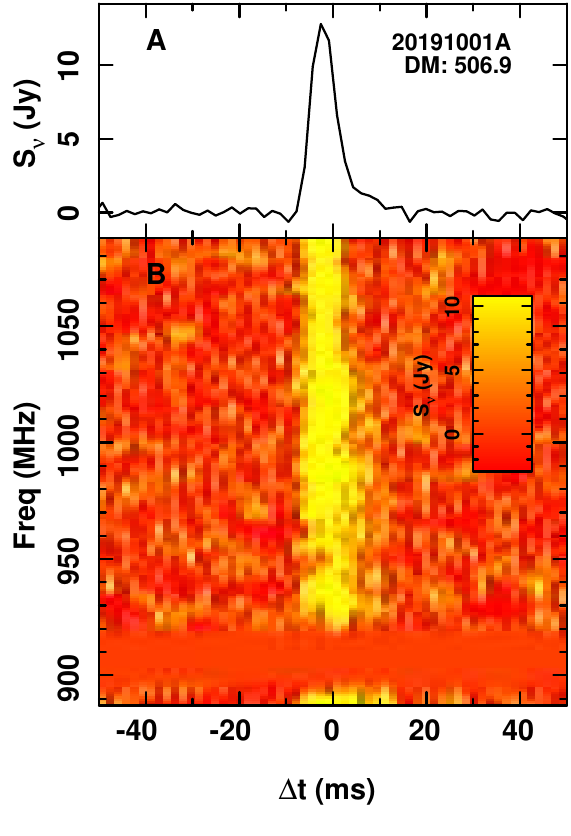} &
\includegraphics[width=2in]{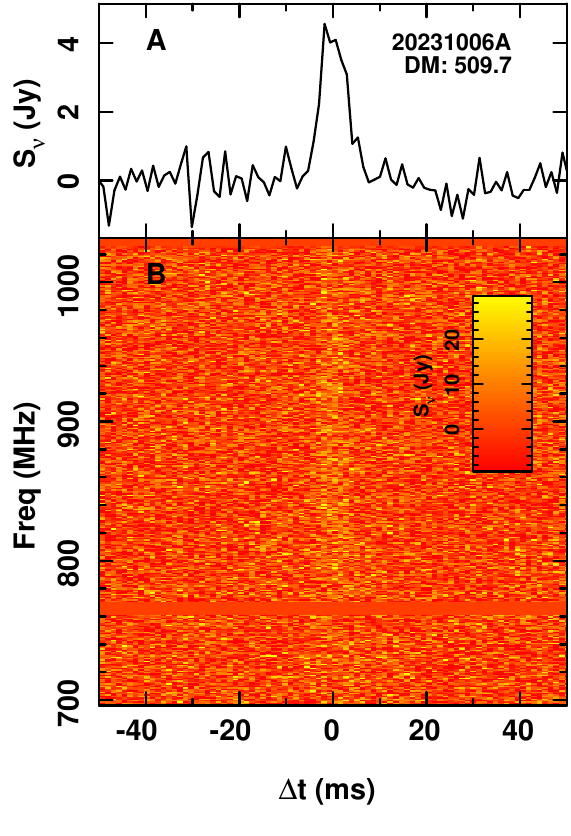}  \\
\end{tabular}
\caption{FRB dynamic spectra (continued). \label{fig:dynspec3}
}
\end{figure*}

\begin{figure*}
 \centering
 \begin{tabular}{ccc}

\includegraphics[width=2in]{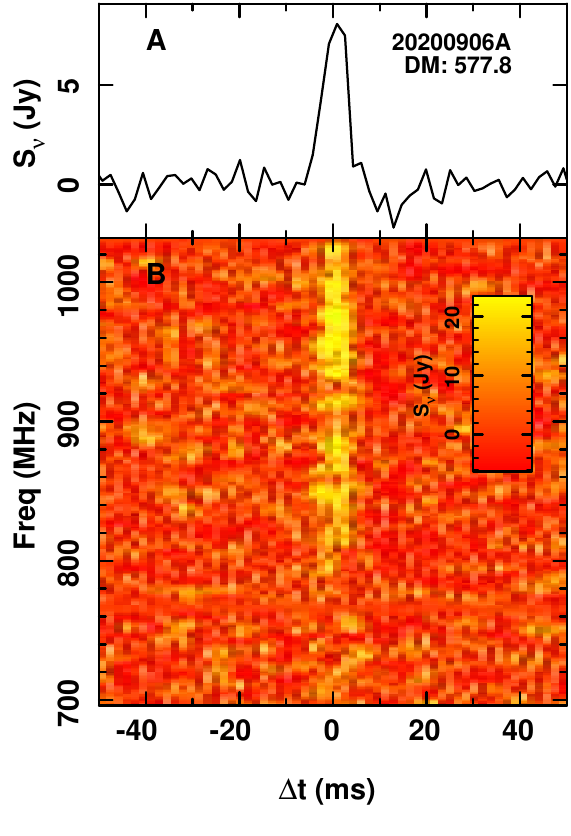} &
\includegraphics[width=2in]{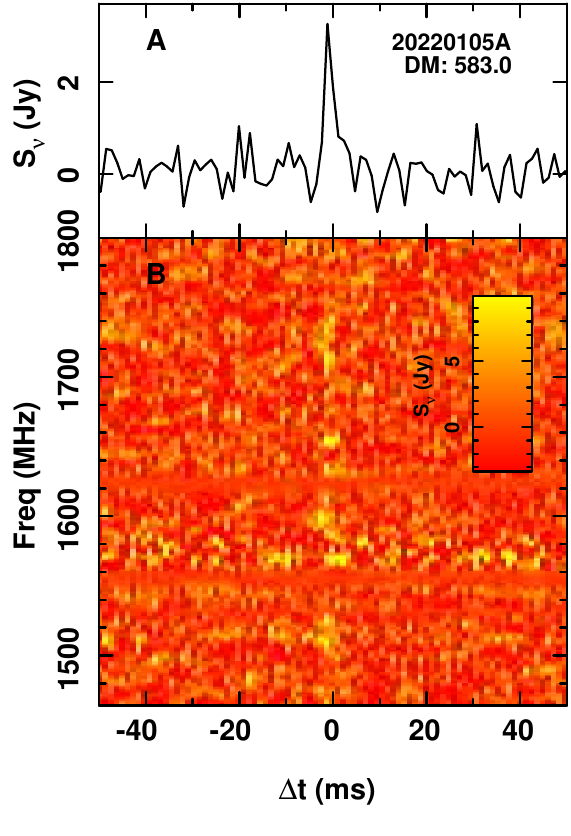} &
 \includegraphics[width=2in]{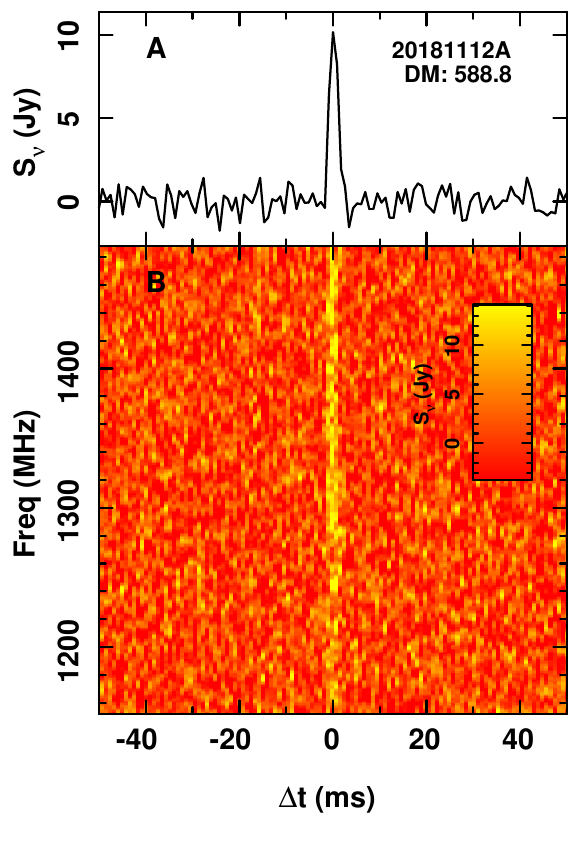} \\

\includegraphics[width=2in]{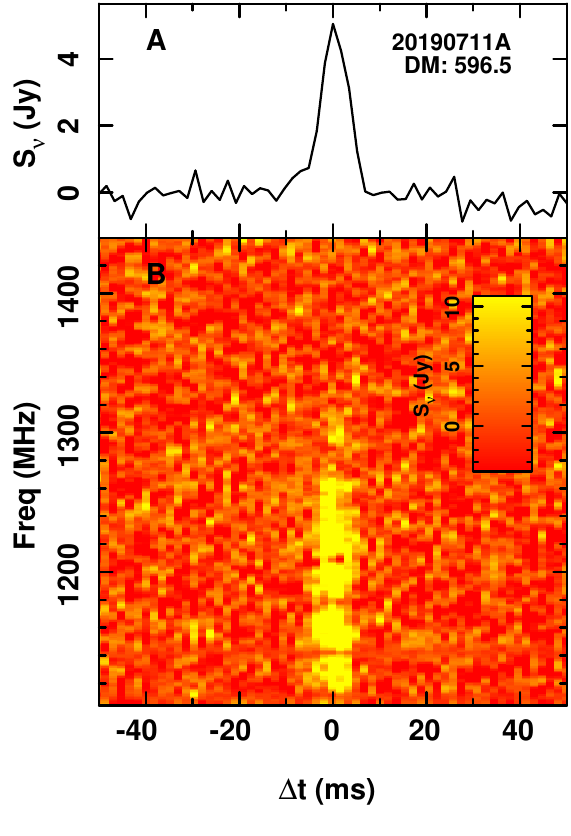} &
\includegraphics[width=2in]{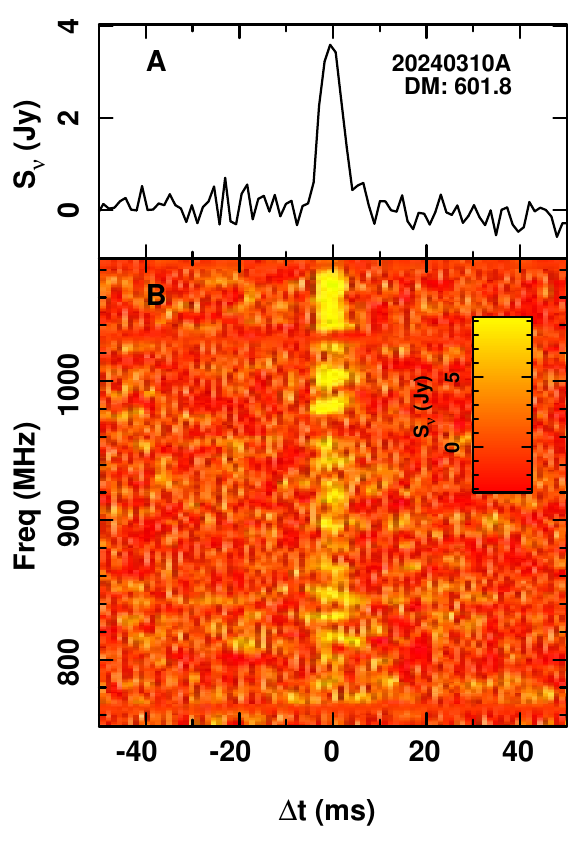} &
\includegraphics[width=2in]{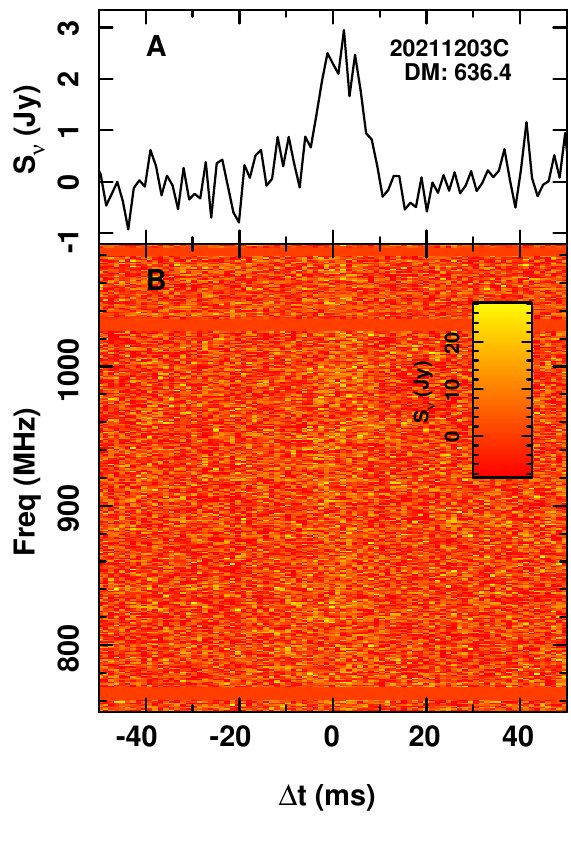} \\

\includegraphics[width=2in]{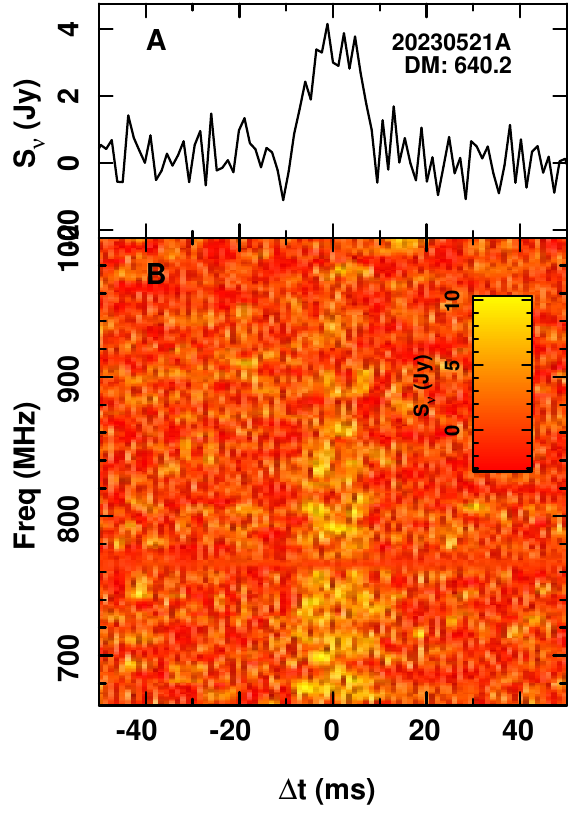} &
 \includegraphics[width=2in]{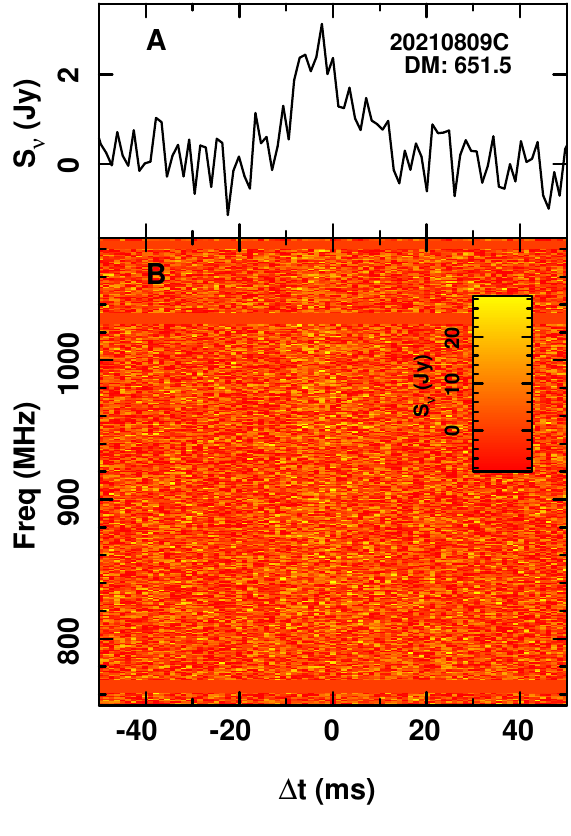} &
 \includegraphics[width=2in]{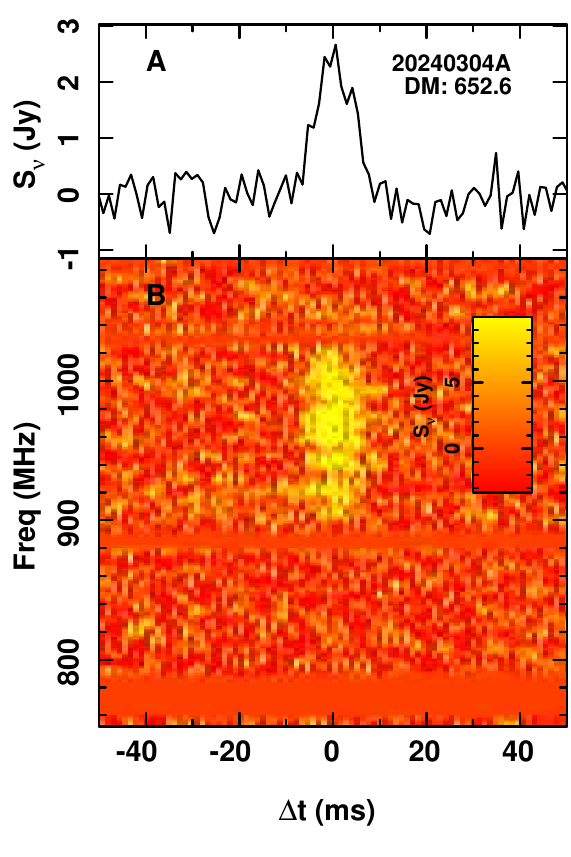}  \\

\end{tabular}
\caption{FRB dynamic spectra (continued). \label{fig:dynspec4}
}
\end{figure*}

\begin{figure*}
 \centering
 \begin{tabular}{ccc}

\includegraphics[width=2in]{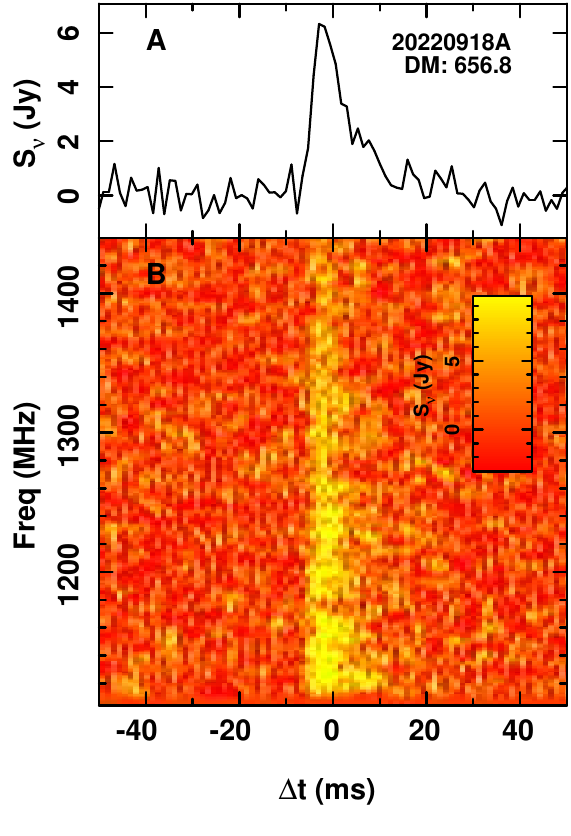} &
\includegraphics[width=2in]{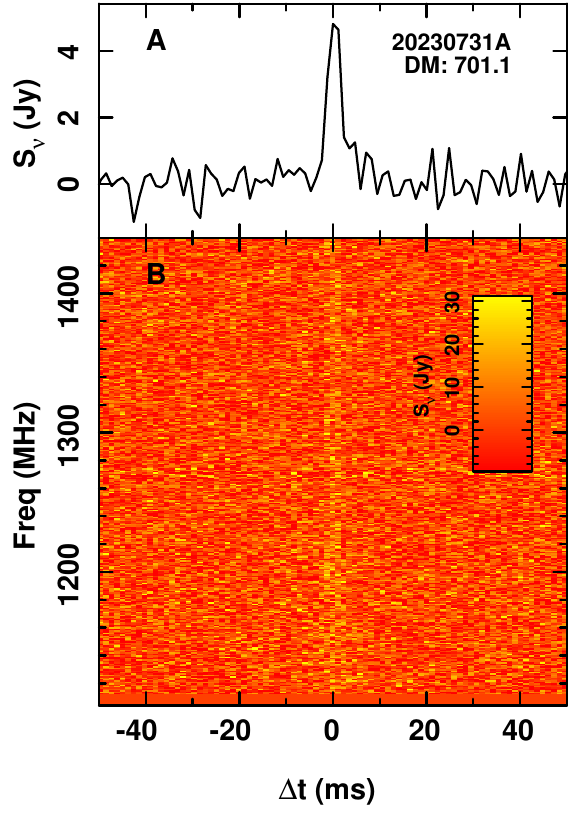} &
\includegraphics[width=2in]{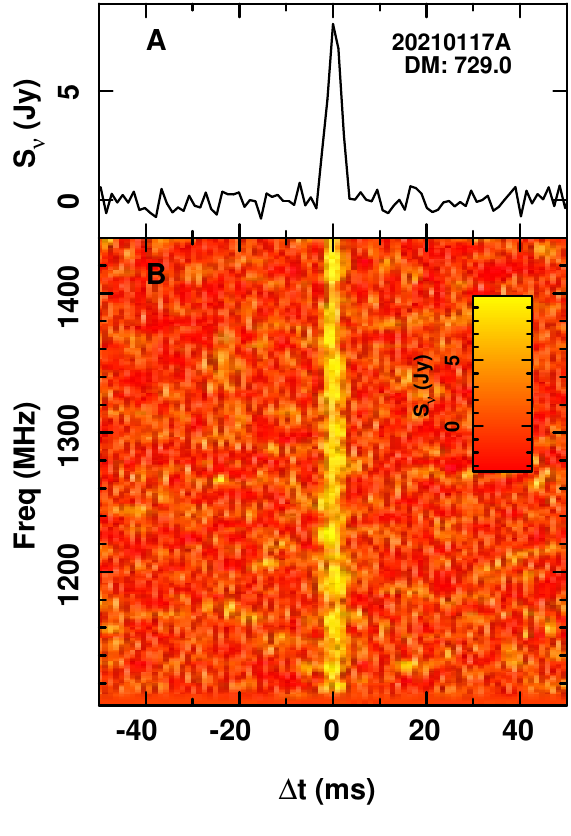}  \\

\includegraphics[width=2in]{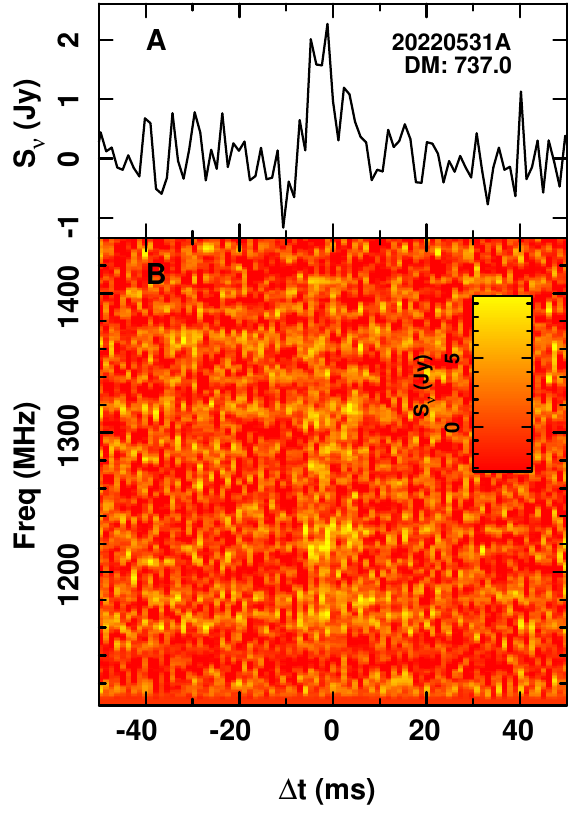} &
 \includegraphics[width=2in]{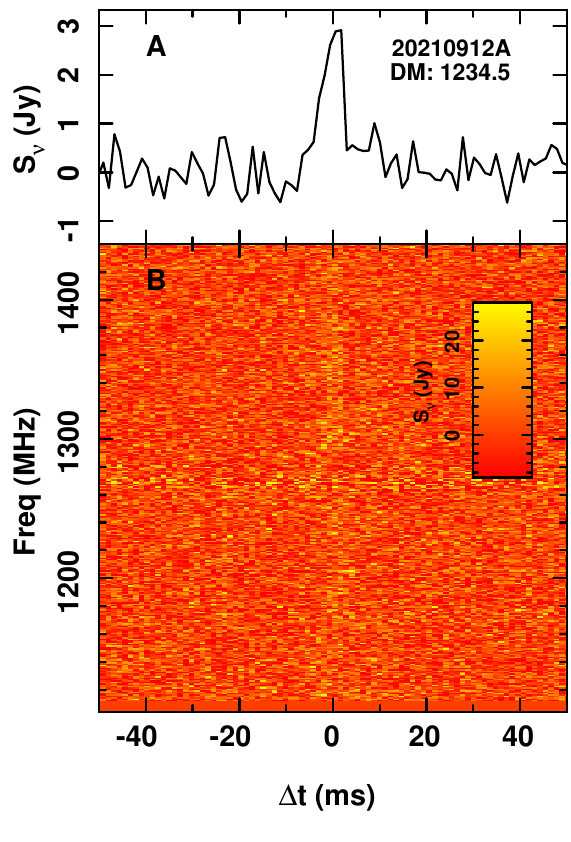} &
\includegraphics[width=2in]{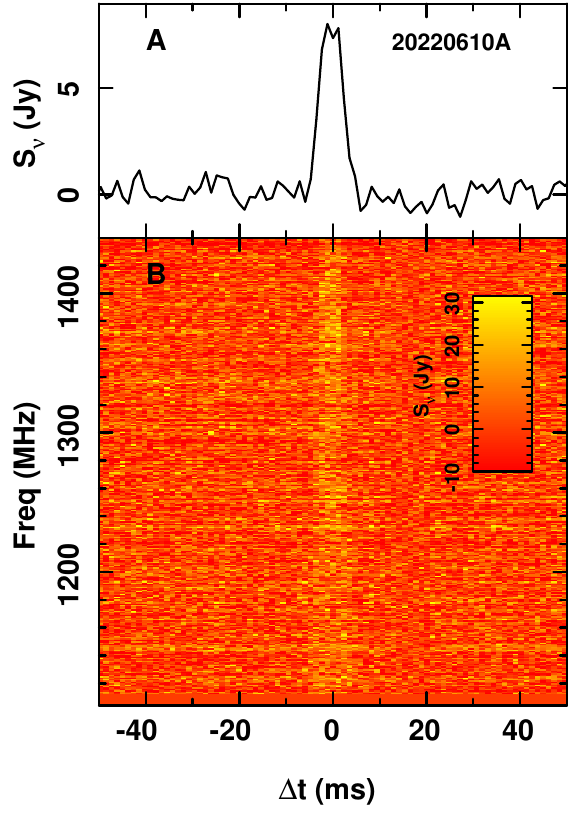}  \\

 &\includegraphics[width=2in]{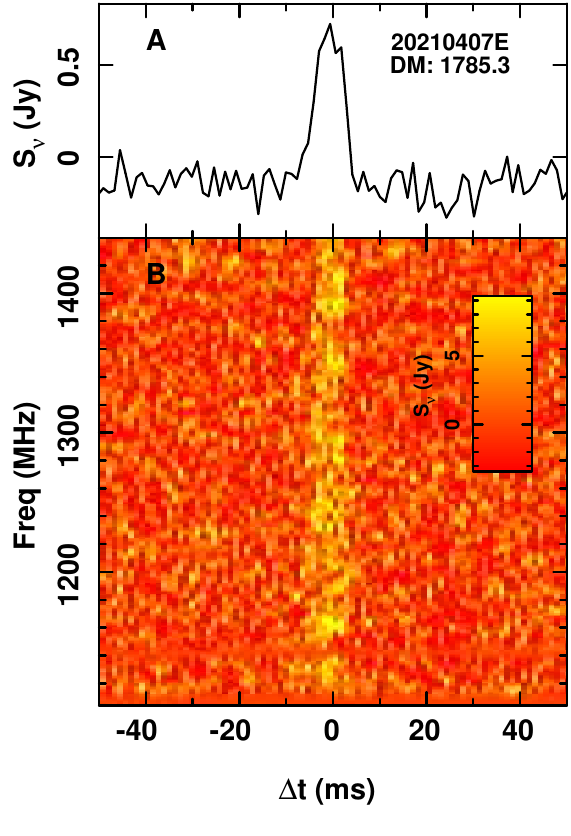}   &   \\
\end{tabular}
\caption{FRB dynamic spectra (continued). \label{fig:dynspec5}
}
\end{figure*}

Figures \ref{fig:dynspec1} to \ref{fig:dynspec5} show the integrated pulse profiles (panel A) and dedispersed dynamic spectra (panel B) of the FRBs discovered in the ICS survey. The bursts have been arranged in order of increasing dispersion measure.  FRB designations are listed in the uppper right corner of each plot.

\section{Host galaxy associations and optical photometry}

Tables \ref{tab:path} to \ref{tab:path4} present the PATH \cite[][]{2021ApJ...911...95A}  probabilities of host-galaxy associations for ASKAP-localised FRBs.

Table \ref{tab:photometry} presented photometry of the FRB host galaxies obtained with VLT/FORS2 and VLT/HAWK-I imaging.

\begin{table*}
\caption{FRB PATH Associations\label{tab:path}.  For each FRB we list the positions of nearby galaxies, their offset  $\theta$ from the burst, angular size  $\phi$, and magnitude. Using this information we have calculated PATH probabilities \PO and \POx.  We note that the automated photometry in the PATH analysis differs from final host galaxy photometry presented in Table \ref{tab:photometry}. We only report  galaxies with \POx $> 10^{-4}$. }
\begin{tabular}{ccccccc}
RA$_{\rm cand}$ & Dec$_{\rm cand}$ 
& $\theta$ 
& \halflight  
& mag 
& \PO & \POx 
\\(deg) & (deg) & ($''$) & ($''$)  
\\ 
\hline 
\multicolumn{7}{c}{FRB20180924B: P(U)=0.0,  P(U|x)=0.0000} \\ 
\hline 
326.1054 & $-40.9002$& 0.85& 1.31& 21.32& 0.8722& 0.9994\\ 
326.1042 & $-40.9002$& 2.91& 0.81& 24.27& 0.0684& 0.0006\\ 
\hline 
\multicolumn{7}{c}{FRB20181112A: P(U)=0.0,  P(U|x)=0.0000} \\ 
\hline 
327.3486 & $-52.9709$& 0.43& 0.67& 21.49& 0.0785& 0.9274\\ 
327.3496 & $-52.9696$& 5.45& 1.06& 19.10& 0.8642& 0.0724\\ 
327.3484 & $-52.9729$& 6.92& 0.57& 22.00& 0.0487& 0.0002\\ 
\hline 
\multicolumn{7}{c}{FRB20190102C: P(U)=0.0,  P(U|x)=0.0000} \\ 
\hline 
322.4149 & $-79.4756$& 0.55& 0.86& 20.73& 0.8426& 1.0000\\ 
\hline 
\multicolumn{7}{c}{FRB20190608B: P(U)=0.0,  P(U|x)=0.0000} \\ 
\hline 
334.0204 & $-7.8988$& 2.87& 1.66& 17.15& 0.9941& 1.0000\\ 
\hline 
\multicolumn{7}{c}{FRB20190611B: P(U)=0.0,  P(U|x)=0.0000} \\ 
\hline 
320.7429 & $-79.3973$& 2.05& 0.49& 22.35& 0.3329& 0.9799\\ 
320.7439 & $-79.3985$& 3.32& 0.26& 24.91& 0.0412& 0.0110\\ 
320.7496 & $-79.3972$& 3.02& 0.27& 25.93& 0.0198& 0.0087\\ 
320.7383 & $-79.3977$& 4.86& 0.35& 23.44& 0.1306& 0.0003\\ 
320.7539 & $-79.3979$& 5.65& 0.67& 23.63& 0.1115& 0.0001\\ 
\hline 
\multicolumn{7}{c}{FRB20190711A: P(U)=0.0,  P(U|x)=0.0000} \\ 
\hline 
329.4194 & $-80.3581$& 0.49& 0.46& 22.93& 0.4779& 0.9986\\ 
329.4187 & $-80.3586$& 2.03& 0.25& 24.88& 0.1009& 0.0014\\ 
\hline 
\multicolumn{7}{c}{FRB20190714A: P(U)=0.0,  P(U|x)=0.0000} \\ 
\hline 
183.9795 & $-13.0212$& 1.16& 0.94& 19.47& 0.7984& 1.0000\\ 
\hline 
\multicolumn{7}{c}{FRB20191001A: P(U)=0.0,  P(U|x)=0.0000} \\ 
\hline 
323.3519 & $-54.7485$& 1.18& 1.36& 17.82& 0.5082& 0.9995\\ 
323.3486 & $-54.7482$& 6.64& 1.43& 17.85& 0.4913& 0.0005\\ 
\hline 
\multicolumn{7}{c}{FRB20191228A: P(U)=0.0,  P(U|x)=0.0000} \\ 
\hline 
344.4307 & $-29.5940$& 0.91& 0.48& 21.92& 0.1473& 1.0000\\ 
\hline 
\multicolumn{7}{c}{FRB20200430A: P(U)=0.0,  P(U|x)=0.0000} \\ 
\hline 
229.7064 & $12.3766$& 0.37& 0.72& 21.18& 0.9389& 1.0000\\ 
\hline 
\multicolumn{7}{c}{FRB20200906A: P(U)=0.0,  P(U|x)=0.0000} \\ 
\hline 
53.4958 & $-14.0833$& 1.56& 1.51& 20.70& 0.8997& 1.0000\\ 
\hline 
\multicolumn{7}{c}{FRB20201124A: P(U)=0.0,  P(U|x)=0.0000} \\ 
\hline 
77.0145 & $26.0605$& 0.71& 0.94& 19.52& 1.0000& 1.0000\\ 
\hline 
\multicolumn{7}{c}{FRB20210117A: P(U)=0.0,  P(U|x)=0.0000} \\ 
\hline 
339.9795 & $-16.1515$& 0.94& 0.51& 22.95& 0.5786& 1.0000\\ 
\hline 
\multicolumn{7}{c}{FRB20210320C: P(U)=0.0,  P(U|x)=0.0000} \\ 
\hline 
204.4589 & $-16.1226$& 0.46& 1.02& 19.23& 0.9146& 0.9992\\ 
204.4580 & $-16.1233$& 3.64& 0.61& 22.13& 0.0529& 0.0008\\ 
\hline 
\multicolumn{7}{c}{FRB20210807D: P(U)=0.0,  P(U|x)=0.0000} \\ 
\hline 
299.2201 & $-0.7623$& 4.83& 2.32& 17.35& 0.7572& 1.0000\\ 
\hline 
\multicolumn{7}{c}{FRB20211127I: P(U)=0.0,  P(U|x)=0.0000} \\ 
\hline 
199.8082 & $-18.8379$& 2.16& 5.07& 15.38& 0.2842& 0.9998\\ 
199.8080 & $-18.8402$& 8.77& 2.28& 18.34& 0.0079& 0.0002\\ 
\hline 
\multicolumn{7}{c}{FRB20211203C: P(U)=0.0,  P(U|x)=0.0000} \\ 
\hline 
204.5626 & $-31.3801$& 0.57& 0.57& 20.29& 0.8243& 1.0000\\ 
\hline 
\end{tabular} 
\end{table*}

\begin{table*}
\caption{FRB PATH Associations (continued) \label{tab:path2}}
\begin{tabular}{ccccccc}
RA$_{\rm cand}$ & Dec$_{\rm cand}$ 
& $\theta$ 
& \halflight  
& mag 
& \PO & \POx 
\\(deg) & (deg) & ($''$) & ($''$)  
\\ 
\hline 
\multicolumn{7}{c}{FRB20211212A: P(U)=0.0,  P(U|x)=0.0000} \\ 
\hline 
157.3509 & $1.3608$& 1.46& 2.72& 16.21& 1.0000& 1.0000\\ 
\hline 
\multicolumn{7}{c}{FRB20220105A: P(U)=0.0,  P(U|x)=0.0000} \\ 
\hline 
208.8038 & $22.4665$& 1.81& 0.90& 21.53& 0.5694& 1.0000\\ 
\hline 
\multicolumn{7}{c}{FRB20220501C: P(U)=0.0,  P(U|x)=0.0000} \\ 
\hline 
352.3792 & $-32.4907$& 0.17& 0.90& 20.57& 1.0000& 1.0000\\ 
\hline 
\multicolumn{7}{c}{FRB20220610A: P(U)=0.0,  P(U|x)=0.0000} \\ 
\hline 
351.0735 & $-33.5137$& 0.73& 0.99& 23.99& 0.2812& 1.0000\\ 
\hline 
\multicolumn{7}{c}{FRB20220725A: P(U)=0.0,  P(U|x)=0.0000} \\ 
\hline 
353.3154 & $-35.9903$& 0.46& 1.77& 17.83& 0.9977& 1.0000\\ 
\hline 
\multicolumn{7}{c}{FRB20220918A: P(U)=0.0,  P(U|x)=0.0000} \\ 
\hline 
17.5917 & $-70.8114$& 0.49& 0.45& 23.60& 0.0770& 0.9965\\ 
17.5901 & $-70.8119$& 3.02& 0.56& 25.34& 0.0201& 0.0035\\ 
\hline 
\multicolumn{7}{c}{FRB20221106A: P(U)=0.0,  P(U|x)=0.0000} \\ 
\hline 
56.7045 & $-25.5696$& 1.33& 2.50& 18.34& 0.9446& 0.9708\\ 
56.7057 & $-25.5701$& 3.24& 1.07& 21.07& 0.0554& 0.0292\\ 
\hline 
\multicolumn{7}{c}{FRB20230526A: P(U)=0.0,  P(U|x)=0.0000} \\ 
\hline 
22.2326 & $-52.7175$& 0.51& 0.78& 21.15& 0.5445& 0.9970\\ 
22.2311 & $-52.7186$& 5.52& 0.95& 21.49& 0.3952& 0.0030\\ 
\hline 
\multicolumn{7}{c}{FRB20230708A: P(U)=0.0,  P(U|x)=-0.0000} \\ 
\hline 
303.1155 & $-55.3563$& 0.14& 0.58& 22.73& 0.0873& 1.0000\\ 
\hline 
\multicolumn{7}{c}{FRB20230902A: P(U)=0.0,  P(U|x)=0.0000} \\ 
\hline 
52.1400 & $-47.3335$& 0.54& 0.68& 21.52& 0.6517& 1.0000\\ 
\hline 
\multicolumn{7}{c}{FRB20231226A: P(U)=0.0,  P(U|x)=0.0000} \\ 
\hline 
155.3639 & $6.1097$& 1.99& 1.79& 19.01& 0.9394& 1.0000\\ 
\hline 
\multicolumn{7}{c}{FRB20240201A: P(U)=0.0,  P(U|x)=0.0000} \\ 
\hline 
149.9072 & $14.0873$& 6.25& 0.72& 26.17& 1.0000& 1.0000\\ 
\hline 
\multicolumn{7}{c}{FRB20240210A: P(U)=0.0,  P(U|x)=0.0000} \\ 
\hline 
8.7770 & $-28.2721$& 9.42& 6.42& 15.13& 0.8425& 1.0000\\ 
\hline 
\multicolumn{7}{c}{FRB20240304A: P(U)=0.0,  P(U|x)=0.0000} \\ 
\hline 
136.3305 & $-16.1662$& 1.84& 0.97& 21.08& 1.0000& 1.0000\\ 
\hline 
\multicolumn{7}{c}{FRB20240310A: P(U)=0.0,  P(U|x)=0.0000} \\ 
\hline 
17.6219 & $-44.4393$& 0.39& 1.06& 20.16& 0.7637& 0.9884\\ 
17.6228 & $-44.4387$& 3.65& 0.93& 21.80& 0.1565& 0.0116\\ 
\hline 
\end{tabular} 
\end{table*} 

\begin{table*}
\caption{Integrated optical and near-IR photometry of ASKAP/CRAFT host galaxies from VLT/FORS2 and VLT/HAWK-I imaging. The measurements are not corrected for Galactic extinction. \label{tab:photometry}}
\begin{tabular}{c|ccccc|ccc}
\hline
\multicolumn{1}{c|}{} & \multicolumn{5}{c|}{FORS2} & \multicolumn{3}{c}{HAWK-I} \\
FRB & $u$ & $g$ & $R$ & $I$ & $z$ & $J$ & $H$ & $K_\mathrm{s}$\\ \hline 
FRB\,20180924B & -- & $21.231(6)$ & -- & $19.875(5)$ & -- & -- & -- & -- \\
FRB\,20181112A & -- & $22.342(9)$ & -- & $21.17(1)$ & -- & -- & -- & -- \\
FRB\,20190102C & $24.0(2)$ & $22.59(1)$ & -- & $21.10(2)$ & $20.83(8)$ & -- & -- & -- \\
FRB\,20190608B & -- & $18.167(5)$ & -- & $17.097(6)$ & -- & -- & -- & -- \\
FRB\,20190611B & -- & $24.02(3)$ & $23.03(2)$ & $22.41(5)$ & -- & -- & -- & -- \\
FRB\,20190711A & -- & $23.87(4)$ & -- & $22.4(1)$ & -- & -- & -- & -- \\
FRB\,20190714A & -- & $21.037(6)$ & -- & $19.618(9)$ & -- & -- & -- & -- \\
FRB\,20191001A & -- & $19.103(4)$ & -- & $17.743(6)$ & -- & -- & -- & -- \\
FRB\,20191228A & -- & $> 23.3$ & -- & $21.90(7)$ & -- & -- & -- & -- \\
FRB\,20200430A & -- & $21.856(8)$ & -- & $20.61(2)$ & -- & -- & -- & -- \\
FRB\,20200906A & -- & $20.910(7)$ & -- & $19.564(7)$ & -- & -- & -- & -- \\
FRB\,20210117A & -- & $23.86(2)$ & -- & $22.68(6)$ & -- & $22.7(1)$ & $> 22.4$ & $> 22.4$ \\
FRB\,20210320C & -- & $20.476(4)$ & -- & $19.194(5)$ & -- & -- & -- & -- \\
FRB\,20210807D & -- & $18.128(1)$ & -- & $16.476(4)$ & -- & -- & -- & -- \\
FRB\,20211127I & -- & $15.819(6)$ & -- & $14.860(5)$ & -- & -- & -- & -- \\
FRB\,20211203C & -- & $20.842(4)$ & $20.258(4)$ & -- & -- & -- & -- & -- \\
FRB\,20211212A & -- & $17.184(6)$ & -- & $16.212(5)$ & -- & -- & -- & -- \\
FRB\,20220105A & -- & -- & $21.317(5)$ & -- & -- & -- & -- & $19.06(8)$ \\
FRB\,20220501C & -- & $21.49(1)$ & -- & $20.47(1)$ & -- & -- & -- & -- \\
FRB\,20220610A & -- & $24.22(5)$ & $23.72(4)$ & -- & -- & $22.12(8)$ & -- & $22.0(1)$ \\
FRB\,20220725A & -- & $18.529(6)$ & $17.843(4)$ & $17.232(4)$ & -- & -- & -- & $16.26(5)$ \\
FRB\,20220918A & -- & -- & $23.63(2)$ & -- & -- & -- & -- & $22.16(7)$ \\
FRB\,20221106A & -- & -- & $18.351(9)$ & -- & -- & -- & -- & $16.40(2)$ \\
FRB\,20230526A & -- & -- & $21.08(1)$ & -- & -- & -- & -- & $17.1(2)$ \\
FRB\,20230708A & -- & -- & $22.65(2)$ & -- & -- & -- & -- & $18.8(1)$ \\
FRB\,20230902A & -- & -- & $21.522(6)$ & -- & -- & -- & -- & $20.36(4)$ \\
FRB\,20231226A & -- & -- & $18.990(6)$ & -- & -- & -- & -- & $18.08(2)$ \\
FRB\,20240201A & -- & -- & $16.97(1)$ & -- & -- & -- & -- & $15.41(5)$ \\
FRB\,20240210A & -- & -- & $14.919(9)$ & -- & -- & -- & -- & $13.82(4)$ \\
FRB\,20240304A & -- & -- & $20.723(7)$ & -- & -- & -- & -- & $20.01(5)$ \\
FRB\,20240310A & -- & -- & $20.143(5)$ & -- & -- & -- & -- & $18.51(7)$ \\
\hline
\end{tabular}
\end{table*}

\end{appendix}



\end{document}